\documentclass[b5paper,10pt,openany]{book}

\usepackage{fancyhdr}
\usepackage{amsmath}
\usepackage{amsthm}
\usepackage{amssymb}
\usepackage{setspace}
\usepackage{graphics}
\usepackage[magyar,english]{babel}
\usepackage{t1enc}
\usepackage[latin2]{inputenc}
\usepackage{boxedminipage}

\newcommand{\bra}[1]{\langle#1\vert}
\newcommand{\ket}[1]{\vert#1\rangle}
\newtheorem{theorem}{Theorem}
    
\newcommand{\clearemptydoublepage}{\newpage{\pagestyle{empty}\cleardoublepage}}

\begin{document}

\pagenumbering{roman}
\pagestyle{empty}
\begin{titlepage}
  \begin{center}
    \vspace*{0.5cm}
    {\Large \bf DEBRECENI EGYETEM}\\
    \vspace*{2mm}
    \begin{center}
    \resizebox{6cm}{!}{
    \includegraphics{pics/de2.epsi}}
    \end{center}
    {\Large \bf TERM\'ESZETTUDOM\'ANYI KAR}\\
    \vspace*{2.5cm}
    {\LARGE \bf Continued fraction representation of quantum mechanical 
    Green's operators}\\
    \vspace*{2cm}
    {\Large Ph.D. thesis} \\
    \vspace*{1cm}
    {\Large by}\\~\\
    {\LARGE Bal\'azs K\'onya}\\
    \vspace*{3cm}
    {{\Large University of Debrecen}\\ 
    {\Large Faculty of Sciences}\\
    {\Large Debrecen, 2000}}    
  \end{center}
\end{titlepage}
\clearemptydoublepage
\newpage
\vspace*{-1cm}
\begin{small}
\thispagestyle{empty}
\begin{center}
  {\large \bf Continued fraction representation of quantum mechanical 
  Green's operators}\\ 
  \vspace{0.7cm}
  \'Ertekez\'es a doktori (Ph.D.) fokozat megszerz\'ese \'erdek\'eben\\
  a fizika tudom\'any\'aban.\\
\end{center} 
\begin{flushleft}  
  \'Irta: K\'onya Bal\'azs okleveles fizikus.\\ 
  K\'esz\"ult a Debreceni Egyetem Term\'eszettudom\'anyi Kar\'anak \\
  fizika doktori programja
  (magfizika alprogramja) keret\'eben. \\ ~\\
\end{flushleft}
\begin{flushleft}
  T\'emavezet\H{o}: Dr. ...................................\\
  ~\\ \hspace*{2.845cm}
  Elfogad\'asra javaslom: 2000 ........................ ......\\
  \vspace*{0.3cm}
  Az \'ertekez\'est b\'{\i}r\'al\'ok\'ent elfogad\'asra javaslom:\\
  ~\\ \hspace*{4cm}
  Dr. ...................................  ...................\\
  ~\\ \hspace*{4cm}
  Dr. ...................................  ...................\\
  ~\\ \hspace*{4cm}
  Dr. ...................................  ...................\\
  ~\\
  Jel\"olt az \'ertekez\'est 2000 ........................ ......-n 
  sikeresen megv\'edte:\\ 
  ~\\
  A b\'{\i}r\'al\'obizots\'ag eln\"oke:\hspace*{0.324cm}
  Dr. ...................................  ...................\\
  \vspace*{0.3cm}
  A b\'{\i}r\'al\'obizots\'ag tagjai:\\
  ~\\ \hspace*{4cm}
  Dr. ...................................  ...................\\
  ~\\ \hspace*{4cm}
  Dr. ...................................  ...................\\
  ~\\ \hspace*{4cm}
  Dr. ...................................  ...................\\
  ~\\ \hspace*{4cm}
  Dr. ...................................  ...................
\end{flushleft}
\vspace*{\stretch{2}}
\begin{center}
  Debrecen -- 2000
\end{center}
\end{small}

\newpage
\begin{small}
\thispagestyle{empty}
\newlength{\myparindent}
\setlength{\myparindent}{\parindent}
\setlength{\parindent}{0cm}

\begin{minipage}[t]{126mm}
Ez a dolgozat a Magyar Tudom\'anyos Akad\'emia Atommagkutat\'o 
Int\'ezete Elm\'eleti Fizika Oszt\'aly\'an k\'esz\"ult 2000-ben.
A dolgozat alapj\'aul szolg\'al\'o 
eredm\'enyek illetve tudom\'anyos k\"ozlem\'enyek 
a MTA ATOMKI Elm\'eleti Fizika Oszt\'aly\'an \'es a 
Karl--Franzens Universit\"at Graz Elm\'eleti Fizika Int\'ezet\'eben
sz\"ulettek 1997. \'es 2000. k\"oz\"ott.
\end{minipage}

\vspace*{2cm}
\begin{boxedminipage}[t]{126mm}
Ezen \'ertekez\'est a Debreceni Egyetem fizika doktori program 
magfizika al\-prog\-ram\-ja keret\'eben k\'esz\'{\i}tettem 1997. 
\'es 2000. k\"oz\"ott \'es ez\'uton beny\'ujtom a Debreceni Egyetem doktori 
Ph.D. fokozat\'anak elnyer\'ese c\'elj\'ab\'ol.

\vspace*{0.7cm}
Debrecen, 2000. \'aprilis 20.
\begin{flushright}
  \parbox{5cm}{
    \centering{K\'onya Bal\'azs}
    }\hspace*{1cm}
\end{flushright}
\end{boxedminipage}

\vspace*{3cm}

\begin{boxedminipage}[t]{126mm}
Tan\'us\'{\i}tom, hogy K\'onya Bal\'azs 
doktorjel\"olt 1997. \'es 2000. k\"oz\"ott
a fent megnevezett doktori alprogram keret\'eben ir\'any\'{\i}t\'asommal 
v\'egezte munk\'aj\'at. Az \'ertekez\'esben foglaltak a jel\"olt \"on\'all\'o
munk\'aj\'an alapulnak, az eredm\'enyekhez \"on\'all\'o alkot\'o 
tev\'ekenys\'eg\'evel meghat\'aroz\'oan hozz\'aj\'arult. Az \'ertekez\'est 
elfogad\'asra javaslom.

\vspace*{0.7cm}
Debrecen, 2000. \'aprilis 20.
\begin{flushright}
  \parbox{5cm}{
    \centering{Dr. Papp Zolt\'an\\
      t\'emavezet\H{o}}
    }\hspace*{1cm}
\end{flushright}
\end{boxedminipage}

\vspace*{\stretch{2}}
\begin{center}
  Debrecen -- 2000
\end{center}
\end{small}

\pagestyle{fancy}
\setcounter{page}{0}
\chapter*{Preface}
\addcontentsline{toc}{chapter}{\numberline{}Preface}
\onehalfspacing

This thesis contains the summary of my research work carried out as a Ph.D.
student at the Theoretical Physics Department of the Institute of Nuclear
Research of the Hungarian Academy of Sciences Debrecen, Hungary and partly
at 
the Theoretical Physics Institute of the Karl-Franzens Universit\"at
Graz within the framework of the fruitful collaboration between the few-body
research groups of Debrecen and Graz. The new results underlying this thesis
have already been presented at international scientific meetings and  published
in four papers appeared in Journal of Mathematical Physics and Physical Review
C. \cite{jacobi_jmp,gezapot_jmp,dirac_jmp,continued_prc}. 

As the title ``Continued fraction representation of quantum mechanical Green's
operators'' implies my research work is concerned with one of the central
concepts of quantum mechanical few-body problems. The exploitation of the
richness of the mathematical theory of  continued fractions has enabled us to
develop a rather general method for evaluating an analytic and
readily computable
representation of Green's operators. This effective representation 
facilitates the solution of  fundamental few-body integral equations.  

Being a theoretical physicist I have always been interested in
mathematics. Therefore as a second year undergraduate physics student I
was enthusiastic to accept the research task on two-point Padé approximants
offered by my present supervisor. Later under his supervision I finished my
diploma work on the second order Dirac equation. By the time 
of my PhD studies I
got completely infected with mathematical physics 
and my attention was 
attracted to the topic of continued fractions and quantum mechanics.
During my work I benefitted a lot from my supervisor's knowledge and experience,
and
I could identify myself with many of his thoughts. He showed me
how mathematics could be used in practice approaching real physical problems.
I hope my thesis may serve as an example for  how
physics has always profited from mathematics.

Here I would like to take the occasion and thank to everybody who helped me
during the time I did my research and wrote my thesis.

Special thanks go to Prof. Zolt\'an Papp my PhD 
supervisor for his guidance and
useful advises and for the excellent 
atmosphere in which we have been working 
together. 

Some of the work covered by this thesis was done and the results were 
published together with Dr.~G.~L\'evai.  I thank him  for the efforts he
invested in our common topic. This collaboration was a very instructive
experience for me, I could learn from him how one could  
always be optimistic even
in the most hopeless moments.   

I am grateful to the Theoretical Physics Department of ATOMKI 
for providing me a
peaceful and pleasant working environment.
I thank Prof. Borbála Gyarmati, who read the manuscript and made numerous
helpful suggestions. 
I am indebted to Prof. R.~G.~Lovas who 
made the completion of my thesis possible by offering a young research fellow
scholarship. 

I am also grateful to the few-body  group of the Theoretical Physics 
Institute of the Karl-Franzens Universit\"at Graz,  especially to Prof. 
W. Plessas, for the vivid scientific atmosphere.

Last but not least I wish to thank all the Professors of the Tuesday and
Thursday 4PM
open-air seminars for their brilliant lectures on tactics, sport diplomacy and
football.
 
\tableofcontents 
\chapter{Introduction}
\pagenumbering{arabic}

\onehalfspacing

The theoretical description of the microscopic world can be approached following
two ``orthogonal'' paths. According to the many-body or field theoretical
approach the microscopic world is considered as an assembly of many or
infinitely many interacting objects, where field theoretical or statistical
methods can be applied in order to describe the  system \cite{fewa,kaku,mandl}.
On the contrary,
few-body physicist tackle physical systems possessing only 
few degrees of freedom being
consisted of a couple of interacting particles and  intend 
 to provide a physically 
complete and mathematically well formulated description.

Few-body systems have
played a crucial role in the development of our understanding of microscopic
world: atomic, nuclear and particle physics heavily rely  on few-body models.
Nowadays however, few-body problem has  an ambiguous reputation of being
a jungle where non-experts are quickly discouraged and specialist enjoy endless
debates on technical improvements concerning equations and mathematical physics
issues. Furthermore  much attention has been 
paid only to computational consequences 
and  little concern about the underlying physics. This is of course a false
impression which originates from the non-trivial nature of the problems studied.
The goal of the few-body physics community, namely giving a complete and
mathematically correct
description of few-body systems, automaticly requires the 
consideration of  mathematical issues, hence this field 
offers an excellent playground for mathematical physics. 
This thesis, following the above ideas, hopes
to  contribute to the magnificent results
achieved by the few-body approach.

Microscopic few-body physics certainly was born together with quantum 
mechanics in the pioneering work of Bohr, Heisenberg and Schr{\"o}dinger trying to
describe simple quantum systems, like the Hydrogen atom.
The field initially developed as a part of nuclear physics as the title of the
first few-body conference (Nuclear forces and the few-nucleon problem 
\cite{first_conf} held in London 1959) suggests.      
Since then few-body physics has expanded to incorporate atomic,
molecular and quark
systems. For example, the two-electron atom had already been attacked by
Hylleraas \cite{hyll} in 1929   and
conquered by Pekeris \cite{pekeris} only in 1959.
The theoretical foundations of few-body quantum physics were laid down by
Lippmann and Schwinger \cite{lippmann},
Gell-mann and Goldberger \cite{gell-mann} by 
developing formal scattering theory of two-particle systems. 
The first attempts to
extend the results to multichannel processes and more 
than two particles led to
unsound mathematical formalism and non-unique solutions.
The rigorous theory of
few-body systems was given by Faddeev \cite{faddeev} who proposed 
a set of coupled integral
equations which have a unique solution for the three-body problem. 
Having the correct few-body theory much effort has been invested into the 
development of numerical methods. The first numerical solution of the Faddeev
equations for three spinless particles interacting with local potentials was
achieved by Humberston {\it at al.} \cite{humberston} in 1968. 
Since then a lot has been achieved due to the unbelievable development in
computational power and the several extremely effective new methods having been
developed. Among others,   configuration and momentum space
Faddeev calculations \cite{config,glocklecikk,spline},
the hyperspherical harmonics expansion method \cite{hyper}, 
the quantum Monte Carlo method \cite{qmc}, 
the Coulomb--Sturmian discrete space Faddeev approach \cite{pzwp}
and the stochastic 
variational method \cite{kalmankonyv} have been used  to
study few-body bound, resonant and scattering phenomena with great success.  

The fundamental equations  governing the dynamics of few-body physical systems,
like the Lippmann--Schwinger equation and the Faddeev equations, are
formulated in terms of integral equations. Integral equation formalism has a
great advantage over the equivalent traditional differential equations
because of the  few-body boundary conditions
are automatically incorporated into the integral equations.
This is the reason why integral equation methods are in favour when scattering
problems with complicated asymptotic behaviour are considered.
Nevertheless for cases, where the  asymptotics of the wave
function is well known, differential equation approach can perform
outstandingly (see i.e. the stochastic variational method \cite{kalmankonyv}).

In spite of the fundamental merits of integral equations, their use in
practical calculations is usually avoided and various approximations to the
few-body Schr{\"o}dinger equation are preferred instead. The reason for this is 
certainly that
the integral equations contain not the usual Hamiltonian operator, but  its
resolvent, the Green's operator in their kernel. The evaluation of the Green's
operator is much more complicated  than the direct treatment of the Hamiltonian
using standard tools of theoretical and mathematical physics.   
The determination of the Green's operator is equivalent to the solution of the
problem characterized by the corresponding Hamiltonian. In standard 
textbooks on quantum mechanics \cite{standard_text} 
Green's operators are introduced  on the level of fundamental equations and
their general properties are extensively studied. Formal scattering theory
\cite{newton,taylor}, thus
few-body quantum mechanics can be formulated upon the notion of Green's
operator. The  Green's operator, similarly to the wave function,
carries all the information
of the physical system. Consequently the determination of the Green's operator,
as the central concept of few-body quantum mechanics,
is of extreme importance. 

The suitable choice of the Hilbert space
representation can facilitate the determination
of the Green's operator. In this respect the 
momentum space representation is
rather appealing as the free Green's operator
 is extremely simple there. This is
the main reason why momentum space 
techniques are so frequently used and also
why they are capable of coping with
 complicated integral equations (for a review
see \cite{glocklecikk}). 
On the other hand the use of discrete 
Hilbert space basis representation is often very
advantageous because it transforms the 
integral equations into matrix equations.
The harmonic oscillator (HO) functions \cite{oszcifuggveny} and the
Coulomb--Sturmian (CS) functions are good examples of discrete Hilbert space
basises. 
The free Green's operator can also be given analytically between harmonic
oscillator states \cite{ho_green}. This allowed the construction of a flexible
method \cite{pse,hopse} for solving the Lippmann-Schwinger equation, which
contains the free Green's operator in its kernel and provides a solution with
correct free asymptotics in HO space.
The CS basis representation of the
Coulomb Green's operator in terms of well computable special functions 
was derived by Papp in Ref.\ \cite{papp1}, where he could perform
complicated analytic integrals of the Coulomb Green's function and the
Coulomb--Sturmian functions making use of the results of Ref.\ \cite{hux_man}.

The CS representation of the Coulomb Green's operator forms the basis of a
quantum mechanical approximation method developed recently
\cite{papp1,papp2,papp3} for describing Coulombic systems. The strength of the
method is that the Coulomb--like interactions in the two-body calculations are
treated  asymptotically correctly, since the Coulomb Green's operator is
calculated analytically and only the asymptotically irrelevant short-range
interaction is approximated. This way  the correct Coulomb asymptotics is
guaranteed. The corresponding computer codes for solving two-body
bound, resonant and scattering state problems were also published \cite{cpc}.
The method has been extended to solve the three-body Coulomb problem in the
Faddeev approach. In this formulation of the Faddeev equations the most crucial
point is  to calculate the resolvent of
the sum of two independent, thus commuting
two-body Coulombic Hamiltonian. This resolvent is given as a convolution
integral of  two-body Green's operators. Therefore the evaluation of the
contour integral requires the analytic knowledge of the two-body Green's
operators. So far good results have been obtained by Papp for bound state
\cite{pzwp} and for below-breakup scattering state 
three-body problems \cite{pzsc}.

In practice there is no general procedure for how to determine  Hilbert space
representation of a Green's operator. All
the Green's operators mentioned before require separate 
and detailed investigation
sometimes based on very specific considerations.
In a recent publication \cite{jacobi_jmp} we have proposed a rather general and
easy-to-apply method for calculating discrete Hilbert space basis representation
of Green's operators belonging to some class of Hamiltonians. We have shown
that if in some basis representation the Hamiltonian possesses an infinite
symmetric tridiagonal (Jacobi) matrix structure,  the corresponding Green's
operator can be given in terms of a continued fraction. The procedure 
necessitates
the analytic calculation of the matrix elements of the Hamiltonian 
as the only input in the analytic evaluation of the
Green's matrix in terms of a convergent continued fraction. 
This method simplifies
the determination of Green's matrices considerably and the representation via
continued fraction provides a readily computable and numerically stable Green's
matrix.
The Green's operator of the non-relativistic two-body Coulomb problem and the
D-dimensional harmonic oscillator problem was evaluated utilizing this method in
Ref. \cite{jacobi_jmp}.
An exactly solvable potential problem, which provides a smooth transition
between the Coulomb and the harmonic oscillator problem, was considered in Ref.
\cite{gezapot_jmp}. A suitable basis was defined in which the Hamiltonian of the
potential problem appeared in tridiagonal form. With the help of the analytically
calculated matrix elements of the Hamiltonian the determination of the 
Green's matrix was straightforward.  
The Hamiltonian of the radial Coulomb Klein--Gordon and second order Dirac
equation was shown to possess infinite symmetric tridiagonal structure in the
relativistic CS basis. This again allowed us to give an analytic representation
of the corresponding relativistic Coulomb Green's operators in terms of
continued fraction \cite{dirac_jmp}.
The continued fraction representation of the Coulomb Green's operator was
utilized to give a unified description of bound, resonant and scattering states
of a model nuclear potential \cite{continued_prc}  using a quantum mechanical
approximation method \cite{pse} devised to solve the Lippmann--Schwinger
equation. 

The layout of this thesis is the following: \\
This introductory chapter is followed by a chapter
 devoted to the concept of the
Green's operator. Here few-body quantum theory is built up around the Green's
operator. In Chapter \ref{chap:math} a review of tridiagonal matrices,
three-term recurrence
relations and continued fractions is given. 
The method of calculating continued
fraction representation of certain class of Green's operators is presented in
\mbox{Chapter \ref{chap:cfr}}. This chapter also covers the calculation of
continued fraction 
representation of the D-dimensional Coulomb,
D-dimensional harmonic oscillator, the generalized Coulomb and 
the relativistic Coulomb Green's
operators.    
In the last chapter two typical applications of the
 Green's operator in few-body physics
are delivered.
First the continued fraction representation of Coulomb Green's
operator is used to present a unified description of bound, resonant and
scattering states of a model nuclear potential  demonstrating the
central role of Green's operators in few-body quantum mechanics.
Then the binding energy of the Helium atom is calculated
 by solving the Faddeev--Merkuriev
equations for the atomic three-body problem in Coulomb--Sturmian
Hilbert space representation.
Chapter \ref{chap:cfr} and \ref{chap:applications} contain
 the new results of the author.
Finally, summary and bibliography closes the thesis.  

\chapter{The quantum mechanical Green's operator}\label{sec:green}

Expressions like Green's operator, Green's function and propagator are
frequently  used
 in different fields of physics and mathematics sometimes with
confusing meanings. For example the many-body Green's function 
$ G_{\alpha \beta}(xt,x't')$ studied in
quantum many-body problem is defined as the ground state expectation value of
fields operators in the second quantized formalism \cite{fewa}.
Whereas the propagator of relativistic quantum field 
theory is often referred to as
the Green's function as well \cite{kaku,mandl}.

The Green's operator of few-body quantum physics corresponds to the definition
of mathematical physics \cite{richtmyer}.

\section{Definition}

Suppose $H$ is a Hermitian differential operator in some 
Hilbert space. The $G(z)$ Green's operator is defined as the 
resolvent operator of $H$ 
\begin{equation}
\label{greendef}
  G(z)(z-H)=(z-H)G(z)={\bf 1} \qquad \mbox{for} \quad  
  z \in {\mathbb C} \slash  \sigma(H),
\end{equation}
where $\sigma(H)$ denotes the spectrum of the Hermitian operator $H$
and ${\bf 1}$ is the identity operator.
So,
the $G(z)=(z-H)^{-1}$ Green's operator is defined on the whole
complex plane  ${\mathbb C}$ except for the spectrum of $H$. 

The $\{\ket{r}\}$ coordinate basis representation of the  \eqref{greendef}
operator equation  
\begin{eqnarray}
  \delta(r-r') = \bra{r} {\mathrm I} \ket{r'}, &&
  \quad  \int \mbox{dr} \ket{r} \bra{r'}= {\bf 1} \\
   G(r,r';z)= \bra{r} G(z) \ket{r'}, && \quad
  \delta(r-r')H(r) = \bra{r} H \ket{r'} 
\end{eqnarray}
leads to an inhomogeneous differential equation 
\begin{equation}
\label{greenfunction}
  [z-H(r)] G(r,r';z)=\delta (r-r')
\end{equation} 
for $G(r,r';z)$. In mathematical physics the Green's 
function corresponding to the linear
Hermitian differential operator $H(r)$, where 
$r \in \Omega $,
and to the complex variable $z$ is
defined as the solution of the inhomogeneous equation \eqref{greenfunction}
subject to certain boundary conditions for $r,r'$ on the surface of the domain
$\Omega$. In fact the Green's function is required to satisfy the same boundary
condition as the wave function. Consequently the $G(r,r';z)$ Green's function
of mathematical physics can be viewed as the coordinate space representation of
the $G(z)$ resolvent operator. 

In what follows, under the notion of few-body
Green's operator the \eqref{greendef}  resolvent of the Hamiltonian is to be
understood. In the next chapters  Hilbert space basis
representation of Green's operators will be investigated.

\section{Basic properties of Green's operator}

The determination of the Green's operator represents the solution of the
underlying differential equation. Therefore the evaluation of the Green's
operator corresponding to the Hamiltonian $H$ is equivalent to the complete
description of the system characterized by $H$. The 
Green's operator carries all the information about the physical system. Once
some representation of $G(z)$ is known one can gain the eigenvalues and the 
eigenfunctions of the Hamiltonian, construct projection operators or determine
the density of states, etc,. The time evaluation operator can also be obtained 
by performing a
Fourier transform of the Green's operator.

Let $ \{ \ket{\phi_n} \}$ and $\{ \alpha_n \}$ denote the complete set of 
orthonormal eigenstate and the corresponding eigenvalues of the 
Hermitian operator $H$
\begin{equation}
  H \ket{\phi_n} = \alpha_n \ket{\phi_n} \ .
\end{equation} 
Utilizing the completeness of the eigenstates, the Green's operator can be
written as
\begin{equation}
\label{greenexp}
  G(z)=\sum_n  \frac{\ket{\phi_n} \bra{\phi_n}}{z-\alpha_n} + 
  \int \mbox{dn}  \frac{\ket{\phi_n} \bra{\phi_n}}{z-\alpha_n}, 
\end{equation}
where $\sum_n$ and $\int\mbox{dn}$ indicates
summation over the discrete and the
continuous spectrum,  respectively.

Since $H$ is Hermitian, all of its $\alpha_n$ eigenvalues are real, which means
that $G(z)$ is an analytic function in the complex $z$-plane except at those
points and intervals of the real axis which corresponds to the spectrum of $H$.
According to \eqref{greenexp} $G(z)$ possesses simple  poles at the
positions of the discrete eigenvalues of $H$, hence the poles of $G(z)$ give the
discrete energy eigenvalues of the Hamiltonian.
The residue at the $\alpha_n$ pole provides information on the corresponding 
non-degenerate $\ket{\phi_n}$ eigenvector according to the projection operator
\begin{equation}
\label{projection}
  \ket{\phi_n} \bra{\phi_n} = \frac 1{2\pi \mbox{i}}\oint_{\alpha_n}
  dz \ G(z),  
\end{equation}
here the contour encircles the $\alpha_n$ and only the $\alpha_n$ eigenvalue.
Equation \eqref{projection} can be used to construct the wave function.

If $\lambda \in {\mathbb R} $ belongs to the continuous part of
the spectrum, then $G(\lambda)$ is usually
defined by a non-unique limiting procedure 
\begin{eqnarray}
  G^{+}(\lambda):=\lim_{\epsilon \to 0^+ } G(\lambda+i\epsilon) \\
  G^{-}(\lambda):=\lim_{\epsilon \to 0^+ } G(\lambda-i\epsilon)
\end{eqnarray} 
because of the existence of the limits. Thus  the continuum spectrum of $H$
appears as a singular line (branch cut) of $G(z)$.
Due to the self-adjointness of the Hamiltonian the Green's operator exhibits
the important property
\begin{equation}
\label{greenadj}
  G(z^{*})=G(z)^{*} \ .
\end{equation}   
Therefore $G^{+}(\lambda)$ and $G^{-}(\lambda)$
are connected by
\begin{equation}
\label{green_anal}
  G^{-}(\lambda)=\left( G^{+}(\lambda) \right)^*,
\end{equation}
from which we have 
\begin{equation}
  {\mbox Re} G^- (\lambda) = {\mbox Re} G^+ (\lambda), \quad 
  {\mbox Im} G^- (\lambda) = -{\mbox Im} G^+ (\lambda) \ .
\end{equation}
The knowledge of  ${\mbox Re} G^{\pm}(\lambda)$ allows us to obtain the density
of states at $\lambda$ 
\begin{equation}
  \Delta (\lambda) = \mp \frac{1}{\pi} {\mbox Im} 
  \{ Tr [G^{\pm} (\lambda)]\} \ .
\end{equation} 

The time development of the $\ket{\psi (t_0)}$  state belonging to the 
Hamiltonian $H$ at $t=t_0$  can be obtained as a solution of the time dependent
Schr{\"o}dinger equation, and it is governed by the $U(t,t_0)$ time evolution
operator according to
\begin{equation}
  \ket{\psi (t)}= U(t,t_0) \ket{\psi (t_0)},
\end{equation}
where the time evolution operator
\begin{equation}
  U(t,t_0)=e^{-\frac{i}{\hbar}H(t-t_0)}
\end{equation}
can be expressed in terms of the Green's operator as
\begin{equation}
\label{propagator}
  U(t,t_0)=i \hbar \int_{-\infty}^{\infty}  \frac{ {\mbox d \omega} }{2\pi}
  e^{-i \omega (t-t_0)} \left[ G^{+} (\hbar \omega) - G^{-}
  (\hbar \omega) \right]
\end{equation}
Equation \eqref{propagator} shows that the $U(t,t_0)$ propagator can be
determined as the Fourier transform of the $G^{\pm}$ Green's operators.

\section{The Green's operator in scattering theory}\label{sec:scattering}

The formulation of quantum scattering theory is heavily relied on the
concept of Green's operator. It is concerned with the
mathematical description of the scattering process during
 which the incident beam
of particles, usually prepared by some accelerator,
 enters into interaction in the
scattering region then the scattered beam is observed by detectors located at
large distance from the interaction area. 

Far from the
scattering region the incoming and the scattered beam
can be thought of as   bunch of particles propagating freely, 
therefore they can be characterized by the $\ket{\psi_{in}}$,
$\ket{\psi_{out}}$ asymptotic states whose time evolution are
governed by the $H^0$ asymptotic Hamiltonian  as
\begin{equation}
  \label{freeev}
  \ket{\psi_{in/out}(t)}=U_0(t)\ket{\psi_{in/out}(t_0)}, \quad 
  U_0(t)=e^{-\frac{i}{\hbar}H^0 t}.
\end{equation}  
On the other hand the time evolution of the scattering process is
 fully determined
 by the $H=H^0+V$ total Hamiltonian involving the scattering 
potential as well:
\begin{equation}
  \label{fullev}
  \ket{\psi(t)}=U(t)\ket{\psi(t_0)}, \quad 
   U(t)=e^{-\frac{i}{\hbar}H t},
\end{equation}  
here $\ket{\psi(t)}$ denotes the actual scattering state vector.
However, during scattering experiments one can only measure the
$\ket{\psi_{in}}$
incoming and the $\ket{\psi_{out}}$ outgoing asymptotes of 
$\ket{\psi}$. Therefore in  scattering theory
 all information about the scattering process
 should be extracted from the
asymptotic behavior of $\ket{\psi}$.

At $t=0$ let $\ket{\phi+}$ ($\ket{\chi-}$) denotes the $\ket{\psi(t)}$ 
scattering state
which evolved from the $\ket{\psi_{in}}=\ket{\phi}$ incoming asymptote
(evolved to the $\ket{\psi_{out}}=\ket{\chi}$ outgoing asymptote).
The  $\ket{\phi+}$ and $\ket{\chi-}$ scattering states are related to their
asymptotes by the equations
\begin{eqnarray}
\label{moller}
   \ket{\phi+}&=& \Omega_{+}\ket{\phi} \\
   \ket{\chi-}&=& \Omega_{-}\ket{\chi} \nonumber,
\end{eqnarray}
where the  $\Omega_{\pm}$ M{\o}ller operators are defined as the limits of
the time evolution operators
\begin{equation}
\label{mollerdef}
  \Omega_{\pm}=\lim_{t \to \mp \infty} U(t)^{\dagger} U^0(t).
\end{equation}      
The existence of the M{\o}ller operator can be proved if the scattering
potential satisfies certain conditions (i.e. not too singular at the origin,
sufficiently short ranged and smooth, see \cite{taylor}). The main goal of the
scattering description is to relate the outgoing asymptote to the incoming one
without reference to the experimentally unknown actual orbit. Utilizing the
$\Omega_{-}^{\dagger}\Omega_{-}= {\bf 1}$ 
isometric property of the M{\o}ller operator
from Eq.\ \eqref{moller} we obtain
\begin{equation}
\label{scatteringopi}
  \ket{\psi_{out}}=\Omega_{-}^{\dagger}\Omega_{+} \ket{\psi_{in}}
  :=S\ket{\psi_{in}},
\end{equation}  
here $S:=\Omega_{-}^{\dagger}\Omega_{+}$ is the scattering operator, which 
carries all the experimentally important information about 
the scattering. Once the
operator $S$ is determined the scattering problem is solved. 

In order to determine the $S$ operator it is convenient to use the
$\{ \ket{\vec{p}}\}$
momentum representation. It is important to note here that the
$\ket{\vec{p}}$
plane wave momentum eigenstates do not represent  physically realizable
states, in fact they are only used in the expansion of physical states of the
scattering process as
\begin{equation}
\label{mom_expansion}
  \ket{\psi}=\int  {\mathrm d^3p} \ \psi (\vec{p}) \ket{\vec{p}}.
\end{equation}
Similarly the scattering state $\ket{\phi+}$ can be expressed as
\begin{equation}
\label{scat_expansion}
  \ket{\phi+} =\Omega_+ \ket{\phi}=
  \int  {\mathrm d^3p} \ \phi (\vec{p}) \Omega_+ \ket{\vec{p}}=
  \int  {\mathrm d^3p} \ \phi (\vec{p}) \ket{\vec{p}+},
\end{equation}
where the $\ket{\vec{p}+}=\Omega_+ \ket{\vec{p}}$ 
improper (i.e. not normalizable)
vector is called the stationary scattering vector corresponding to the 
$\ket{\vec{p}}$ incoming plane wave.

In  momentum representation a careful mathematical analysis leads to the
very important result
\begin{equation}
\label{impres}
\begin{split}
  \Omega_+ \ket{\phi} &=\ket{\phi+} =\lim_{t \to -\infty} U(t)^{\dagger}
   U_0(t) \ket{\phi} \\
  &=\ket{\phi} + i \int_{0}^{- \infty} \mbox{d} \tau \
  U(\tau)^{\dagger} V U^0(\tau) \ket{\phi} \\
  &=\ket{\phi} + i\lim_{\epsilon \to 0^{+}} \int_{0}^{- \infty} \mbox{d} \tau \
   e^{+ \epsilon \tau}  U(\tau)^{\dagger} V U^0(\tau) \ket{\phi} \\
  &= \ket{\phi} + i\lim_{\epsilon \to 0^{+}} \int {\mathrm d^3 p} \
  G(E_p+i\epsilon) V \ket{\vec{p}} \left< \vec{p} | \phi \right>
\end{split}
\end{equation}
which establishes the connection between the $\Omega_+$ M{\o}ller and the
$G^{+}(z)$ Green's operator.  
After expanding $\ket{\phi+}$ and $\ket{\phi}$ in Eq.\ \eqref{impres} 
in terms of the improper states
$\ket{\vec{p}}$ and $\ket{\vec{p}+}$
 we  obtain  for the stationary scattering vector $\ket{\vec{p}\pm}$
\begin{equation}
\label{scattering_state}
  \ket{\vec{p}\pm}= \ket{\vec{p}} +  \lim_{\epsilon \to 0^{+}}
  G(E_p \pm i\epsilon) V \ket{\vec{p}} \ .
\end{equation}
Making use of the 
\begin{equation}
\label{opii}
  A^{-1}= B^{-1} + B^{-1}(B-A)A^{-1}
\end{equation} 
operator identity with $A=z-H$ and $B=z-H^{0}$, then with their interchange,
we can derive the 
\begin{eqnarray}
\label{opidentiti}
  G(z)=G^0(z)+G^0(z)VG(z)  \\
  G(z)=G^0(z)+G(z)VG^0(z) \nonumber
\end{eqnarray}  
resolvent equations, which relate the $G^{0}$ and the $G$ operators.
With these equations Eq.\ \eqref{scattering_state}
can be recast in the form
\begin{equation}
\label{lippmann_sch}
   \ket{\vec{p}\pm}= \ket{\vec{p}} +  \lim_{\epsilon \to 0^{+}}
   G^0(E_p \pm i\epsilon) V \ket{\vec{p}\pm}.
\end{equation}   
This type of integral equations were first formulated by Lippmann and Schwinger
\cite{lippmann}, consequently Eq.\ \eqref{opidentiti} and 
Eq.\ \eqref{lippmann_sch}
are known as the Lippmann--Schwinger equations for $G$ and $\ket{\vec{p}\pm}$,
respectively.  
According to \eqref{scattering_state} and \eqref{lippmann_sch}
the evaluation of the
stationary scattering vectors necessitates the determination of the full Green's
operator or the solution of the \eqref{lippmann_sch} 
Lippmann--Schwinger equation. 

Being able to calculate $\ket{\vec{p}\pm}$,
one can obtain any scattering information.
For example, the $ \bra{\vec{p}^{\: \prime}} S \ket{\vec{p}}$ scattering 
matrix can be written as
\begin{equation}
\label{scattering_matrix}
  \bra{\vec{p}^{\: \prime}} S \ket{\vec{p}} =\delta_3 
  (\vec{p}^{\: \prime}-\vec{p})-2\pi i \delta(E_{p'}-E_p) \ 
  t(\vec{p}^{\: \prime} \gets \vec{p}).
\end{equation} 
Here $t(\vec{p}^{\: \prime} \gets \vec{p})$ can be calculated using the 
$\ket{\vec{p}\pm}$ scattering states
\begin{equation}
\label{tpp}
  t(\vec{p}^{\: \prime} \gets \vec{p})=
  \bra{\vec{p}^{\: \prime}} V \ket{\vec{p}+}=
  \bra{\vec{p}^{\: \prime} \! -} V \ket{\vec{p}},
\end{equation}   
or using the Green's operator
\begin{equation}
\label{tpp2}
    t(\vec{p}^{\: \prime} \gets \vec{p})=
    \bra{\vec{p}^{\: \prime}} V + VGV \ket{\vec{p}}.
\end{equation}   
The $(V+VGV)$ operator is called the $T$ operator. Its  
$t(\vec{p}^{\: \prime} \gets \vec{p})$ matrix element 
is referred to as the on-shell $T$
matrix, because in Eq.\ \eqref{scattering_matrix} 
it appears together with the $\delta(E_{p'}-E_p)$ delta
function, and thus defined only for the ${p'}^{2}=p^2$ "on-shell" energy case.

The experimentally off important differential scattering cross section 
$d\sigma / d\Omega$ can be derived from the \eqref{scattering_matrix} formula
for the scattering matrix, and has the form
\begin{equation}
\label{cross_section}
    \frac{d\sigma}{d\Omega}=| f(\vec{p}^{\: \prime} \gets \vec{p})| ^2 ,
\end{equation}    
where $f(\vec{p}^{\: \prime} \gets \vec{p})$ is the scattering amplitude,
which is just a trivial factor times the on-shell $T$ matrix
\begin{equation}
\label{sa}
   f(\vec{p}^{\: \prime} \gets \vec{p})=
   -(2\pi)^2m \  t(\vec{p}^{\: \prime} \gets \vec{p}).
\end{equation}

Finally we quote here the asymptotic form of the coordinate representation of
the stationary scattering vector
\begin{equation}   
\label{crssv}
  \lim_{r \to \infty}  \left< \vec{r} | \vec{p}+ \right> =
  (2\pi)^{-\frac{3}{2}} \left[ e^{i \vec{p} \vec{r}} +
  f(p \hat{r} \gets \vec{p}) \frac{e^{ipr}}{r} \right],
\end{equation}  
which reflects the traditional definition of scattering states as the sum of an
incident plane wave and a spherically spreading scattered wave, whose   
$f(p \hat{r} \gets \vec{p})$ amplitude determines the scattering cross section
via \eqref{cross_section}.
 
In this short review it was demonstrated that the knowledge of the Green's
operator enables one to determine the scattering states, which is equivalent to
the solution of the scattering problem and makes possible the calculation of
any scattering quantities.

\section{Dunford--Taylor integrals of  Green's operators}
\label{sec:convolution}

From the theory of linear operators \cite{dunford} we know that the analytic 
function of a bounded self-adjoint  operator $A$ can be
defined as the Dunford--Taylor integral representation
\begin{equation}
\label{cauchy}
  f(A)=\frac{1}{2\pi \mathrm{i}} \oint_C dz \;f(z) (z-A)^{-1},
\end{equation}
where $C$ encircles the spectrum of $A$ in positive direction and $f$
should be analytic on the domain encircled by $C$. This definition 
can be considered as  the
operator equivalent of  the Cauchy integral formula.

In few-body physics one often encounters the problem of determining Green's
operators corresponding to Hamiltonians composed of two independent 
operators with known resolvents. 
This is the typical situation when the Green's operator of  separable
subsystems  are known (e.g. think of a three-dimensional motion decomposed to
the sum of a two- and a one-dimensional motion).
The question, whether it is possible to determine the Green's operator of
the composite system making use of the sub-Green's operators naturally emerges.
The answer can be given by utilizing the \eqref{cauchy} Dunford--Taylor
integral representation.

\begin{figure}
\begin{center}
\resizebox{10cm}{!}{
\includegraphics{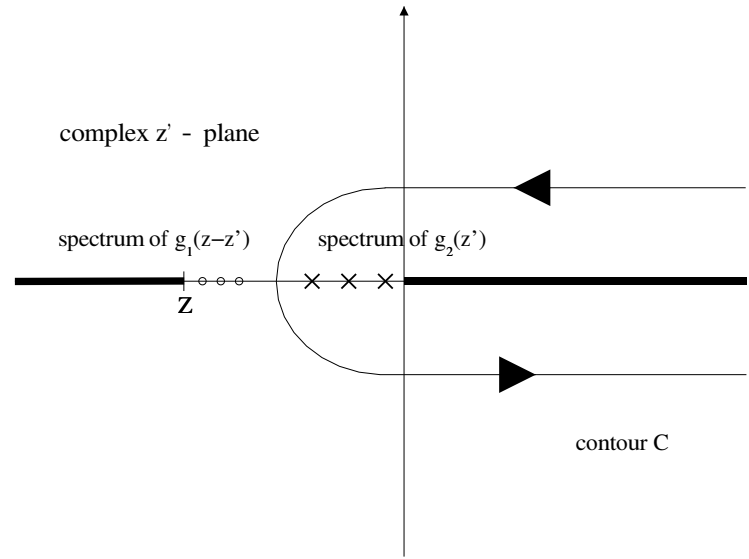} 
} 
\end{center}
\caption{\label{konturabra} The integration path $C$ in the  
complex $z'$-plane
for the convolution integral of the \eqref{convolution} Green's operator.}
\end{figure}

Let $h_1,h_2$ be two independent (hence commuting) bounded linear operators 
with the analytically known $g_1(z)=(z-h_1)^{-1}$, $g_2(z)=(z-h_2)^{-1}$ 
resolvents.
Then applying the \eqref{cauchy} definition  for 
$f(\lambda)=(z-h_1-\lambda)^{-1}$ and  $A=h_2$,
the  $G(z)=(z-H)^{-1}$ Green's operator of the 
$H=h_1+h_2$ composite Hamiltonian can be evaluated  by performing 
a convolution integral of the  $g_1(z)$, $g_2(z)$ sub-Green's operators
\begin{equation}
\label{convolution} 
  G(z) =\frac 1{2\pi \mbox{i}}\oint_C
  dz^\prime \,g_1(z-z^\prime)\;
  g_2(z^\prime),
\end{equation}
where the contour $C$ should encircle, in counterclockwise direction,
the spectrum (discrete and continuous) of $h_2$ without penetrating into the
spectrum of $h_1$. In Fig.\ \ref{konturabra} a typical
integration scenario in the complex
$z'$-plane is shown for a bound state system ($z<0$).
We note  that among others, Bianchi and Favella \cite{bianchi}  
proposed a convolution integral of the type
of \eqref{convolution} for determining the resolvent of the sum of two
commuting Hamiltonian.

The \eqref{cauchy} integal representation  can also be used to  
construct other well-known operators as a Dunford--Taylor
integral of the Green's operator.  
For example, the identity operator
\begin{equation}
\label{di}  
  1=\frac{1}{2\pi\mbox{i}}
  \oint_C dz\: \: (z-h)^{-1},
\end{equation}
the Hamilton operator
\begin{equation}
\label{dh}
  h =\frac{1}{2\pi\mbox{i}}
  \oint_C dz\: z \: (z-h)^{-1} \ ,
\end{equation}  
or the time evolution operator
\begin{equation}
\label{dt}
  \exp(-\mbox{i}h t) =\frac{1}{2\pi\mbox{i}}
  \oint_C dz\: \exp(-\mbox{i} z t) \: (z-h)^{-1}
\end{equation}  
can be given as analytic integrals of the $g=(z-h)^{-1}$ Green's operator.

\chapter{\mbox{Tridiagonal matrices,  re\-curren\-ce}
\mbox{re\-la\-ti\-ons and continued
fractions}}\label{chap:math}

Based on the mathematical literature \cite{golub,rozsa,lorentzen,jones} 
this chapter is dedicated to the survey of the mathematical 
machinery being used in the forthcoming part of the thesis.
First a truncated inverse of an infinite  symmetric tridiagonal
matrix is presented.
Tridiagonal matrices inherently imply three-term recurrence relations 
for the matrix elements of their inverse.
For this reason the  basic properties of  
three-term recurrence relations are outlined. The final section 
discusses the fundamentals of the theory of continued  fractions.
The analytic continuation and an important type of transformation
of  continued fractions is investigated. The intimate relation of 
continued fractions and three-term recurrence relations is revealed by
Pincherle's theorem.

\section{Infinite tridiagonal matrices}\label{sec:tridiagonal}

Infinite rank matrices naturally emerge in quantum mechanics when  physical
observables are represented on a discrete
infinite dimensional basis. 
Moreover the matrix representation of many physical operators
are tridiagonal in a certain basis. In fact, some computational methods,
like
the Lanczos method \cite{lanczosm,golub}, are based on the scheme of creating a
basis that renders a given Hamiltonian tridiagonal.
However we are  always limited  to 
consider finite rank truncations of infinite matrices. 
Here we derive  a  formula \cite{jacobi_jmp} for the n-th leading submatrix 
of the  inverse of
an  infinite symmetric tridiagonal matrix. The underlying idea originally
came up  in the work of P\'al \cite{husi}, and we give here the proof.

Let $A:=[a_{ij}], \: i,j=1,2,\ldots,\infty$, be an infinite symmetric   
tridiagonal  (Jacobi) matrix:
\begin{equation}
\label{tridiag}
  a_{i i-1}=\beta_{i-1}, \quad a_{i i}=\alpha_{i}, \quad a_{i i+1}
  =\beta_{i}, \quad
  a_{ij}=0 \quad \mbox{for} \quad |i-j|>1.
\end{equation}
Let $B:=[b_{ij}], \: i,j=1,2,\ldots,\infty$ denote the 
infinite rank matrix, for which
\begin{equation}
\label{def}
  {\bf 1}=AB=BA, \quad \delta_{ij}=\sum_{k=1}^{\infty} a_{ik}b_{kj}.
\end{equation}
Then the following theorem can be stated.
\begin{theorem}\label{theorem:inverz}
Let us define the $n$-th leading submatrices of  $A$ and of $B$   as \\
$A_n:=[a_{ij}]$, $B_n:=[b_{ij}], \:i,j=1,2,\ldots,n$, respectively.  
Then 
\begin{equation}
\label{tetel}
  B_n=(A_n+F_n)^{-1},
\end{equation}
where $F_n:=[f_{ij}], \quad i,j=1,\ldots,n$, is given by
\begin{equation}
\label{farok}
  f_{ij} =
    \left\{
       \begin{array}{ll}
        0   &   \quad i,j < n \\
        0   &   \quad i=n, j<n \\
        0   &   \quad j=n, i<n \\
        a_{n n+1} \frac{ \displaystyle b_{1 n+1}}{\displaystyle b_{1 n}}
        & \quad i=j=n
       \end{array}
    \right.
\end{equation}
\end{theorem}
\begin{proof}[Proof of Theorem \ref{theorem:inverz}]
Utilizing the (\ref{tridiag}) tridiagonal property of $A$, 
 the (\ref{def}) definining relation  of $B$
for $i,j\leq n$ can be written  as
\begin{equation}
\label{p3}
  \delta_{ij}=\sum_{k=1}^{n} a_{ik}b_{kj} + \delta_{in}a_{n n+1}b_{n+1 j},
  \quad i,j\leq n,
\end{equation} 
which we can rewrite in the form
\begin{equation}
\label{p4}
  \delta_{ij}=\sum_{k=1}^{n} \left(a_{ik}b_{kj} + \delta_{in}a_{i n+1}b_{n+1 j}
  \delta_{k n} \frac{b_{k j}}{b_{n j}}\right)= \sum_{k=1}^{n} \left(a_{ik}
  + \delta_{in}\delta_{k n}a_{i n+1}\frac{b_{n+1 j}}{b_{n j}}\right)b_{k j}.
\end{equation}
The inverse of a symmetric tridiagonal matrix possesses the following 
property \cite{rozsa}:
\begin{equation}
\label{rozsa}
  b_{ij}=\left\{ 
\begin{array}{r@{\quad\mbox{if}\quad}l}
  p_iq_j, & i\le j \\ 
  p_jq_i, & j\le i
\end{array}
\right. .  
\end{equation}
Therefore 
\begin{equation}
\label{p5}
  \frac{b_{n+1 j}}{b_{n j}}=\frac{p_{j}q_{n+1}}{p_{j}q_n}
  =\frac{p_{1}q_{n+1}}{p_{1}q_n}=\frac{b_{1 n+1}}{b_{1 n}},
\end{equation}
and follows the statement of the theorem:
\begin{equation}
\label{p6}
  \delta_{ij}=\sum_{k=1}^{n} \left(a_{ik}
  + \delta_{in}\delta_{kn}a_{i n+1}\frac{b_{1 n+1}}{b_{1,n}}\right)b_{k j}
  =\sum_{k=1}^{n}(a_{ik}+f_{ik})b_{kj}. \quad 
\end{equation}
\end{proof}

Finally we give a schematic picture of the theorem.
The infinite symmetric  tridiagonal matrix $A$ and its n-th leading
submatrix $A_n$ can be illustrated as
\begin{equation}
  A=
  \left(
  \begin{array}{cccc}
  A_n& \cdot &  &  \\
  \cdot & \cdot & \cdot &  \\
  & \cdot & \cdot & \cdot \\
  &  & \cdot & \cdot
  \end{array}
  \right)_{\infty \bf{x} \infty} \quad 
  A_n=  
  \left(
  \begin{array}{cccc}
  \cdot & \cdot &  &  \\
  \cdot & \cdot & \cdot &  \\
  & \cdot & \cdot & \cdot \\
  &  & \cdot & \cdot
  \end{array}
  \right)_{n \times n}
\end{equation}
Similarly the infinite $B$ and the truncated $B_n$ general matrices can be
plotted 
\begin{equation}
  B=
  \left(
  \begin{array}{cccc}
  B_n& \cdot  & \cdot & \cdot \\
  \cdot & \cdot  & \cdot & \cdot \\
  \cdot& \cdot & \cdot & \cdot \\
  \cdot& \cdot & \cdot & \cdot
  \end{array}
  \right)_{\infty \bf{x} \infty} \quad 
  B_n=  
  \left(
  \begin{array}{cccc}
  \cdot & \cdot & \cdot & \cdot \\
  \cdot & \cdot & \cdot & \cdot \\
  \cdot& \cdot & \cdot & \cdot \\
  \cdot& \cdot & \cdot & \cdot
  \end{array}
  \right)_{n \times n}
\end{equation}     
With the above schematic representation of matrices, the \eqref{tetel}
statement of Theorem \ref{theorem:inverz} can be recast in the form
\begin{equation}
\label{illustr}
  \left( \begin{array}{cccc} \cdot & \cdot & \cdot & \cdot\\ 
  \cdot & \cdot & \cdot & \cdot \\ 
  \cdot & \cdot & \cdot & \cdot \\ 
  \cdot & \cdot & \cdot & \cdot \end{array} \right)_{ n \times n} = 
  \left[ \left( \begin{array}{cccc} \cdot & \cdot & & \\ 
  \cdot & \cdot & \cdot & \\ & \cdot & \cdot & \cdot \\
  & & \cdot & \cdot \end{array} \right)_{ n \times n} 
  + \left( \begin{array}{cccc} & & & \\ 
  & & & \\ & & & \\ & & & f_{n,n} \end{array} \right)_{ n \times n}
  \right]^{-1}\ 
\end{equation}
where $f_{n n}=a_{n n+1}b_{1 n+1}/b_{1 n}$.

According to Eq.\ \eqref{illustr} 
the truncated inverz $B_n$ is readily calculable from the $a_{ij}$ tridiagonal
matrix elements, provided the $b_{1 n+1}/b_{1 n}$ quotient 
is at our disposal.

\section{Three-term recurrence relations}\label{sec:three-term}

Due to Eq.\ \eqref{def} the matrix elements $b_{ij}$ of the inverz of an
infinite  symmetric tridiagonal matrix 
automatically satisfy a three-term recurrence relation in both indexes
\begin{eqnarray}
\label{rec1}
  \delta_{ij}=\beta_{i-1} b_{i-1 j}+\alpha_i b_{i j}+\beta_i b_{i+1 j}, 
  \quad \mbox{for}
  \quad i=1,\ldots,\infty,   \quad j \; \mbox{fixed},  \\ 
  \delta_{ij}=\beta_{j-1} b_{i j-1}+\alpha_j b_{i j}+\beta_j b_{i j+1}, \quad
  \mbox{for} \quad j=1,\ldots,\infty,   \quad i \;  \mbox{fixed}. \label{rec2}
\end{eqnarray}
Let us consider the (homogeneous) three-term recurrence relation
\begin{equation}
\label{3term}
   x_{n+1}=b_nx_n+a_nx_{n-1}, \quad a_n\neq 0, \qquad n=1,2,3,\ldots ,
\end{equation}
where $a_n,b_n$  complex numbers are  the coefficients
of the recurrence relation.
A sequence of complex numbers $\{x_n\}^{\infty}_{n=0}$
is called a solution of \eqref{3term} if its $x_n$
elements satisfy the equation for all $n\in {\mathbf N}$.
The set of all solutions $\{x_n\}^{\infty}_{n=0}$ of \eqref{3term} form a
linear vector space of dimension two over the field of complex numbers, and the
zero vector of this space is the $ \{0\}^{\infty}_{n=0}$ trivial solution.
A distinguished type of solution is the minimal solution. If there exists a
non-trivial solution $\{m_n\}^{\infty}_{n=0}$ and another solution 
$\{d_n\}^{\infty}_{n=0}$ of \eqref{3term} such that  
\begin{equation}
\label{minimal}
  \lim_{n\rightarrow \infty }\frac{m_n}{d_n}=0,
\end{equation}
then  $\{m_n\}$ is called the minimal (subdominal)  solution. A solution which 
is not minimal, like $\{d_n\}$, is called the dominant solution.
In general, a recurrence relation may or may not have a minimal solution, and it
is clear that if $\{m_n\}$ is a minimal solution then 
$\{m'\}:=c\{m\}$ for  $\forall c \neq 0$ is a minimal solution as well.
Moreover any other non-trivial $\{y_n\}$ solutions are dominant since they can
be written
\begin{equation}
\label{dominant}
  \{y_n\}=c_1\{m_n\}+c_2\{d_n\}, \quad c_2\neq 0,
\end{equation}
where $\{d_n\}$ is a dominant solution. The existence of a minimal solution is
strongly related to the convergence of a continued fraction (see Section
\ref{sec:cont_frac}).

Suppose the $x_0$ and $x_1$ elements of the  \eqref{3term} three-term
recurrence relation are at our disposal and we want to apply the
recurrence relation in a direct way to determine $\{x_n\}_{n=0}^{\infty}$
recursively.
However, for a minimal solution this method does fail in practice.
Gautschi \cite{gautschi} pointed out that  a direct 
recursive calculation of a minimal solution of a three-term 
recurrence relation is
numerically unstable. A surprising demonstration of 
this instability can be found
in \cite{lorentzen} on page 219.

We note here that three-term linear recurrence relations are closely 
linked to linear difference equations of order two.

\section{Continued fractions}\label{sec:cont_frac}

The study of continued fractions, i.e. mathematical expressions of the form
\begin{equation}
\label{cf}
	b_o+
  \cfrac{a_1} {b_1+ 
  \cfrac{a_2} {b_2+ 
  \cfrac{a_3} {b_3+ \cdots + 
  \cfrac{a_n} {b_n}}}}
\end{equation}
started in the 16th century with a work of 
Bombelli \cite{bombelli} in 1572, who
wrote down the first finite continued fraction
\begin{equation}
\label{bombeli}
 3+ \cfrac{4} {6+
    \cfrac{4} {6}}
 =3.6
\end{equation}
to approximate $\sqrt{13} \approx 3.6055$.
The traditional  form \eqref{cf} of a continued fraction  is often written 
more economically as
\begin{equation}
\label{cf_notation}
  b_0(z)+
  \frac{a_1}{b_1}
  { \atopwithdelims.. +}
  \frac{a_2}{b_2}
  { \atopwithdelims.. +}
  \frac{a_3}{b_3} 
  { \atopwithdelims.. +} \cdots { \atopwithdelims.. +}
  \frac{a_n}{b_n}.
\end{equation}
The first continued fractions were constructed from integers and were used to
approximate various algebraic numbers, among others the $\pi$.
From the 17th century on the continued fraction evolved into an important
tool of number theory. First Schwenter, Huygens, Wallis latter 
Lagrange, Legendre, Gauss and Galois made extensive use of regular 
continued fractions, continued fractions with integer coefficients,
in their number theoretical studies.
    
Euler extended the notion of a continued fraction by generalizing its
coefficients as functions of complex variables rather than simple numbers.
This way continued fraction expansions could be used as special tools 
in the analytic approximations of special classes of analytic functions.
Following Euler's results a long list of brilliant  
mathematicians from the 19th
century,  names like
Laplace, Jacobi, Riemann, Stieltjes, Frobenius and Tchebycheff 
made flourish the analytic theory of continued fractions
within the framework of the classical 19th century analysis.

Mathematical physics had also discovered the continued fractions. Continued
fraction expansion of special functions, solution of three-term
recurrence relations and  Padé approximants represent important field of
applications. The available continuously increasing computational power has 
also contributed to the present-day
success of continued fractions in mathematical physics.

Rigorously a continued fraction is defined as an ordered pair
of the type 
\begin{equation}
\label{cfdef}
  ((\{a_n(z)\},\{b_n(z)\}),\{f_n(z)\}),
\end{equation}
where $\{a_n(z)\}_1^\infty$
and $\{b_n(z)\}_0^\infty$, with all $a_n(z)\neq 0$, are 
two sequences of complex
valued functions defined on the region $D$ of the complex plane.
The complex values of $a_n(z)$ and $b_n(z)$ are called the $n$-th 
partial numerator and partial denominator of the continued fraction,
or simply the 
coefficients or elements. The $\{f_n(z)\}$ sequence of complex functions
is given by 
\begin{equation}
  f_n(z)=S_n(0,z), \qquad n=0,1,2,\ldots,
\end{equation}
where
\begin{equation}
  S_n(w_n,z)=S_{n-1}(s_n(w_n,z),z),\qquad S_0(w_0,z)=s_0(w_0,z),
\end{equation}
with the 
\begin{equation}
\label{linfractrafo}
  s_n(w_n,z)=\frac{a_n(z)}{b_n(z)+w_n},\quad n\geq 1,\qquad
  s_0(w_0,z)=b_0(z)+w_0,
\end{equation}
linear fractional transformation.   

Here $S_n(w_n,z)$ is called the $n$-th approximant of the continued fraction
with respect to the $\left\{ w_n\right\} _{n=0}^\infty $ complex series. 
Applying recursively the \eqref{linfractrafo} linear fractional transformation
the $n$-th approximant of the continued fraction can be written as
\begin{equation}
\label{approxi}
  S_n(w_n,z)=b_0(z)+\frac{a_1(z)}{b_1(z)}
  {\atopwithdelims.. +}
  \frac{a_2(z)}{b_2(z)}
  {\atopwithdelims.. +} \cdots {\atopwithdelims.. +} 
  \frac{a_n(z)}{b_n(z)+w_n} \ .
\end{equation}
Similarly, for the $S_n(0,z)$  approximant we got
\begin{equation}
\label{nth_approxi}
  S_n(0,z)=
   b_0(z)+\mathop{K}_{i=1}^n \left( 
  \frac{a_i(z)}{b_i(z)}\right)
  =b_0(z)+\frac{a_1(z)}{b_1(z)}
  {\atopwithdelims.. +}
  \frac{a_2(z)}{b_2(z)}
  {\atopwithdelims.. +} \cdots
  {\atopwithdelims.. +}\frac{a_n(z)}{b_n(z)}, 
\end{equation}
where the new notation
$ \mathop{K}_{i=1}^n (a_i / b_i)$ 
was introduced for the approximant.
Subsequently for the $ ((\{a_n(z)\},\{b_n(z)\}),\{f_n(z)\})$
continued fraction one of the 
\begin{equation}
\label{conti}
  b_0(z)+\frac{a_1(z)}{b_1(z)}%
  {\atopwithdelims.. +}
  \frac{a_2(z)}{b_2(z)}
  {\atopwithdelims.. +} \cdots,
  \quad  \mbox{or} \quad 
  b_0(z)+\mathop{K}_{n=1}^\infty \left( 
  \frac{a_n(z)}{b_n(z)}\right)
\end{equation}
notations will be used.

The convergence of a continued fraction means the convergence of the
sequence of approximants $S_n(w_n,z)$ to an extended complex number 
\begin{equation} 
\label{cfdef2}
  f(z)=\lim_{n\rightarrow \infty }S_n(w_n,z)  \ .
\end{equation}
If $f(z)$ exists for two different sequences of 
$\{w_n\}$ then $f(z)$ is unique. For a detailed discussion
of 
convergence results see Chapter~IV of \cite{jones}.

There are several algorithm to compute the 
$n$-th approximant of a continued fraction. The backward recurrence algorithm
consists of summing-up the  \eqref{approxi} fraction starting at its tail.
This method is proved to be numerically stable. The main drawback of the 
backward recurrence algorithm is that it needs to be recalculated for
each  approximant. An alternative method, the forward recurrence 
algorithm is based upon the following 
\begin{equation}
\label{forward_rec1}
  S_n(w_n,z)=\frac{A_n(z)+ A_{n-1}(z)w}{B_n(z)+B_{n-1}(z)w}
\end{equation}
representation of $S_n(w_n,z)$,
where  $A_n(z),B_n(z)$ are called the $n$-th numerator, denominator,
respectively. The   $n$-th numerator and denominator  satisfy a recurrence
relation namely
\begin{eqnarray}
\label{forward_rec2}
   A_n(z)=b_n(z)+A_{n-1}(z)+a_n(z) A_{n-2}(z) \\
   B_n(z)=b_n(z)+B_{n-1}(z)+a_n(z) B_{n-2}(z),
\end{eqnarray}
with $A_{-1}=1,A_0=b_0,B_{-1}=0,B_0=1$. In contrast with the forward algorithm,
one can easily obtain $S_{n+1}(w_{n+1},z)$ from
$S_n(w_n,z)$ using the backward algorithm,
but on the other hand, this method is less stable than the first one.

A special class of continued fractions for which the limits 
\begin{equation}
  \lim_{n\rightarrow \infty }a_n(z)=a(z)\qquad \text{and}\qquad
  \lim_{n\rightarrow \infty }b_n(z)=b(z)
\end{equation}
exist for all $z\in D$ is called the limit 1-periodic continued fraction.
The more general case is the limit-k periodic continued fraction for which we
have
\begin{equation}
  \lim_{n\rightarrow \infty }a_{kn+p}(z)=a_p(z)\quad \text{and}\quad
  \lim_{n\rightarrow \infty }b_{kn+p}(z)=b_p(z) 
  \quad \mbox{for} \quad p=1,2, \ldots,k \ .
\end{equation}
Limit periodic continued fractions play an important role in the analytic theory
of continued fractions, since most of the continued fraction representation of
special functions are  limit periodic continued
fractions. 
The convergence properties of limit periodic continued fractions are determined
by the behavior of their tail by means of the 
$w_{\pm }(z)$ fixed points of the limit linear fractional transformation 
\begin{equation}
  s(w,z)=\lim_{n\rightarrow \infty }s_n(w_n,z)=\frac{a(z)}{b(z)+w}.
\end{equation}
The $w_{\pm }(z)$ fixed points are given as the solutions of the  
$s(w)=w$ fixed point equation
\begin{equation}
\label{fixed}
  w=\frac{a(z)}{b(z)+w} \quad \Rightarrow \quad 
  w_{\pm }(z)=-b(z)/2 \pm \sqrt{(b(z)/2)^2+a(z)}.
\end{equation}
The fixed point with smaller modulus is called attractive fixed point,
while the other one is named as the  repulsive one. Since $w_{\pm}(z)$
represent the tail of a \mbox{limit 1-periodic} continued fraction we can 
accelerate
the convergence using the attractive fixed point in
the approximant $S_n(w,z)$ \cite{accel}.

\subsection{Analytic continuation of continued fractions}
\label{sec:analytic_conti}

The idea of analytic continuation 
of a continued fraction
$b_0(z)+\mathop{K}(a_n(z)/b_n(z))$ by means of an appropriate 
choice of $w_n(z)$ for the
$S_n(w_n(z))$ approximant was proposed by Waadeland \cite{waad66} and later
recalled by Masson \cite{masson}.
By examining limit periodic continued fractions they sought a modification
of the tail of a continued fraction which led to its analytic extension.

If a continued fraction converges in a certain
complex region $ D$, then in many cases it is possible to extend the
region of convergence to a larger domain $D^{*}\supseteq D$, where $D^{*}$
depends  on the choice of the functions $w_n(z)$. 

In the case of limit
1-periodic continued fractions the analytic continuation $f_{D^{*}}(z) $
of the continued fraction  $f(z)$ in \eqref{cfdef2} is defined with the
help of the fixed points $w_{\pm }(z)$ of Eq.\ (\ref{fixed}) as 
\begin{equation}
\label{analcont}
   f_{D^{*}}(z)=\lim\limits_{n\to \infty }S_n(w_{\pm }(z),z).  
\end{equation}
In Eq.\ \eqref{analcont} the analytic expressions for the   $w_{\pm }(z)$
fixed points are continued analytically and then employed to sum up
the continued fraction, this way providing the analytic continuation
of the continued fraction.

\subsubsection{Bauer--Muir transformation}\label{bauer-muir}

The numerical computation of  approximants $S_n(w_{\pm }(z),z)$ 
might be unstable,
specially for  $z$ belonging to the extended region, which leads to 
unsatisfactory convergence. This problem can
be overcome by using the Bauer--Muir transformation of a continued fraction
(see eq. \cite{lorentzen}) We note that the method dates
back to the original work of Bauer \cite{bauer} and Muir \cite{muir}
in the 1870's.
 
The Bauer--Muir transform of a continued fraction 
$b_0(z)+K\left(a_n(z)/b_n(z)\right) $ with respect to a sequence 
of complex numbers $\left\{ w_n\right\}_{n=0}^\infty $
is defined as the continued fraction $d_0(z)+K\left(c_n(z)/d_n(z)\right)$,
whose ``classical'' approximants $S_n(0,z)$ are
equal to the modified approximants $S_n(w_n,z)$ of the original continued
fraction. The transformed continued fraction exists and can be calculated as 
\begin{eqnarray}
  d_0 &=&b_0+w_0,\quad c_1=\lambda _1,\quad d_1=b_1+w_1,  \label{bauermuir} \\
  c_i &=&a_{i-1}q_{i-1,}\quad d_i=b_i+w_i-w_{i-2}q_{i-1}, \quad i\geq 2, 
 \nonumber \\
  \lambda _i &=&a_i-w_{i-1}(b_i+w_i),\quad q_i= \lambda _{i+1}/\lambda _i\quad
  i\geq 1,  \nonumber
\end{eqnarray}
if and only if $\lambda _i\neq 0$ for $i=1,2,\ldots .$

\subsection{Pincherle's theorem}
Continued fractions and three-term recurrence relations are intimitaly 
connected.
The existence of a minimal solution of a \eqref{3term} three-term
 recurrence relation
is strongly related to the convergence
of a continued fraction constructed from the coefficients of the recurrence
relation. This connection is revealed by Pincherle in 1894 \cite{pincherle}.
Below we state Pincherle's theorem without its proof which can be found
in \cite{lorentzen,jones}. 

\begin{theorem}[Pincherle's Theorem]
Let $a_n$ and $b_n$ be a sequence of complex numbers with $a_n\neq 0 $
for $n=1,2, \dots$\\
{\bf (A)} The three-term recurrence relation
\begin{equation}
\label{pinch_3term}
   x_{n+1}=b_nx_n+a_nx_{n-1}, \qquad n=1,2,3,\ldots ,
\end{equation}
with coefficients $a_n$ and $b_n$ has a minimal
solution $\{m_n\}$ if and only if the continued fraction 
\begin{equation}
\label{pincherle1}
  \mathop{K}_{n=1}^\infty \left(\frac{a_n}{b_n}\right)=
  \frac{a_1}{b_1}
  {\atopwithdelims.. +}
  \frac{a_2}{b_2}
  {\atopwithdelims.. +} \cdots {\atopwithdelims.. +}
  \frac{a_n}{b_n}
  {\atopwithdelims.. +} \cdots 
\end{equation}
converges.\\
{\bf (B)} Provided \eqref{pinch_3term} has a minimal solution $\{m_n\}$,
then for $N=0,1,2,\ldots, $
\begin{equation}
\label{pincherle2}
  \frac{m_{N+1}}{m_N}=
  -K_{n=1}^\infty \left(\frac{a_{n+N}}{b_{n+N}}\right)=
  - \frac{a_{1+N}}{b_{1+N}}
  {\atopwithdelims.. +}
  \frac{a_{2+N}}{b_{2+N}}
  {\atopwithdelims.. +} \cdots {\atopwithdelims.. +}
  \frac{a_{n+N}}{b_{n+N}}
  {\atopwithdelims.. +} \cdots 
\end{equation}
\end{theorem}
\noindent Remarks: 
\begin{itemize}

\item  Equation \eqref{pincherle2}  asserts that the ratio of  two arbitrary
successive elements of the
minimal solution can be calculated by a continued fraction.
\item The connection between continued fractions and three-term recurrence
relations provides the link between continued fractions
and special functions (\mbox{i.e.} hypergeometric functions 
and orthogonal polynomials).
\item The genious Indian mathematican Ramanujan left behind, written 
in his notebook \cite{ramanujan}, many  interesting and original contributions
to modern mathematics among which several dealt with the 
theory of continued fractions. Unfortunately he did not give proofs 
of his ideas.
It is worth noting that 
mathematicians have found Pincherle's theorem as a useful tool to
prove some of Ramanujan's formulae. 

\end{itemize}

\chapter[Continued fraction representation of \ldots]
{\mbox{Continued fraction rep\-re\-sen\-ta\-tion} of Green's operators}
\label{chap:cfr}

In this chapter we present  a rather general and easy-to-apply
method for calculating discrete Hilbert space basis representation
of those Green's operators that correspond to Hamiltonians  having
infinite symmetric tridiagonal (i.e. Jacobi) matrix form.
The procedure necessitates the evaluation of the Hamiltonian matrix on this
basis.  The analytically
calculated elements of the Jacobi matrix are used to construct a 
continued fraction, which is utilized in the evaluation of the
Green's matrix. The constructed continued fraction representation of the
Green's operator is convergent in the bound state energy region 
and can be continued analytically
to the whole complex energy plane. 

Our method of calculating Green's matrices ensures a complete analytic
and readily computable representation of the Green's operator on the  
whole complex plane. Furthermore this is achieved at a very little cost:
in practice only
the matrix elements of the Hamiltonian are required.

After the expose of the method  the continued fraction representation of 
specific Green's operators are given.
The Coulomb Green's operator,  relativistic
Green's operators, the Green's operator corresponding to the D-dimensional
harmonic oscillator and the Green's operator of the
generalized Coulomb potential is  considered.
The convergence and the analytic continuation of the continued fraction is
illustrated with the example of the Coulomb Green's operator.
The numerical accuracy of the method is demonstrated via the calculation of the
relativistic energy spectrum of hydrogen-like atoms.

\section{The method}\label{sec:method}

In order to determine a matrix representation of the Green's operator, first a 
suitable basis has to be defined. Let us consider the set of  states 
$\{|i\rangle \}$ and $\{|\tilde{i}\rangle \}$,
with $i=0,1,2,\ldots $, which form a complete biorthonormal basis, i.e. 
\begin{eqnarray}
\label{biort}
  \langle \tilde{i}|j\rangle =\langle i|\tilde{j}\rangle =\delta _{ij} \\
  \nonumber
  {\bf 1}=\sum _{i=0}^{\infty }|\tilde{i}\rangle 
  \langle i|=\sum _{i=0}^{\infty }|i\rangle \langle \tilde{i}|
\end{eqnarray}
and render the $E-H$ operator symmetric tridiagonal
\begin{equation}
\label{jacobi}
  {J}_{i^{\prime }i}=\langle i^{\prime }|(E-{H})|i\rangle 
  =\left( \begin{array}{cccc} \cdot & \cdot & & \\ 
  \cdot & \cdot & \cdot & \\ & \cdot & \cdot & \cdot \\ 
  & & \cdot & \cdot \end{array} \right)_{\infty \times \infty} . 
\end{equation}
Here $E$ is a complex parameter and $H$ denotes the Hamilton operator.
According to Eq.\ \eqref{jacobi} we say that the Hamiltonian exhibits a Jacobi
matrix structure.

The  Green's operator   corresponding
to the Hamiltonian $ {H} $  satisfies the operator equation
\begin{equation}
\label{gdef}
  {\bf 1}={G}(E)(E-{H}).
\end{equation}
In the \eqref{biort} discrete biorthonormal basis representation 
this relation takes the form of the following matrix equation
\begin{equation}
\label{gginv}
  \delta _{ji}=\sum _{i^{\prime }=0}^{\infty } 
  \langle \tilde{j}|{G}(E)|\tilde{i}^{\prime }\rangle 
  \langle i^{\prime } |(E-{H})|i  \rangle
  =\sum _{i^{\prime }=0}^{\infty }G_{ji^{'}}J_{i^{'}i}.
\end{equation}
Since the basis is chosen such that $E-H$ possesses a tridiagonal matrix 
representation, the Green's matrix appears as the inverz of a symmetric
infinite tridiagonal matrix.
 
In Section \ref{sec:tridiagonal} the inverz of an infinite symmetric
tridiagonal matrix was studied and a formula for its $n$-th leading submatrix
was derived. Let us denote the $n$-th leading submatrix of the infinite 
Green's matrix by  $G^{(n)}$.
According to Theorem \ref{theorem:inverz},
$G^{(n)}$ can be written as 
\begin{equation}
\label{invn}
  {G}_{ij}^{(n)}=\left[ {J}_{ij}+ \delta _{jn}\, \, \delta _{in}\, \, 
  {J}_{n n+1}\, \, \frac{{G}_{0n+1}}{{G}_{0n}} \right] ^{-1}\ .
\end{equation}
Eq. (\ref{invn}) asserts that the  Green's matrix elements $G^{(n)}_{ij}$
can be calculated from the 
Jacobi matrix elements $J_{ij}$ provided the
${G}_{0n+1}/{G}_{0n}$ quotient is at our disposal.

On the other hand, from the tridiagonality of $E-H$ and from
Eq.\ \eqref{gginv} it automatically follows
that the matrix elements ${G}_{ji}=\langle \tilde{j}|{G}(E)|\tilde{i}\rangle$
satisfy a three-term recurrence relation 
\begin{equation}
\label{3termg}
  \delta _{ji}={G}_{ji-1}{J}_{i-1i}+{G}_{ji}{J}_{ii}+{G}_{ji+1}{J}_{i+1i}.
\end{equation}
Therefore ${G}_{0n+1}/{G}_{0n}$ represents a ratio of two consecutive elements
of a solution of a three-term recurrence relation.
It is known, that the solutions of  three-term recurrence relations
span a two-dimensional space and a special type of solution, called the 
minimal solution, is intimately connected to continued fractions.
(see Section \ref{sec:three-term} and \ref{sec:cont_frac}). 
According to Pincherle's Theorem, if the Green's matrix
elements represent a minimal solution of the \eqref{3termg} 
three-term recurrence relation,
then the ratio of its two consecutive  elements is given 
in terms of a convergent continued fraction
\begin{equation}
\label{frakk}
  \frac{G_{0n+1}}{G_{0n}}=-
  \cfrac{u_n}    {d_n+
  \cfrac{u_{n+1}}{d_{n+1}+ 
  \cfrac{u_{n+2}}{d_{n+2} + \cdots }}}  ,
\end{equation}
with coefficients 
\begin{equation}
\label{egyutth}
  u_i=-\frac{J_{i,i-1}}{J_{i,i+1}},
  \quad d_i=-\frac{J_{i,i}}{J_{i,i+1}}  .
\end{equation}

The Green's matrix elements $G_{ij}$ and the \eqref{3termg} three-term
recurrence relation, by the \eqref{gdef} definition,
obviously depend on the (complex) energy parameter $E$. 
First we show that there is a  region of the complex $E$-plane where
the physically
relevant solution of the (\ref{3termg}) recurrence relation  for the Green's
matrix is the minimal one, which makes possible a convergent continued
fraction representation of the Green's operator on this region by utilizing 
Eqs.\ \eqref{invn} and  \eqref{frakk}. Afterwards the analytic
expression of the convergent continued fraction is extended to other domain of
the $E$-plane, where the physical Green's matrix is not the minimal
solution of the recurrence relation.

On the $\Re E < 0$ region of the complex plane
and in case of short-range potentials the coordinate space representation of the
Green's operator can be constructed as \cite{newton} 
\begin{equation} 
\label{g}
  G(r,r^\prime,k)=\varphi_l(k,r_<) f_l^{(+)}(k,r_>)/{\cal F}(k), 
\end{equation}
where $\varphi_l(k,r)$ is the regular solution, ${\ f}_l^{(+)}(k,r)$ is the
Jost solution, ${\cal F}(k)$ is the Jost function and $k$ is the wave
number. The Jost solution is defined by the relation 
\begin{equation}
\label{jostdef}
  \lim\limits_{r\to\infty} \mbox{e}^{\mp i kr} {\ f}_l^{(\pm)}(k,r)=1.
\end{equation}
Let us define a ``new'' Green's function as 
\begin{equation}
  \widetilde{G}(r,r^\prime,k)=\varphi_l(k,r_<) f_l(k,r_>)/{\cal F}(k),
\end{equation}
where $f_l$ is a linear combination of ${\ f}_l^{(+)}$ and ${\ f}_l^{(-)}$.
If $\Re E < 0$ ${\ f}_l^{(+)}$ is exponentially decreasing and ${\ f}_l^{(-)}
$ is exponentially increasing. Thus, for any $\widetilde{G}$ we have 
\begin{equation} 
\label{gto0}
  \lim\limits_{r^\prime \to\infty} \frac{G(r,r^\prime,k)}
   {\widetilde{G}
   (r,r^\prime,k)} =0, \qquad \mbox{if\ \ } \Re E < 0. 
\end{equation}
We note, that both $G$ and $\widetilde{G}$ satisfy the defining equation
Eq.\ (\ref{gdef}), but only $G$ of Eq.\ (\ref{g}) is the physical Green's
function. The above considerations, with a slight modification in Eq.\ (\ref
{jostdef}), are also valid for the Coulomb case.
An interesting result of the study of Ref.\ \cite{l2_g} is that the Green's
matrix from Jacobi-matrix Hamiltonian,
has an analogous structure to Eq.\ (\ref{g}) 
\begin{equation}
\label{gJ}
  G_{ii^\prime}(k)=({\varphi_l})_{i_<}(k) 
  (f_l^{(+)})_{i_>}(k)/ {\cal F}(k),
\end{equation}
where $({\varphi_l})_{i}(k)=\langle \varphi_l(k) \vert \widetilde{i} \rangle$
and $(f_l^{(+)})_{i}(k)=\langle f_l^{(+)}(k) \vert \widetilde{i} \rangle$.
Similarly, we define 
$(f_l)_{i}(k)=\langle f_l(k) \vert \widetilde{i} \rangle$
and 
\begin{equation}
\label{gJt}
  \widetilde{G}_{ii^\prime}(k)= ({\varphi_l})_{i_<}(k) (f_l)_{i_>}(k)/ {
  \cal F}(k).  
\end{equation}
On the $\Re E < 0$ region of the complex plane as 
$r\to\infty$ $f_l (k,r)$
exponentially dominates over $f_l^{(+)}(k,r)$, thus for their $L^2$
representation the following relation holds 
\begin{equation}
\label{fgto0}
  \lim\limits_{i \to\infty} \frac{ (f_l^{(+)})_{i}(k)} {(f_l)_{i}(k)} =0,
  \qquad \mbox{if\ \ } \Re E < 0.  
\end{equation}
This implies a similar relation for the Green's matrices 
\begin{equation}
\label{ggto0}
  \lim\limits_{i^\prime \to\infty} \frac{G_{ii^\prime}(k)} 
  {\widetilde{G}_{ii^\prime}(k)} =0,
  \qquad \mbox{if\ \ } \Re E < 0.  
\end{equation}
So, we can conclude that in the $\Re E<0$ region of complex $E$-plane 
the physically relevant
Green's matrix $G_{ii^{\prime }}$ appears as the minimal solution of the
(\ref{3termg}) recurrence relation.
Therefore our Green's matrix  for bound-state energies 
can be determined by a convergent continued fraction.

In other regions of the complex energy plane (i.e.\ the region of 
scattering energies and resonances) 
the continued fraction fails to converge and
the recurrence relation does not have a minimal solution. On the other 
hand, $G_{0N+1}/G_{0N}$ in \eqref{frakk} is an analytic function of the complex 
energy parameter, and there is  a domain of the complex plane 
where this function is represented by a 
convergent continued fraction, thus values on other domains can be 
obtained by the analytic continuation of a continued fraction 
(see Section \ref{sec:analytic_conti}). 

Let us suppose  we have  a limit
1-periodic continued fractions, that is the 
$u_i$ and $d_i$ coefficients in (\ref{egyutth})
have the following limit property
\begin{equation}
\label{ud}
  \lim_{i\rightarrow \infty }u_i=u \hspace{.3cm} \mbox{and}
  \ \lim_{i\rightarrow \infty} d_i=d . 
\end{equation}  
In this case the continued fraction  (\ref{frakk}) takes the form
\begin{equation}
\label{cflim}
  \cfrac{G_{0 n+1}}{G_{0n}}=-
  \cfrac{u_n}    {d_n+
  \cfrac{u_{n+1}}{ d_{n+1}+ \cdots +
  \cfrac{u}      {  d +  
  \cfrac{u}      {  d +  \cdots }}}} \ .
\end{equation}
The $w$ tail of this continued fraction  satisfies the implicit
relation
\begin{equation}
\label{tail2}
  w=\frac{u}{d+w}\ , 
\end{equation}
which is solved by
\begin{equation}
\label{wpm}
  w_{\pm}=-d/2 \pm \sqrt{(d/2)^2+u}\ .
\end{equation}
Replacing the tail of the continued fraction by its explicit analytic 
form $w_{\pm}$, we can speed up the convergence and, which is more important, 
we can perform an analytic continuation. 
This implies the usage of
$w_{\pm}=w_{\pm}(z)$
in the \eqref{approxi} formula for the  $S_n(w,z)$
 continued fraction approximants
 also at those $z$ values,
where the continued 
fraction  fails to converge. 

The $w_{+}$ choice gives an analytic continuation to the physical sheet,
while $w_{-}$, which also converges, gives an analytic continuation to the
unphysical sheet.
This observation can be taken by considering
 the formula for the Green's operator on the physical sheet \cite{taylor} 
\begin{equation}
  \langle \widetilde{i} \vert G(E+\mbox{i}0)\vert \widetilde{i} \rangle -
  \langle \widetilde{i} \vert G(E-\mbox{i}0) \vert \widetilde{i} \rangle =
  -2\pi \mbox{i} \langle \widetilde{i} \vert \psi(E) \rangle \langle \psi(E)
  \vert \widetilde{i} \rangle,
\end{equation}
where $\psi(E)$ is the scattering wave function,
and utilizing the \eqref{green_anal} analytic properties
of Green's operators. Then we can readily obtain that the imaginary part of
$\langle \widetilde{i} \vert G(E+\mbox{i}0)\vert \widetilde{i} \rangle$
should be negative and this condition can only be fulfilled with the choice of 
$w_{+}$.

We have shown that
Eqs.\ (\ref{invn}) and (\ref{frakk}) together with the theory of analytic
continuation of a  continued fraction supply a complete discrete 
Hilbert space representation of the Green's operator on the whole complex
energy plane.
The Green's matrix can be obtained from the Jacobi Hamiltonian
for arbitrary complex energies by
simply evaluating a continued fraction and performing a matrix inversion.

\section{D-dimensional Coulomb Green's operator}\label{sec:coulomb}

Here we use the Coulomb--Sturmian basis and show that on this 
basis the D-dimensional Coulomb Hamiltonian
possesses a Jacobi-matrix structure. 
Then the analytically derived $J$-matrix elements are utilized, following the
method of the previous section, to calculate the continued fraction
representation of the Green's operator. The 
convergence of the continued fraction
and the technique of the analytic continuation is demonstrated in practice.

Let us consider the $D$-dimensional radial Coulomb Hamiltonian 
\begin{equation}
\label{coulham}
   H^{\rm C}=-\frac{\hbar^2}{2m}\left
   (\frac{\mbox{d}^2 }{\mbox{d} r^2}
   + \frac{1}{ r^2}\left(l+\frac{D-3}
  { 2}\right) \left(l+\frac{D-1}{ 2}\right)
  \right) 
   + \frac{Z{\rm e}^2}{ r}\ ,
\end{equation}
where $m$, $l$, ${\rm e}$ and $Z$ stand for the mass, angular momentum,
electron charge and charge number, respectively.
(See, e.g. Ref. \cite{knt85} and references.) 
The bound state energy spectrum is given by 
\begin{equation}
\label{hen}
  E_{n_r l}= -\frac{mZ^2e^4}{ 2\hbar^2(n_r+l+\frac{D-1}{ 2})^2}, 
\end{equation}
and the corresponding wave functions are 
\begin{equation}
\label{hwf}
  \psi_{n_r l}(r)= a_0 \left(\frac{r_0 \Gamma(n_r+1)} 
  {2\Gamma(n_r+2l+D-1)}\right)^{1/2} \exp(-\frac{a_0}{2}r)
  (a_0 r)^{l+\frac{D-1}
  {2}} L_{n_r}^{(2l+D-2)}(a_0 r),  
\end{equation}
where we used the notation $a_0= ((n_r+l+\frac{D-1}{2})r_0)^{-1}$
and $r_0=\hbar^2/(2mZe^2)$.

The Coulomb--Sturmian (CS) functions, the Sturm--Liouville
solutions of the Hamiltonian (\ref{coulham}), appear as
\begin{equation} 
\label{csf}
  \phi_{n l}(b,r) = \left( \frac{
  \Gamma(n+1)}{ \Gamma(n+2l+D-1)}\right)^{1/2} 
  \exp(-br) (2br)^{l+\frac{D-1}{2}} L_n^{(2l+D-2)}(2br), 
\end{equation}
where $b$ is a scale parameter, $n$ is the radial quantum number and 
$L_n^{(\alpha)}$ denotes the generalized Laguerre polynomials \cite{as}. 
The CS functions of \eqref{csf} are  the generalizations of the
corresponding CS functions for the three-dimensional case \cite{rotenberg}.
Introducing the notation
$\langle r\vert n\rangle \equiv\phi_{n l}(b,r)$ and
$\langle r\vert \widetilde{ n} \rangle \equiv \phi_{n l}(b,r)/r$
for the CS function and  its biorthonormal partner respectively,
we can express the orthogonality and completeness of
these functions as 
\begin{eqnarray}
\label{csort}
   \langle n \vert \widetilde{ n^{\prime}} \rangle 
   =\delta_{nn^{\prime}} \\
   {\bf 1}=\sum _{n=0}^\infty | \widetilde{n} \rangle \langle n |
   =\sum_{n=0}^\infty | n \rangle \langle \widetilde{n} |,
\end{eqnarray}
confirming that they form a discrete biorthonormal
basis in the sense of \eqref{biort}.
The overlap of two CS functions can be written in terms of a three-term
expression 
\begin{equation}
\begin{split}
\label{overlap_cs}
  \langle n|n^{\prime }\rangle =(2b)^{-1} \bigg[  
  &+\delta_{nn^{\prime}}(2n+2l+D-1)   \bigg. \\
  &-\delta _{nn^{\prime}-1} ((n+1)(n+2l+D-1))^{1/2} \\
  & \bigg. -\delta _{nn^{\prime }+1}(n(n+2l+D-2))^{1/2} \bigg]  .  
\end{split}
\end{equation}
A similar expression holds for the matrix elements of $H^C$ 
\begin{equation}
\begin{split}
\label{hmatrix_coul}
  \langle n|H^C|n^{\prime }\rangle =\frac{\hbar ^2b}{4m}\bigg[
  &+\delta_{nn^{\prime }}\left( 2n+2l+D-1- \frac{2}{r_0b}\right) \bigg. \\
  &+\delta _{nn^{\prime }-1}((n+1)(n+2l+D-1))^{1/2} \\
  &\bigg.+\delta_{nn^{\prime }+1}(n(n+2l+D-2))^{1/2}\bigg].  
\end{split}
\end{equation}

Let us denote the Coulomb Green's operator
as $G^{\rm C} (E)=(E-H^{\rm C})^{-1}$ and  determine its CS
matrix elements $G^{\rm C}_{i j}=
\langle \widetilde{ i } | G^{\rm C}_l |
\widetilde{ j } \rangle$ by applying the general method described previously.
The starting point in this procedure is the  
observation that the $J_{i j}=
\langle i |(E- H^{\rm C})|j  \rangle$ matrix possesses an
infinite symmetric tridiagonal i.e. Jacobi structure, and  the nonzero
elements of this tridiagonal matrix are obtained  
immediately from Eqs.\ \eqref{overlap_cs} and \eqref{hmatrix_coul} as
\begin{eqnarray}
\label{jacobi_coul}
  J_{ii}&=& (2i+2l+D-1) (k^2-b^2 )
  \frac{\hbar^2}{4mb}- Z{\rm e}^2  \\
  J_{ii-1}&=&-[i(i+2l+D-2)]^{1/2} (k^2+b^2 )
  \frac{\hbar^2}{4mb} \ , \nonumber
\end{eqnarray}
where $k=(2m E/\hbar^2)^{1/2}$ is the wave number.

Then the $n$-th leading submatrix 
$G^{{\rm C} (n)}_{ij}$ of the infinite Green's matrix
is represented by
\begin{equation}
\label{invn_c}
  G^{{\rm C} (n)}_{ij}=[J_{ij}+ 
  \delta _{jn}\, \, 
  \delta _{in}\, \, J_{nn+1}\, \, F ]^{-1}\ ,
\end{equation}
where $F$ is a  continued fraction
\begin{equation}
\label{frakk_c}
  F=-
  \cfrac{u_n}     {d_n+
  \cfrac{u_{n+1}}{d_{n+1}+ 
  \cfrac{u_{n+2}}{d_{n+2} + \cdots }}} \ ,
\end{equation}
with coefficients 
\begin{equation}
\label{egyutth_c}
  u_i=- {J_{i,i-1}}/{J_{i,i+1}}, 
  \quad d_i=- {J_{i,i}}/{J_{i,i+1}} \ .
\end{equation}

The above continued fraction $F$, as it stands, 
is only convergent  for negative energies, 
but since it is  a limit-$1$ periodic continued fraction,
i.e. its coefficients $u_i$ and $d_i$ possess the limit properties
\begin{eqnarray}
\label{coulomb_limits}
  u & \equiv & \lim_{i\rightarrow \infty }u_i=-1  \\
  d & \equiv & \lim_{i\rightarrow \infty} d_i= 2(k^2 -  b^2)/ 
 ( k^2 +   b^2)\ ,  \nonumber
\end{eqnarray}  
it can be continued analytically to the whole complex energy plane 
by replacing its tail  with 
\begin{equation}
\label{wpm_c}
  w_{\pm}=(b \pm  {\mathrm{i}}k )^2/(b^2+k^2)  \ ,
\end{equation}
according to Section \ref{sec:method}.
This way Eq.\ \eqref{invn_c} provides the CS basis
representation of the Coulomb Green's operator on the whole complex
energy plane.  

We note here that the Coulomb--Sturmian representation
of the Coulomb Green's operator  $G^{{\rm C}}_{ij}$ has already been 
calculated \cite{papp1,papp2,papp3,cpc} 
by evaluating a complicated integral 
of the coordinate space Green's operator and
the \eqref{csf} CS functions. This integral, for example in the case of
$ G^{{\rm C} }_{00}=\langle \widetilde{0} |G^{\rm C}| \widetilde{0} \rangle$ 
could be performed analytically leading to the result 
\begin{equation}
\begin{split}
\label{g00}
   G^{\rm C}_{0 0} &=-\frac{4mb}{\hbar^2(b-\mbox{i}k)^2} \frac{1}{l+(D-1)/2+
   \mbox{i}\gamma}   \\
   &\times \;_2F_1\left(-l-\frac{D-3}{2}+\mbox{i}\gamma, 1;l+\frac{D+1}{2}+
   \mbox{i}\gamma+2;
   \left(\frac{b+\mbox{i}k} {b-\mbox{i} k}\right)^2\right) 
\end{split}
\end{equation}
containing the $\mbox{}_2F_1$
hypergeometric function \cite{as}. Afterwards $G^{\rm C}_{0 0}$ together with
a three-term recurrence relation  could be applied 
in order to calculate other matrix elements recursively.

We point out  that with the choice 
of $Z=0$ the D-dimensional Coulomb Hamiltonian (\ref{coulham})
reduces to the D-dimensional kinetic 
energy operator and our formulas provide the CS basis representation of the 
Green's operator of the free particle as well.

\subsection{Convergence of the continued fraction}

Below we demonstrate the convergence and the numerical accuracy of the
\eqref{invn_c} continued fraction representation of
the Coulomb Green's operator.
We calculate the  $G_{00}^C(E)$ matrix element of the D-dimensional
Coulomb Green's operator for  $l=0$ and $D=3$ case at $E=(-100,0)$ and
$E=(1000,1)$
energy values from  the bound and scattering state regions of the
complex $E$-plane.
For comparison, the numerical value of $G_{00}^C(E)$ 
calculated by the analytic expression \eqref{g00} is also quoted here and
denoted by $G_A$.

The  continued fraction \eqref{frakk_c} has been evaluated 
by calculating its  $n$-th approximants with respect to 
different $\{w_n\}$ series.  
For  bound state energies  the convergence of the continued fraction
with respect to the different choice of $\{w_n\}$,  while for the
scattering case the effect of the analytic continuation and the
Bauer-Muir transformation has been examined.

In Table \ref{kotott}  we can observe excellent convergence of the continued
fraction to the exact value in all cases. In case of $\Re E \le 0$
the Green's matrix elements represent the minimal solution of a three-term
recurrence relation, thus due to Pincherle's theorem the
continued fraction is convergent.
The different choice of $\{w_n\}$, e.g.\ we took
the $w_n=0$, $w_n=w_{+}$ and $w_n=w_{-}$ choices,
influences only the speed of convergence.

In Table \ref{szort} the continued fraction approximants of $G_{00}(E)$
are shown for  a scattering energy case.  
In the region of $\Re E \ge 0$, in complete accordance with
Pincherle's theorem, the original continued fraction (\ref{frakk_c}) diverges,
only its analytic continuations are convergent. We recall here that
$w_n=w_{+}$ provides analytic continuation to the physical,
while $w_n=w_{-}$ to the unphysical sheet.
In Table \ref{szort} only approximants with respect to  $w_{+}$
are given. 
However, as the first column  shows, the convergence is rather
poor. This can be considerably  improved by the repeated application of the
Bauer--Muir transformation (see Section \ref{bauer-muir}).
In fact, as can be seen in the last column, an accuracy similar 
to the bound state case
can be easily  reached here with e.g.\ an eightfold Bauer--Muir transform.

Examining Table \ref{kotott} and \ref{szort}  we can draw 
the conclusion that the general and easily computable
continued fraction method of Section \ref{sec:method} 
provides a convergent, numerically stable and
accurate   representation on the whole complex plane.

\begin{table}[tbp]
\begin{center}
{\small 
\begin{tabular}{r|ccc|}
& \multicolumn{3}{c|}{$E=(-100,0)$} \\ 
$n$ & $G^{(0)}$ & $G^{(w_{+})}$ & $G^{(w_{-})}$ \\ \hline
1 & (-5.44922314793965,0) & (-5.59142801316938,0) & (-0.92906408986331,0) \\ 
2 & (-5.54075476366523,0) & (-5.56131039101044,0) & (-4.70080363351349,0) \\ 
3 & (-5.55501552420656,0) & (-5.55812941271530,0) & (-5.39340957492282,0) \\ 
4 & (-5.55726787304507,0) & (-5.55775017508704,0) & (-5.52662100417805,0) \\ 
5 & (-5.55762610832912,0) & (-5.55770176067796,0) & (-5.55192403535264,0) \\ 
6 & (-5.55768333962797,0) & (-5.55769530213874,0) & (-5.55663951922343,0) \\ 
7 & (-5.55769251168083,0) & (-5.55769441374319,0) & (-5.55750389878413,0) \\ 
8 & (-5.55769398510141,0) & (-5.55769428874846,0) & (-5.55766025755765,0) \\ 
9 & (-5.55769422223276,0) & (-5.55769427085427,0) & (-5.55768824220786,0) \\ 
10 & (-5.55769426045319,0) & (-5.55769426825710,0) & (-5.55769320762502,0)
\\ 
11 & (-5.55769426662100,0) & (-5.55769426787592,0) & (-5.55769408236034,0)
\\ 
12 & (-5.55769426761735,0) & (-5.55769426781946,0) & (-5.55769423553196,0)
\\ 
13 & (-5.55769426777843,0) & (-5.55769426781103,0) & (-5.55769426221577,0)
\\ 
14 & (-5.55769426780450,0) & (-5.55769426780976,0) & (-5.55769426684377,0)
\\ 
15 & (-5.55769426780872,0) & (-5.55769426780957,0) & (-5.55769426764335,0)
\\ 
16 & (-5.55769426780940,0) & (-5.55769426780954,0) & (-5.55769426778103,0)
\\ 
17 & (-5.55769426780951,0) & (-5.55769426780954,0) & (-5.55769426780465,0)
\\ 
18 & (-5.55769426780954,0) &  & (-5.55769426780870,0) \\ 
19 & (-5.55769426780954,0) &  & (-5.55769426780939,0) \\ 
20 &  &  & (-5.55769426780950,0) \\ 
21 &  &  & (-5.55769426780954,0) \\ 
22 &  &  & (-5.55769426780954,0) \\ \hline
& \multicolumn{3}{c|}{$G_{A} =(-5.55769426780954, 0)$}
\end{tabular}
}
\end{center}
\caption{Convergence of the continued fraction for the $G_{00}$ element of the
Green's matrix at $\Re E <0$. The first, second and
third column contain approximants of the continued fraction with $w_n=0$, $%
w_n=w_+$ and $w_n=w_-$, respectively. For comparison we also give the $G_A$
\eqref{g00} analytic result. All the $G_{00}$ values are scaled with $10^2$.
\label{kotott}}
\end{table}

\begin{table}[tbp]
\begin{center}
{\small 
\begin{tabular}{r|cccc|}
& \multicolumn{4}{c|}{$E=(1000,1)$} \\ 
$n$ & $G^{(w_{+})}(0)$ & $G^{(w_{+})}(1) $ & $G^{(w_{+})}(5) $ & 
$G^{(w_{+})}(8) $ \\ \hline
1 & (1.076,-0.678) & (1.8072,-0.3293) & (-0.4129544,-0.14238595) & 
(-0.2321154,-0.073120618) \\ 
5 & (1.074,-0.279) & (1.1225,-0.3783) & (4.29352799,-1.63424931) & 
(-1.4408861,-0.350899497) \\ 
10 & (1.198,-0.325) & (1.1425,-0.3162) & (1.13445656,-0.32244006) & 
(1.20667672,0.237375310) \\ 
15 & (1.110,-0.346) & (1.1497,-0.3353) & (1.14598003,-0.33019962) & 
(1.14702383,-0.332562329) \\ 
20 & (1.160,-0.307) & (1.1415,-0.3287) & (1.14512823,-0.33023791) & 
(1.14511731,-0.330140243) \\ 
25 & (1.141,-0.354) & (1.1478,-0.3298) & (1.14523860,-0.33015825) & 
(1.14523597,-0.330179581) \\ 
30 & (1.139,-0.312) & (1.1437,-0.3311) & (1.14522511,-0.33018993) & 
(1.14522395,-0.330182552) \\ 
35 & (1.157,-0.341) & (1.1458,-0.3290) & (1.14522377,-0.33017859) & 
(1.14522539,-0.330181124) \\ 
40 & (1.131,-0.325) & (1.1451,-0.3312) & (1.14522642,-0.33018226) & 
(1.14522527,-0.330181563) \\ 
45 & (1.158,-0.327) & (1.1448,-0.3294) & (1.14522458,-0.33018138) & 
(1.14522524,-0.330181440) \\ 
50 & (1.135,-0.337) & (1.1457,-0.3305) & (1.14522559,-0.33018133) & 
(1.14522527,-0.330181470) \\ 
55 & (1.149,-0.320) & (1.1446,-0.3300) & (1.14522512,-0.33018161) & 
(1.14522525,-0.330181465) \\ 
60 & (1.145,-0.340) & (1.1456,-0.3300) & (1.14522528,-0.33018135) & 
(1.14522526,-0.330181464) \\ 
65 & (1.140,-0.322) & (1.1449,-0.3304) & (1.14522527,-0.33018153) & 
(1.14522525,-0.330181466) \\ 
70 & (1.152,-0.334) & (1.1453,-0.3298) & (1.14522523,-0.33018143) & 
(1.14522526,-0.330181465) \\ 
75 & (1.137,-0.329) & (1.1452,-0.3304) & (1.14522528,-0.33018147) & 
(1.14522526,-0.330181466) \\ 
80 & (1.152,-0.327) & (1.1450,-0.3296) & (1.14522524,-0.33018146) & 
(1.14522526,-0.330181465) \\ 
85 & (1.140,-0.335) & (1.1454,-0.3302) & (1.14522527,-0.33018145) & 
(1.14522526,-0.330181465) \\ 
90 & (1.147,-0.323) & (1.1450,-0.3301) & (1.14522525,-0.33018147) & 
(1.14522526,-0.330181465) \\ 
95 & (1.146,-0.336) & (1.1453,-0.3300) & (1.14522526,-0.33018145) & 
(1.14522526,-0.330181465) \\ \hline
& \multicolumn{4}{c|}{$G_{A} =(1.14522526,-0.330181465) $}
\end{tabular}
}
\end{center}
\caption{Convergence of the continued fraction for the $G_{00}$ element of the
Green's matrix at $\Re E > 0$. 
All columns contain approximants of the analytic continuation of
the continued fraction with respect to $w_+$.
While the first, second, third
and fourth column contain approximants  without,
with one-fold, with five-fold and with eight-fold
Bauer-Muir transform, respectively.
For comparison we also give the $G_A$
\eqref{g00} analytic result. All the $G_{00}$ values
are scaled with $10^2$. \label{szort}}
\end{table}

\subsection{Numerical test}

\begin{figure}
\begin{center}
\resizebox{10cm}{!}{
\rotatebox{-90}{
\includegraphics{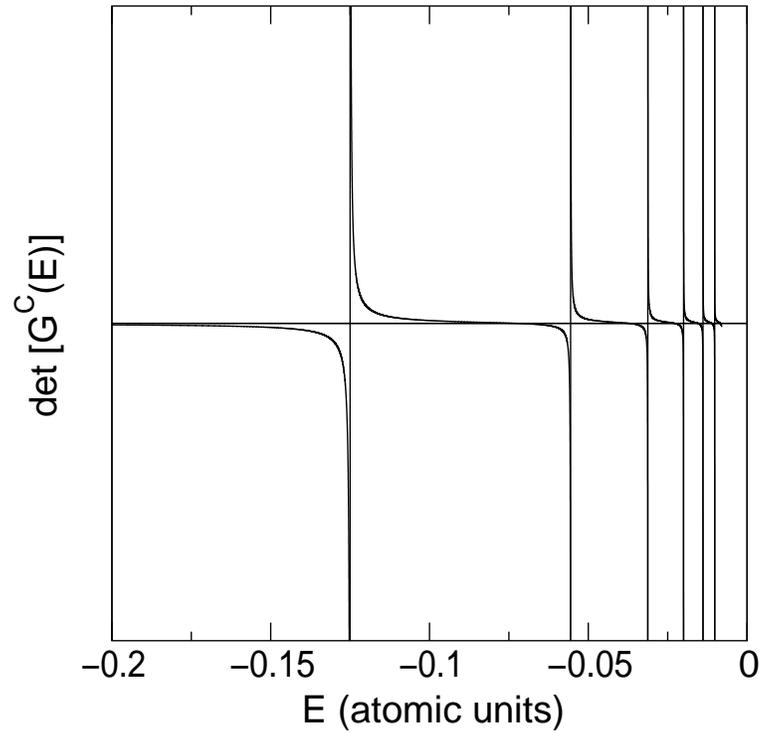}
}} 
\end{center}
\caption{The determinant of a $3 \times 3 $  three-dimensional 
Coulomb Green's matrix
${\rm det} [ G^C(E)]$ as the function of
the energy $E$ for $l=1$.
The bound states of the Coulomb problem are
located at energies where the vertical lines cross the 
horizontal axis. (These lines are shown only for demonstrative
purposes, they do not correspond to the values of
${\rm det} [G^C(E)]$.) Atomic units of $m={\rm e}=\hbar=1$ 
and $Z=-1$ were used.
For the sake of clarity only the first 6 energy levels are shown.
These are located at $E_n=-1/[2(n+l+1)]^2$ with $n\le 5$ according 
to \eqref{hen}. 
\label{coulfig}}
\end{figure}

We can immediately test the analytic properties of $G^C$ by determining  
the \eqref{hen} eigenvalues  of the attractive
Coulomb interaction in
three-dimension as the poles of the Green's matrix 
(see Section \ref{sec:green}). 
Figure \ref{coulfig} shows $\det [G^C(E)] $,
the determinant of the Green's matrix as the function of the $E$ energy
parameter.  The poles coincide with the exact
Coulomb energy levels 
up to machine accuracy. We stress that, from the point of view of 
determining the energy eigenvalues, the rank  of the matrix 
and the specific choice of the CS basis parameter are irrelevant.
An arbitrary-rank matrix  representation
of the Coulomb Green's operator exhibits all the properties
of the system and our Green's matrix contains all the infinitely many
eigenvalues. This is especially interesting if we compare with the usual
procedure of calculating eigenvalues of a finite Hamiltonian matrix
of rank $N$,  which could
only provide an upper limit for the $N$ lowest eigenvalues. Our procedure 
does not truncate the Coulomb Hamiltonian, since
all the higher $J_{ij}$  matrix elements are implicitly contained 
in the continued fraction.

In order to have a more stringent test we have performed the contour
integral 
\begin{equation}
\label{contourint2}
  I(C)=\frac 1{2\pi i}\oint_C\;\mbox{d}E \;G_{00}(E ).
\end{equation}
If the domain surrounded by $C$ does not contain any pole, then 
$I(C) = 0$. If this domain contains a single bound state pole, 
$I(C)=\langle \widetilde{0}|\psi \rangle \langle \psi | \widetilde{0}\rangle$
must hold,
while if $C$ circumvents the whole spectrum then 
$I(C)=\langle \widetilde{0}|\widetilde{0}\rangle$ 
is expected. With appropriate selection of Gauss
integration points we could reach 12 digits accuracy in all cases. This
indicates that the calculation of the Green's matrix from a J-matrix
via the continued fraction method is accurate on the whole complex plane.

\section{Relativistic Coulomb Green's operators}\label{sec:relgreen}

In this section we specify our
method  for relativistic Coulomb
Green's operators:  the Coulomb Green's operator of the Klein-Gordon
and of the second order Dirac equations. 
The latter is physically equivalent to the
conventional Dirac equation but seems to have several advantages from the
mathematical point of view.
For details see  Ref.\ \cite{hostler_rel} and references therein.

The  Hamiltonian of the radial Coulomb Klein--Gordon and second order
Dirac equations are shown to possess an infinite symmetric tridiagonal matrix 
structure on the relativistic Coulomb--Sturmian basis. 
This allows  us  to 
give  an analytic representation of  the corresponding
Coulomb Green's operators in terms of continued fractions. 
The poles of the Green's matrix reproduce the exact relativistic
hydrogen spectrum.

It is noted here that the Coulomb--Sturmian matrix elements 
of the second order Dirac equation 
has already been obtained by Hostler \cite{hostler_rel}
via evaluating complicated
contour integrals. Our derivation, however is much simpler, it relies only
on the Jacobi-matrix structure of the Hamiltonian, and the 
result obtained is also better suited for numerical calculations. 
In Hostler's paper the results appear in terms  of $\Gamma$ and
hypergeometric functions, while our procedure results in an easily  
computable and analytically continuable continued fraction.

The radial Klein-Gordon and second order Dirac  equation for Coulomb
interaction are given by
\begin{equation}
\label{raddirac}
  H_{u}  \left| \xi ^{u }\right\rangle =0, 
\end{equation}
where 
\begin{equation}
\label{relham}
  H_{u} = \left( \frac E{\hbar c}\right) ^2-\mu ^2+
  \frac{2\alpha Z}{\hbar c}\frac 
  Er+\frac{d^2}{dr^2}-\frac{u \left(u +1\right) }{r^2}.
\end{equation}
Here $\mu=mc/\hbar$, $\alpha=\mbox{e}^2/\hbar c$, $m$ is the mass
and $Z$ denotes the charge.
For the Klein-Gordon case $u$ is given by
\begin{equation}
  u =-\frac12 +\sqrt{\frac14+l(l+1)-(Z\alpha)^2},
\end{equation} 
and in the case of the second order Dirac equation
for the different spin states we have
\begin{equation}
  u_{\pm}= -\frac 12 \mp \frac 12 + 
  \sqrt{\left( j+\frac 12\right) ^2- (Z\alpha)^2}.
\end{equation} 
The relativistic Coulomb Green's operator 
is defined as the inverse  of the  Hamiltonian $H_{u}$
\begin{equation}
\label{relgreendef}
  H_{u }G_{u }=G_{u }H_{u }={\mathbf{1}}_{u },
\end{equation}
where ${\mathbf{1}}_{u }$ denotes the unit operator of the radial 
Hilbert space ${\cal H}_{u }$.

In complete analogy with the non-relativistic case we can define
the relativistic Coulomb--Sturmian functions as solutions of 
the Sturm-Liouville problem
\begin{equation}
\label{relsturmdef}
  \left( -\frac{d^2}{dr^2}+\eta ^2+\frac{u \left( u +1\right) }
  {r^2}-\frac{2\eta (n+u +1)}r\right) S_{n;\eta }^{u }(r)=0,
\end{equation}   
where  $\eta$ is a real parameter and $n=0,1,2, \ldots,\infty$ is
the radial quantum number. 
They take the form
\begin{equation}
\label{sturm}
  \left\langle r\left| n u;\eta \right.
  \right\rangle   \equiv S_{n;\eta }^{u }(r)  
  =\left[ \frac{n!}{\left( n+2u+1\right) !}\right]
  ^{\frac 12}\left( 2\eta r\right) ^{u+1}e^{-\eta r}L_n^{2u+1}(2\eta r), 
\end{equation}
where $L$ is a  Laguerre-polinom.
The relativistic Coulomb--Sturmian   functions,
together with their  biorthonormal partner
$\left\langle r\right.
 \widetilde{\left| n u;\eta \right\rangle }
=1/r\cdot \left\langle r\right. 
\left| n u;\eta \right\rangle $,
form a basis: i.e., they are orthogonal 
\begin{equation}
\label{orto}
  \widetilde{\left\langle n u;\eta \right. } \left| m u;\eta
  \right\rangle =\left\langle n u;\eta \right. 
  \widetilde{\left| m u;\eta \right\rangle }=\delta _{nm}, 
\end{equation}
and form a complete set in ${\cal H}_{u}$
\begin{equation}
\label{completeness}
  \sum_{n=0}^\infty {\left| n u;\eta \right\rangle }
  \widetilde{\left\langle n u;\eta \right| }=\sum_{n=0}^\infty
   \widetilde{ \left| n u;\eta \right\rangle } 
   \left\langle n u;\eta \right| 
  ={\mathbf{1}}_{u}. 
\end{equation}
A straightforward calculation yields
\begin{equation}
\label{reloverlap}
\begin{split}
  \left\langle n u;\eta \right| \!\left.m u;\eta \right\rangle \!\!
  &=\!\!\frac 1{2\eta }\!\left[ 
  \!\delta_{nm}\!\left( 2u+2n+2\right) -\delta _{nm-1\!}
  \sqrt{\left( n+1\right)\!\left( n+2u+2\right) }\right. \\ 
  &\left. -\delta _{nm+1}\sqrt{n\left( 2u +n+1\right) }\right]. 
\end{split}
\end{equation}
Utilizing  this relation and considering Eq.\ (\ref{relsturmdef}) we 
can easily calculate the Coulomb--Sturmian matrix elements of $H_{u}$,
\begin{equation}
\label{diracmatrix2}
\begin{split}
   H_{nm}&:=\left\langle nu;\eta \right| H_{u}\left| 
   m u;\eta \right\rangle = \\ 
   &+\delta _{nm}\left( \frac{2\alpha zE}{\hbar c}-2(u+n+1)\eta +2(u+n+1)
   \frac{(E/ \hbar c ) ^2-\mu ^2+\eta ^2}{2\eta }\right) \\ 
   &-\delta _{nm-1}\left( \frac{(E/ \hbar c ) ^2-\mu ^2+\eta ^2}{2\eta }
   \sqrt{(n+1)(n+2u+2)}\right) \\ 
   &-\delta _{nm+1}\left( \frac{(E/ \hbar c ) ^2-\mu ^2+\eta ^2}
   {2\eta }\sqrt{n(n+2u+1)}\right),
\end{split}
\end{equation}
which happen to possess a Jacobi-matrix structure. 

Now the Coulomb--Sturmian matrix elements of the relativistic
Coulomb Green's operators 
\begin{equation}
\label{greenmatrix}
  \left( G_{u}\right) _{nm}\equiv \widetilde{\left\langle
  n u;\eta \right| }G_{u}\widetilde{\left| m u;\eta
  \right\rangle }  
\end{equation}
corresponding to  Hamiltonians \eqref{relham}, 
can be straightforwardly determined by using the continued fraction method
of Equations \eqref{invn} and
\eqref{frakk} with the coefficients
\begin{equation}
\label{coef_rel}
  u_i  =  -\frac{H_{i i-1}}{H_{i i+1}}, \quad  
  d_i =  -\frac{H_{ii}}{H_{ii+1}}.
\end{equation}
Again, the continued fraction representation  convergent for bound-state 
energies and can be continued analytically to the whole complex energy plane.

\subsection{Relativistic energy spectrum}
In Table \ref{rel_spectrum} we demonstrate the numerical precision of 
our Green's matrix by evaluating the
ground and some highly excited sates of  
relativistic  hydrogen-like atoms, which, in fact, correspond to the poles
of the Dirac Coulomb Green's matrix. In particular, the
poles of the determinant of (\ref{greenmatrix}) were located.
Here we repeat again   that irrespective of the rank $N$ of the Green's matrix
the poles should provide the exact Dirac results. 
In  Table \ref{rel_spectrum}
we have taken $2\times 2$ matrices. Indeed, the results of
this method, $E_{\text{cf}}$, agree with the exact one 
in all cases, practically up to  machine accuracy, 
this way making possible  the study of the fine structure splitting.

\begin{table}[h]
\begin{center} 
\begin{tabular}{|l|l|l|l|l|}
\hline
  & energy levels & $E_{\mbox{cf}}$ & $E_{\mbox{D}}$ & 
$E_{\mbox{S}}$ \\ \hline
hydrogen & $1$S$_{1/2}$ & $-0.5000066521$ & $-0.5000066521$ & $%
-0.5 $ \\ \cline{2-5}
$ $ & $2$P$_{1/2}$ & $-0.1250020801$ & $-0.1250020801$ & $-0.125$ \\ 
\cline{2-5}
& $2$P$_{3/2}$ & $-0.1250004160 $ & $-0.1250004160 $ & $-0.125$ \\ 
\cline{2-5}
& $50$P$_{1/2}$ & $-0.0002000002 $ & $-0.0002000002 $ & $ 
-0.0002 $ \\ \cline{2-5}
& $50$P$_{3/2}$ & $-0.0002000001 $ & $-0.0002000001 $ & $ 
-0.0002 $ \\ \hline
uranium & $1$S$_{1/2}$ & $-4861.1483347 $ & $-4861.1483347 $ & $-4232$
\\ \cline{2-5}
& $100$D$_{3/2}$ & $-0.4241695002$ & $-0.4241695002$ & $-0.4232$ \\ 
\cline{2-5}
& $100$D$_{5/2}$ & $-0.4238303306$ & $-0.4238303306$ & $-0.4232$  \\
\hline
\end{tabular}
\end{center}
\caption{Energy levels of hydrogen-like atoms in atomic units.
$E_{\mbox{cf}}$ is the relativistic spectrum calculated via continued fraction,
 $E_{\mbox{ D }}$ and $E_{\mbox{\text S}}$ are 
textbook values of the relativistic Dirac
and the non-relativistic Schr\"odinger spectrum, respectively.
\label{rel_spectrum}}
\end{table}

\section{D-dimensional harmonic oscillator}

The Hamiltonian of the D-dimensional harmonic oscillator reveals
a Jacobi matrix representation in the harmonic oscillator
basis. Therefore the analytically calculated Jacobi-matrix
elements can be utilized as the input of the continued fraction
method for calculating harmonic oscillator basis representation of the
Green's operator corresponding to the harmonic oscillator potential.
 
The radial Hamiltonian describing the D-dimensional 
harmonic oscillator problem has the form 
\begin{equation}
\label{hho}
  H^{HO}= \left[ -\frac{\hbar^2}{2m}\left( 
  \frac{\mbox{d}^2 }{\mbox{d} r^2} -\frac{1}{ r^2}\left(l+\frac{D-3}{ 2}
  \right)\left( l+\frac{D-1}{ 2}\right)\right) 
  +\frac{1}{2}m\omega^2 r^2 \right],
\end{equation}
where $\omega$ is the harmonic oscillator parameter.
The energy eigenvalues are 
\begin{equation}
\label{heno}
  E_{n l}=\hbar\omega\left(2n +l+\frac{D}{2}\right)  
\end{equation}
and the corresponding eigenfunctions can be written as 
\begin{equation}
\label{hwfo}
  \langle r \vert \omega,nl \rangle =
   v^{\frac{1}{4}}\left(\frac{2\Gamma(n +1)}
   { \Gamma(n +l+\frac{D}{2})}\right)^{1/2} \exp(-\frac{v}{2}r^2) 
   (vr^2)^{\frac{l}{2}+\frac{D-1}{4}} L_{n} ^{(l+\frac{D}{2}-1)}(vr^2), 
\end{equation}
where $v=m\omega /\hbar$. The harmonic oscillator functions \eqref{hwfo} 
with fixed $\omega$ are orthonormal and form
a complete set in the usual sense
\begin{eqnarray}
\label{ho_orto}
  \langle  \omega,n'l  \vert  \omega,nl  \rangle =\delta_{n n'} \\ 
  {\bf 1}=\sum_{n=0}^{\infty} \vert \omega,nl
  \rangle \langle  \omega,nl \vert.
\end{eqnarray}
The harmonic oscillator Hamiltonian \eqref{hho} with parameter 
$\omega $ on the basis of the
harmonic oscillator functions with different parameter $\omega ^{\prime }$
takes a Jacobi-matrix form 
\begin{equation}
\begin{split}
\label{matrix_ho}
  \langle \omega ^{\prime },nl|H^{HO}(\omega) &| 
  \omega ^{\prime },n^{\prime}l\rangle  =  \\
  &+\delta_{n n^{\prime }}
  \left( \hbar \frac{\omega ^2+{\omega^{\prime }}^2}{2\omega ^{\prime }}
  \left(2n^{\prime }+l+\frac D2\right)\right) \\
  & - \delta_{n n^{\prime }-1 } \hbar \frac{\omega 
  ^2-{\omega ^{\prime }}^2}{2\omega^{\prime }}\left( n^{\prime }
  \left(n^{\prime }+l+ \frac D2-1\right)\right) ^{1/2}   \\
  & - \delta_{n n^{\prime }+1 } \hbar 
  \frac{\omega ^2-{\omega ^{\prime }}^2}
  {2\omega^{\prime }}\left( (n^{\prime }+1)\left(n^{\prime }+l 
  +\frac D2\right)\right) ^{1/2}  
\end{split}
\end{equation}
The general method of Section \ref{sec:method} requires the knowledge of
the matrix elements $J_{ij}=\langle \omega ^{\prime },il|E-H^{HO}(\omega) 
| \omega ^{\prime },jl\rangle$ which readily follows from \eqref{ho_orto}
and \eqref{matrix_ho} according to
\begin{equation}
\label{ho_jacobi}
  J_{ij}=E \times \delta_{ij}-  \langle \omega ^{\prime },il|H^{HO}(\omega) 
  | \omega ^{\prime },jl\rangle .
\end{equation}  
The calculation of the Green's matrix 
$G_{ij}(E)=\langle \omega ^{\prime },il|G(E,\omega) 
| \omega ^{\prime },jl\rangle$ goes similarly to the previous sections
making use of formulae   \eqref{invn} and \eqref{frakk}.

It is impossible to overestimate the importance of the harmonic oscillator 
in theoretical physics. Here I would like only to mention one exotic
topic, the physics of anyons, which are  quantum mechanically 
indistinguishable particles following fractional statistics,
where the harmonic  oscillator potential plays an important role
\cite{anyons}.

\section{The generalized Coulomb potential}\label{sec:gezapot}

Quantum mechanical models and practical calculations often rely 
on some exactly solvable models like the Coulomb and the 
harmonic oscillator potentials.
The actual example we consider here is the generalized Coulomb potential 
introduced by L\'evai and Williams \cite{lgbw93},
which is the member of the Natanzon confluent 
potential class \cite{cs91}. 
This potential is Coulomb-like asymptotically, while its short-range 
behavior depends on the parameters: it can be finite or singular 
as well at the origin. Its shape therefore can approximate various 
realistic problems, such as nuclear potentials with relatively 
flat central part, or atomic potentials that incorporate the effect of 
inner closed shells by a phenomenological repulsive core. 
Another interesting feature of the  D-dimensional generalized 
Coulomb potential is that it contains
the Coulomb and harmonic oscillator potentials as  limiting cases, 
this way providing a smooth
transition between the Coulomb and the harmonic oscillator problems
in various dimensions. 

More and more interactions can be modelled by making 
advantage of the rather flexible potential shapes offered by 
exactly solvable potentials.
Virtually all quantum mechanical 
methods rely in some respect on
analytically solvable potentials. Very often their wave function 
solutions are used as  Hilbert space bases.
More powerful methods can be constructed if we select a basis
which allows the exact analytical calculation of the Green's
operator of an analytically solvable potential.

In this section we show that an appropriate Sturm--Liuville basis
can be defined on which the
matrix elements of the
Hamiltonian exhibit a Jacobi matrix. The corresponding Green's matrix
then follows from the method of Section \ref{sec:method}.

\subsection{The potential}\label{gcoul} 

Let us consider the radial Schr\"odinger equation in D spatial 
dimensions with a potential $V(r)$ that depends only on the radial 
variable $r$ 
\begin{equation}
\label{sch}
  H \psi(r)\equiv
  \left(
  -\frac{ {\rm d}^2 }{ {\rm d} r^2} 
  +\frac{1}{  r^2}(l+\frac{D-3}{ 2})(l+\frac{D-1}{ 2}) 
  + v(r) \right) \psi(r)=\epsilon\psi(r)\ , 
\end{equation}
where $v(r)\equiv 2m\hbar^{-2} V(r)$ and $\epsilon \equiv  
2m\hbar^{-2} E$. 
We define the generalized Coulomb potential \cite{lgbw93}  in D-dimension
as   
\begin{equation}
\begin{split}
\label{pot}
  v(r)  = & -\frac{1}{  r^2}(l+\frac{D-3}{ 2})(l+\frac{D-1}{ 2}) 
  + (\beta-\frac{1}{2})(\beta-\frac{3}{2}) \frac{C}{4h(r)(h(r)+\theta)}  \\
  & -\frac{q}{h(r)+\theta} - \frac{3C}{16 (h(r)+\theta)^2} 
  +\frac{5C\theta}{16(h(r)+\theta)^3} \ ,  
\end{split}
\end{equation}
where $h(r)$ is defined  in terms of its inverse function 
\begin{equation}
\label{rh}
  r=r(h)=C^{-\frac{1}{2}}\left[ \theta \tanh^{-1}\left(\left(
  \frac{h}{h+\theta}\right)^{\frac{1}{2}}\right)
  +(h(h+\theta))^{\frac{1}{2}}\right] \ .
\end{equation}
The $h(r)$ function maps the $[0,\infty)$ half axis onto 
itself and can be approximated with $h(r)\simeq C^{\frac{1}{2}}r$ 
and $h(r)\simeq Cr^2/(4\theta)$ in the $r\rightarrow\infty$ and 
$r\rightarrow 0$ limits, respectively. 
 
Bound states are located at 
\begin{equation}
\label{en}
  \epsilon_n = -\frac{C}{4}\rho_n^2 \ ,
\end{equation}
where
\begin{equation}
\label{rho}
  \rho_n= 
  \frac{2}{\theta}\left[\left((n+\beta/2)^2 +\frac{q\theta}{C}
  \right)^{\frac{1}{2}}-(n+\beta/2)\right] \ ,
\end{equation}
and the bound-state wave functions can be written in terms of associated 
Laguerre--polynomials as 
\begin{equation}
\begin{split}
\label{bswf}
  \psi_{n}(r)&=C^{\frac{1}{4}}\rho_n^{\frac{\beta+1}{2}}
  \left(\frac{\Gamma(n+1)}{ 
  \Gamma(n+\beta)(2n+\beta+\rho_n \theta)}\right)^{1/2}  \\
  & \times (h(r)+\theta)^{\frac{1}{4}} (h(r))^{\frac{2\beta-1}{4}}
  \exp(-\frac{\rho_n}{2}h(r)) L_n^{(\beta-1)}(\rho_n h(r)) \ .
\end{split}
\end{equation}

Potential (\ref{pot}) clearly carries angular momentum dependence: 
its first term merely compensates the centrifugal term arising 
from the kinetic term of the Hamiltonian. Its second term also 
has $r^{-2}$-like singularity (due to $h^{-1}(r)$) and
it cancels the angular momentum dependent term in the two 
important limiting cases that recover the $D$-dimensional Coulomb 
and the harmonic oscillator potentials. The third term of (\ref{pot}) 
represents an asymptotically Coulomb--like interaction, while the 
remaining two terms behave like $r^{-2}$ and $r^{-3}$ for large 
values of $r$. 
The long-range behavior of potential (\ref{pot}) suggests its use 
in problems associated with the electrostatic field of some charge 
distribution. The deviation from the Coulomb potential close to 
the origin can be viewed as if  the point-like charge could be replaced with 
an extended charged object. The relevant charge density is readily 
obtained from the potential using 
\begin{equation}
\label{chd}
  \rho(r) = -\frac{\hbar^2}{8\pi m{\rm e}} \Delta v(r)\ .
\end{equation}

\begin{figure}
{\centering \begin{tabular}{|c|c|c|}
\hline 
\resizebox*{3.84cm}{7cm}{\rotatebox{-90}
{\includegraphics{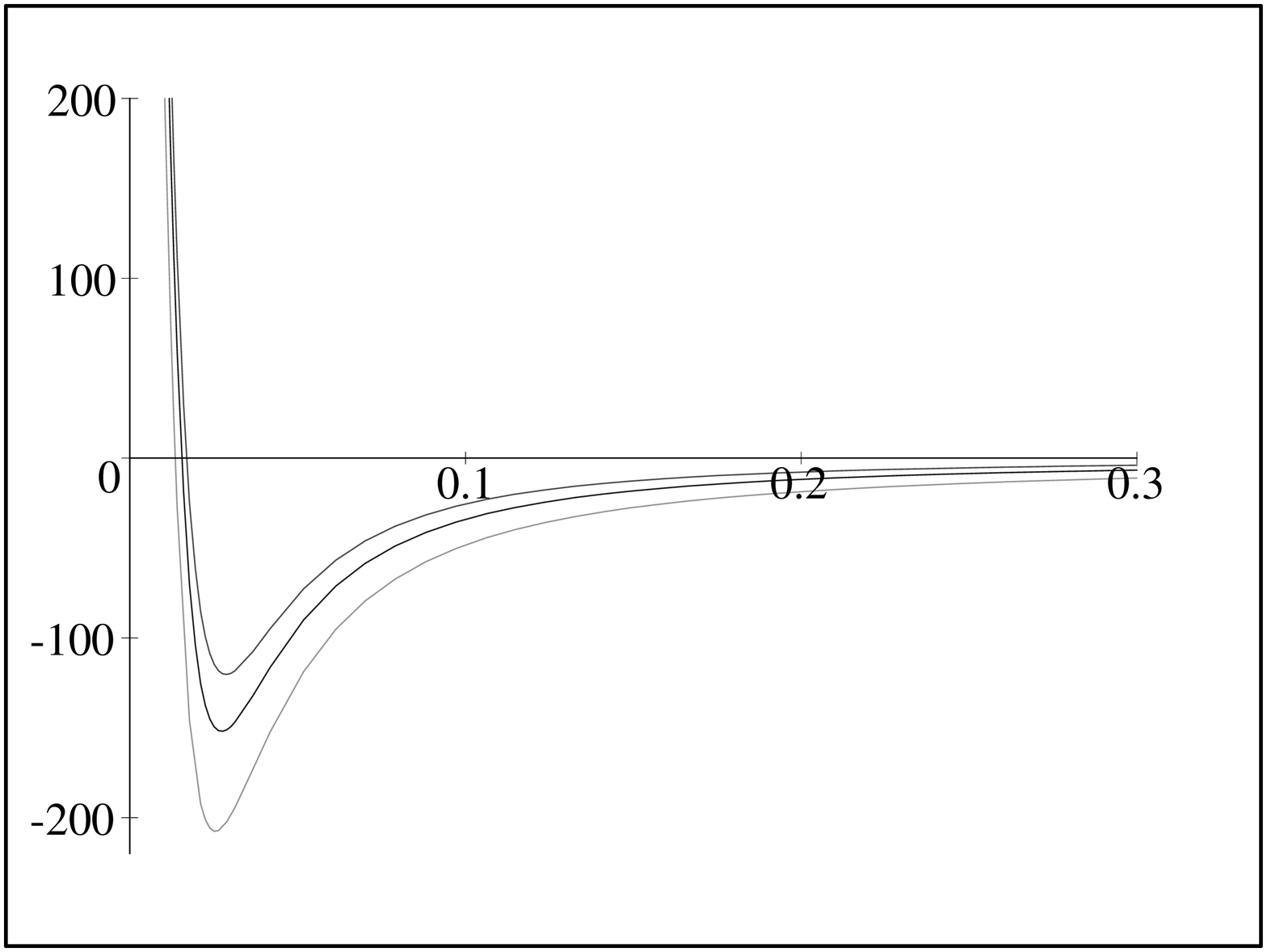}}} &
\resizebox*{3.84cm}{7cm}{\rotatebox{-90}
{\includegraphics{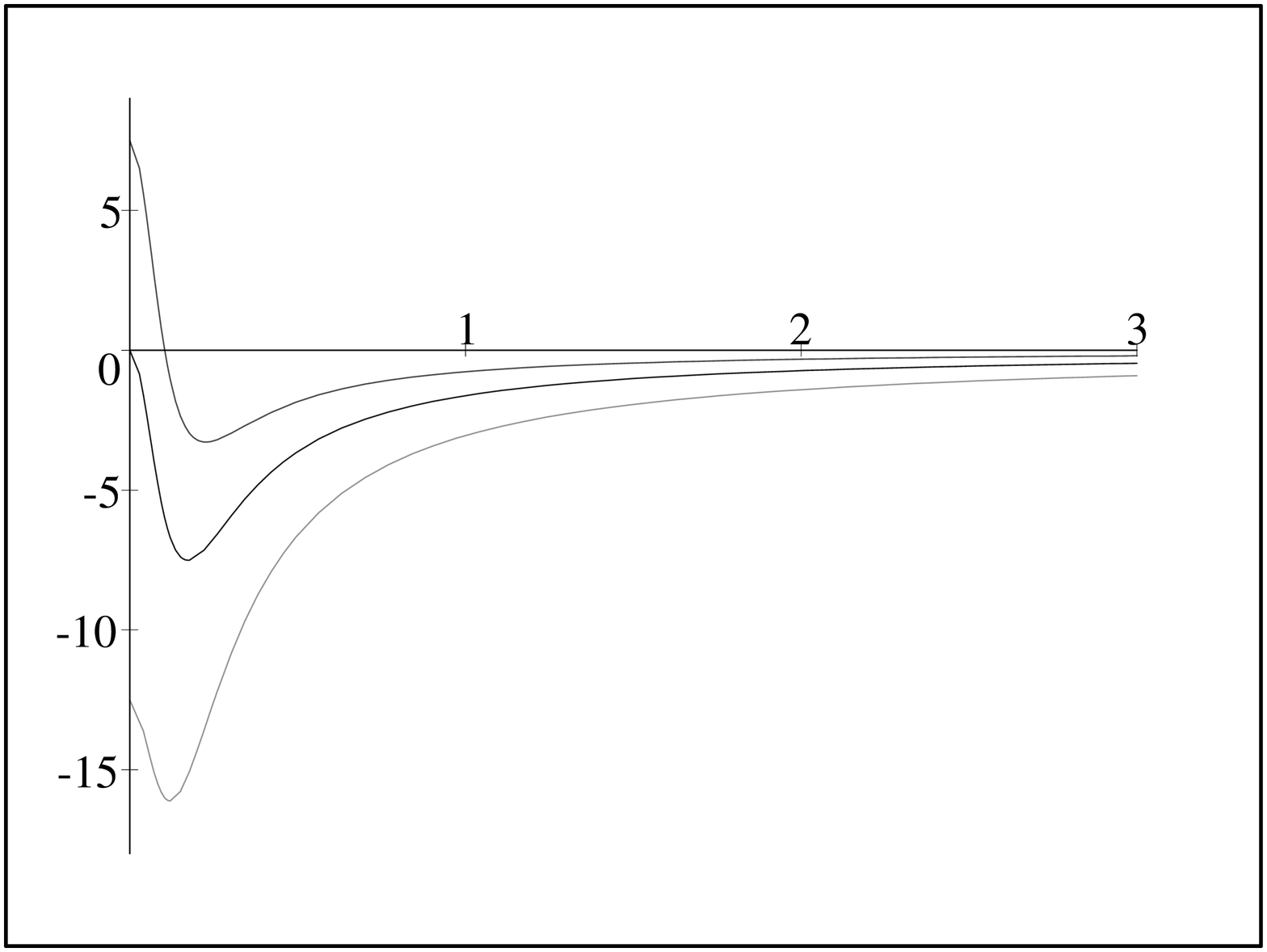}}} &
\resizebox*{3.84cm}{7cm}{\rotatebox{-90}
{\includegraphics{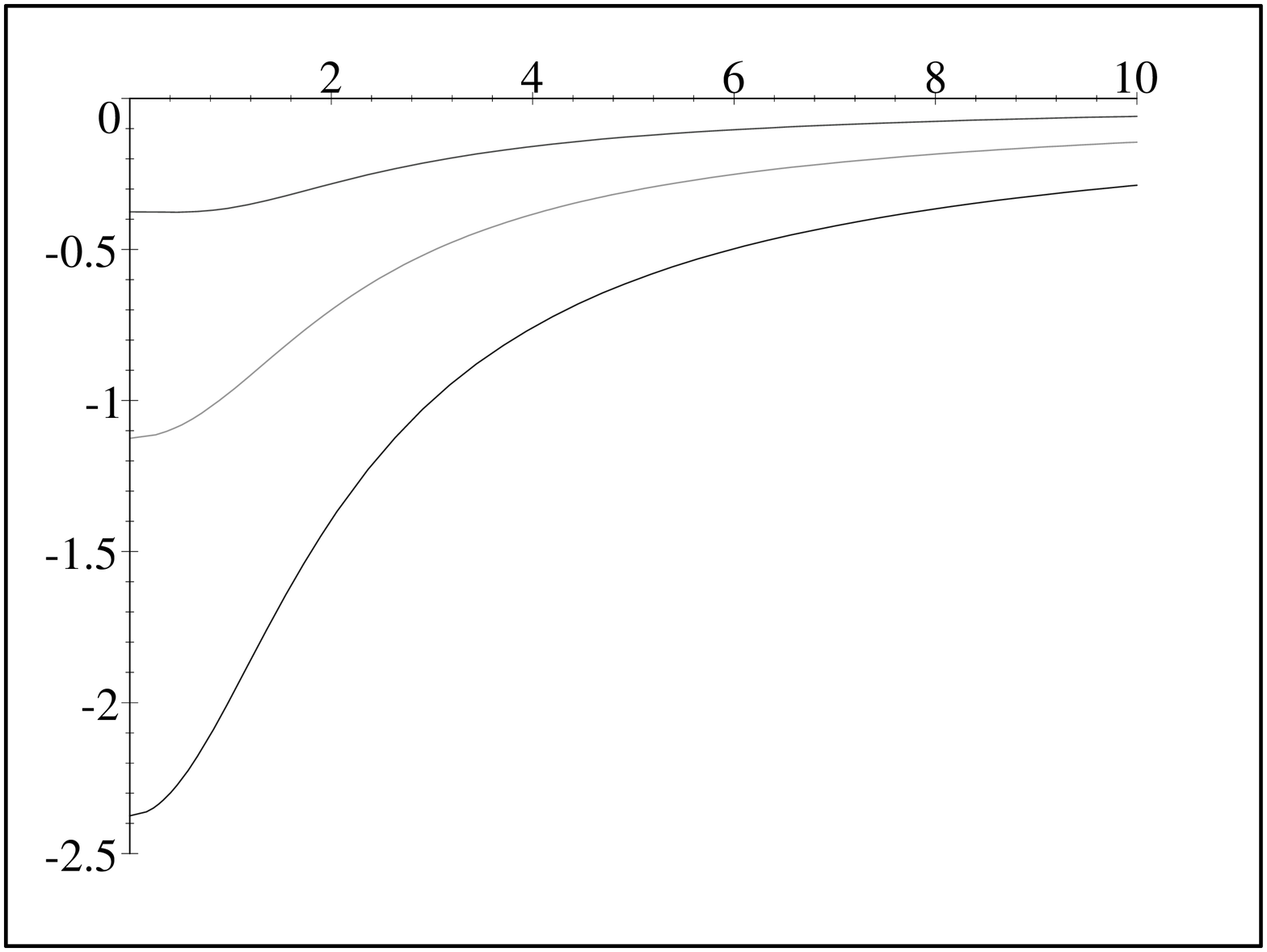}}} \\
$\theta=0.01$&
$\theta=0.1$&
$\theta=1$\\
\hline 
\end{tabular}\par}
\caption{The generalized Coulomb potential for $q$=0.5, 1.25, 2.5;
$\theta$=0.01, 0.1, 1; $C$=1 and $\beta=3/2$. $l=0$ and $D=3$ 
is also implied. 
In each panel the largest $q$ corresponds to the 
lowest curve. 
Note the different scales of the horizontal 
($r$) and the vertical ($v(r)$) axes. \label{fig:gezapot}}
\end{figure}

\begin{figure}
{\centering
\resizebox*{8cm}{!}{\rotatebox{-90}
{\includegraphics{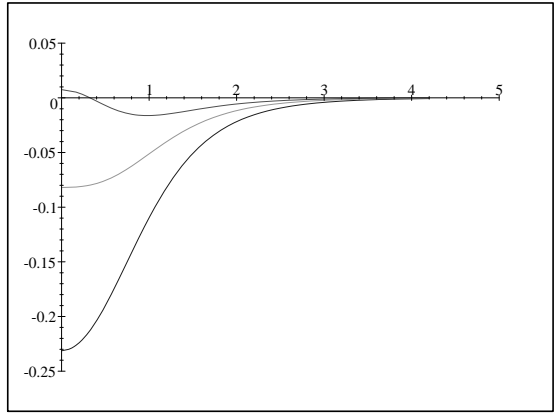}}} \par}
\caption{The charge distributions corresponding to the potentials displayed
in Fig. \ref{fig:gezapot} for $\theta$=1 and $q$=0.5, 1.25, 2.5. 
(The largest $q$ corresponds to the lowest curve.)
\label{fig:charge_dist}}
\end{figure}

In Figures \ref{fig:gezapot} and \ref{fig:charge_dist}
we present examples for the actual shape of 
potential 
(\ref{pot}) and the corresponding charge distribution (\ref{chd}) 
for various values of the parameters. It can be seen that 
this potential is suitable for describing the Coulomb field 
of extended objects. It is a general feature of potential (\ref{pot}) 
that for small values of 
$\theta$ a (finite) positive peak appears near the origin, which 
also manifests itself in a repulsive ``soft core'', corresponding 
to a region with positive charge density (see Fig. \ref{fig:charge_dist}). 

\subsubsection{The Coulomb and harmonic oscillator limits}

The special limits of the generalized Coulomb potential can be realized by
specific choices of the parameters in Eq.\ (\ref{rh}):

The $D$-dimensional Coulomb limit follows from the $\theta\rightarrow 0$ 
limit and it is recovered from Eq.\ (\ref{pot}) by the  
$\beta=2l+D-1$ and $C^{-\frac{1}{2}}q=2mZ{\rm e}^2/\hbar^2$, 
choices: the third term of (\ref{pot}) becomes the Coulomb term, 
the fifth one vanishes, while the other three terms  becoming 
proportional with $r^{-2}$  cancel out completely.   

In order to reach the oscillator limit one has to take 
$\theta\rightarrow \infty$ keeping $C/\theta\equiv \tilde{C}$ 
constant together with the 
redefinition of  the potential (\ref{pot}) and the energy eigenvalues 
by adding $q/\theta$ to both.  This choice simply represents 
resetting the energy scale: $\epsilon=0$ corresponds to 
$v(r\rightarrow\infty)$ 
for the Coulomb problem, and to $v(r=0)$ for the harmonic oscillator. 
(Note that the energy eigenvalues also have different signs in the 
two cases.) Besides $C/\theta =\tilde{C}$, the 
$\tilde{q} \equiv q/\theta^2$ 
parameter also has to remain constant in the    
$\theta\rightarrow\infty$ transition here. The potential thus 
adapted to the harmonic oscillator limit reads 
\begin{equation}
\begin{split}
\label{potho}
  \tilde{v}(r) & \equiv v(r)+q\theta  = \\  
  & \quad -\frac{1}{  r^2}\left(l+\frac{D-3}{ 2}\right)\left(l+
  \frac{D-1}{ 2}\right) 
  + \left(\beta-\frac{1}{2}\right)\left(\beta-\frac{3}{2}\right)
  \frac{\tilde{C}}{4h(r)(1+\frac{h(r)}{\theta})}  \\
  & \quad -\frac{\tilde{q} h(r)}{1+\frac{h(r)}{\theta}} 
  - \frac{3\tilde{C}}{16 \theta} 
  \frac{1}{\left(1+\frac{h(r)}{\theta}\right)^2} 
  +\frac{5\tilde{C}}{16\theta}\frac{1}{\left(1+\frac{h(r)}{\theta}
  \right)^3} \ .
\end{split}  
\end{equation}
The harmonic oscillator potential is recovered from (\ref{potho}) 
by the $\beta=l+D/2$ and $\tilde{C} \tilde{q}=(2m\omega/\hbar)^2$ 
choice. The 
two last terms in (\ref{potho}) vanish, the first and the second 
cancel out, while the third one reproduces the harmonic oscillator 
potential. The new form of the energy eigenvalues is 
\begin{equation}
\label{enho}
  \tilde{\epsilon}_n \equiv \epsilon_n +q/\theta = 
  \tilde{C}(2n+\beta)\left[\left( \frac{1}{\theta^2}(n+
  \frac{\beta}{2})^2 
  +\frac{\tilde{q}}{\tilde{C}}\right)^{\frac{1}{2}} -\frac{1}{\theta}
  (n+\frac{\beta}{2})\right] ,
\end{equation}
which indeed, reduces to the 
$\tilde{\epsilon}_n=(2m\omega/\hbar)(2n+l+D/2)$ 
oscillator spectrum in the $\theta\rightarrow\infty$ limit. 
The wave functions (\ref{bswf}) are unchanged, except for the 
redefinition of the parameters.

\subsection{The matrix elements of the Green's operator}\label{green}

We define the generalized Coulomb--Sturmian basis as the solution of the 
generalized Sturm--Liouville equation.
The Sturm--Liouville  equation, which depends on $n$ as a parameter and
corresponds to the  generalized Coulomb potential (\ref{pot}), reads
\begin{multline}
\label{gcst}
  \Bigg[
  -\frac{{\rm d}^2}{{\rm d} r^2} 
  -\frac{3C}{16 (h(r)+\theta)^2} 
  +\frac{5C\theta}{16(h(r)+\theta)^3} 
  +\frac{C(\beta-\frac{1}{2})(\beta-\frac{3}{2})}{4h(r)(h(r)+\theta)}  \\
  -\left(\frac{\rho^2\theta}{4}+\rho(n+\frac{\beta}{2})\right)
  \frac{C}{h(r)+\theta} + \frac{C}{4}\rho^2 \Bigg] \phi(\rho,r)=0\ ,
\end{multline}
and is solved by the generalized Coulomb--Sturmian (GCS) functions 
\begin{multline}
\label{gcsf}
  \langle r \vert n \rangle \equiv \phi_{n}(\rho,r)=  \\
  \left( \frac{\Gamma(n+1)}{ 
  \Gamma(n+\beta)}\right)^{1/2} (\rho h(r)+\rho\theta)^{\frac{1}{4}} 
  (\rho h(r))^{\frac{2\beta-1}{4}} \exp(-\frac{\rho}{2}h(r))
  L_n^{(\beta-1)}(\rho h(r)) \ . 
\end{multline}
Here $\rho$ is a parameter 
characterizing the generalized Coulomb--Sturmian basis. The GCS 
functions, being solutions of a Sturm--Liouville problem,
have the property of being orthonormal with 
respect to the weight function $C^{\frac{1}{2}}(h(r)+\theta)^{-1}$. 
Introducing the notation $\langle r \vert \widetilde n \rangle 
\equiv \phi_n(\rho,r) C^{\frac{1}{2}}(h(r)+\theta)^{-1}$ 
the orthogonality and completeness relation of the GCS functions 
can be expressed as 
\begin{equation}
\begin{split}
\label{gcs_orto}
  \langle n' \vert \widetilde n \rangle & =\delta_{n' n} \\
  {\bf 1}=\sum_{n=0}^{\infty} \vert \widetilde n \rangle \langle n \vert 
  &=\sum_{n=0}^{\infty} \vert  n \rangle \langle \widetilde n \vert \ . 
\end{split}
\end{equation}

Analytic  calculations yield that both the overlap of 
two GCS functions and the 
$H_{n' n} = \langle n' \vert H \vert n \rangle $ 
Hamiltonian matrix possesses  a tridiagonal form, 
therefore the matrix elements of the $\epsilon- H$ operator 
also have this feature  
\begin{equation}
\begin{split}
\label{eminh0}
  \langle n \vert \epsilon- H \vert n' \rangle  = 
  & +\delta_{nn'}\left[ \frac{\epsilon}{C^{\frac{1}{2}}\rho} (2n+\beta -
  \rho\theta ) -\frac{C^{\frac{1}{2}}\rho}{4} \left( -\frac{4q}{C\rho} 
  +(2n+\beta)  \right) \right] \\
  & - \delta_{n n'+1} \left( n(n+\beta-1)\right)^{\frac{1}{2}} 
  \left(\frac{\epsilon}{C^{\frac{1}{2}}\rho} + \frac{C^{\frac{1}{2}}
  \rho}{4}\right)  \\
  & -  \delta_{n n'-1} \left( (n+1)(n+\beta)\right)^{\frac{1}{2}} 
  \left(\frac{\epsilon}{C^{\frac{1}{2}}\rho} + \frac{C^{\frac{1}{2}}
  \rho}{4}\right) \ . 
\end{split}
\end{equation}
This means, that similarly to the previous sections the 
matrix elements of the 
Green's operator in the GCS basis, 
$G_{ij}= \langle \widetilde n \vert  G \vert \widetilde n \rangle $,
can be determined by using continued fractions, 
as described in Section \ref{sec:method} utilizing the analytically known
Jacobi-matrix elements of \eqref{eminh0}.

\chapter{Applications}\label{chap:applications}

The  continued fraction method for calculating 
Green's matrices on the whole
complex energy plane together with  methods
for solving integral 
equations in discrete Hilbert space basis representation
provide  a  rather general and easy-to-apply  
quantum mechanical approximation scheme.

In the first part of this chapter  the continued fraction 
representation of the
Coulomb--Sturmian space  
Coulomb Green's operator (Section \ref{sec:coulomb}) is used for solving the 
two-body Lippmann--Schwinger equation 
with a potential modelling the interaction of two $\alpha$ particles
in order to find bound, resonance and scattering solutions.

In the second part of this chapter 
our Green's operator is applied for solving the Coulomb
three-body bound state  problem in the Faddeev--Merkuriev integral
equation approach. In particular, the binding energy of the Helium atom is 
determined by solving the Faddeev--Merkuriev equations 
in the Coulomb--Sturmian space representation.

Both solution schemes have been devised by Papp in
Refs.\ \cite{papp1,papp2,papp3,cpc} and 
Refs.\ \cite{pzwp,pzsc}, respectively.
We demonstrate here that the continued fraction
representation of the Coulomb Green's operator
in practice is as good as the original one given
by Papp in terms of hypergeometric functions.  

The two examples of this chapter are  intended  to show the importance of the
analytic representation of the Green's operators through the 
efficiency of the discrete Hilbert space expansion method for solving 
fundamental integral equations.

\section{Model nuclear potential calculation}

In this section we apply the method of Refs.\ \cite{papp1,papp2,papp3,cpc} 
together 
with the continued fraction
representation of the Coulomb Green's operator in order to 
calculate bound, resonant and scattering state
solutions of a potential problem in a unified manner. The 
particular example we consider here is a potential modelling  the 
interaction of two $\alpha$ particles. 
This example is thoroughly discussed in the pedagogical work 
\cite{schmid} in the context of a conventional approach based on 
the numerical solution of the Schr\"odinger equation.
 
The interaction of  two $\alpha$ particles can be approximated 
by the potential  
\begin{equation}
\label{be8pot}
  {V}_{\alpha-\alpha}(r)=
  -A \exp(-\beta r^2) + \frac{Z^2{\rm e}^2}{r}\ {\rm erf}(\gamma r) , 
\end{equation}
where ${\rm erf} (z)$ is the error function \cite{as}. 
This potential is a composition of a bell-shaped deep, attractive 
nuclear potential, and a repulsive electrostatic field between two 
extended  charged objects. 
The units used in the Hamiltonian of this system 
are suited to nuclear physical applications,  i.e.\ 
the energy and length scale are measured in MeV
and fm, respectively. In these units 
$\hbar/(2m)=10.375$ MeV fm$^2$ ($m$ is the reduced mass of 
two $\alpha$ particles) and ${\rm e}^2= 1.44$ MeV fm. The other 
parameters are $A=$122.694 MeV, $\beta=0.22$ fm$^{-2}$, $\gamma=0.75$ 
fm$^{-1}$  and $Z=2$ (the charge number of the $\alpha$ 
particles).

Our radial Hamiltonian $H_l$ containing the model potential \eqref{be8pot}
can be split into two terms
\begin{equation}
   H_l=  H_l^{\rm C} + V_l \ .
\end{equation}
Here $V_l$  is the asymptotically irrelevant short-range potential and
$H_l^{\rm C}$ denotes the asymptotically 
relevant radial Coulomb Hamiltonian \eqref{coulham}.
Since the $\alpha-\alpha$ potential possesses a Coulomb tail,
the short-range potential is  defined by 
\begin{equation}
  V_l(r)=V_{\alpha-\alpha}(r)-\frac{Z^2{\rm e}^2}{r} = 
  -A \exp(-\beta r^2) - \frac{Z^2{\rm e}^2}{r}\ {\rm erfc}(\gamma r) ,
\end{equation} 
with ${\rm erfc} (z)=1- {\rm erf} (z)$.

The bound, resonant and scattering state solutions of the potential problem
characterized by the Hamiltonian $H_l$ 
can be obtained by solving the Lippmann--Schwinger integral equation.
The bound and resonant state  wave functions 
satisfy the homogeneous Lippmann--Schwinger  equation
\begin{equation}
\label{LS1}
  |\Psi_{l }  \rangle = G_l^{\rm C}(E ) V_l |\Psi_{l }\rangle  
\end{equation}
at real negative and complex $E$ energies, respectively.
While the  wave function $|\Psi^{(\pm)}_l \rangle$ describing a 
scattering process satisfies the inhomogeneous Lippmann--Schwinger 
equation (Section \ref{sec:scattering})
\begin{equation}
\label{LS}
  |\Psi^{(\pm)}_l  \rangle =|\Phi^{(\pm)}_l \rangle 
  +G_l^{\rm C}(E \pm {\rm i}0) V_l |\Psi^{(\pm)}_l \rangle \ ,  
\end{equation}
where $|\Phi^{(\pm)}_l \rangle $ is the solution to the Hamiltonian 
$H^{\rm C}_l$ with scattering asymptotics. 
In Equations \eqref{LS1}, \eqref{LS} $G^{\rm C}_l(z)$ denotes the
radial Coulomb Green's operator defined as
$G^{\rm C}_l(z)=(z-H^{\rm C}_l)^{-1}$.

We are going to solve these equations by using a discrete Hilbert space basis
representation
in a unified way by approximating
only the potential term $V_l$. For this purpose we write the unit operator
in the form 
\begin{equation}
\label{biort3}
  {\bf 1}=\lim_{N\to\infty} {\bf {1}}_{N}\ ,
\end{equation}
where 
\begin{equation}
\label{1N}
  {\bf {1}}_{N}= \sum _{n=0}^{N }|\tilde{n}\rangle \sigma_n^N
  \langle n|=\sum _{n=0}^{N }|n\rangle \sigma_n^N \langle \tilde{n}|\ .
\end{equation}
In this case the $\{|n\rangle ,|\tilde{n}\rangle\}$ 
biorthonormal basis is specified as the
Coulomb--Sturmian basis \eqref{csf}.
The $\sigma$ factors have the properties $\lim_{n\to\infty} \sigma_n^N =0$
and $\lim_{N\to\infty} \sigma_n^N =1$, and render the limiting
procedure in (\ref{biort3})
smoother. They were introduced originally for improving the convergence 
properties of truncated trigonometric series \cite{lanczos}, 
but they turned out to be also very efficient  in solving integral equations 
in discrete Hilbert space basis representation \cite{borbely}. 
The choice of $\sigma_n^N$ 
\begin{equation}
\label{sigma}
  \sigma_n^N = \frac{1-\exp\{-[\alpha(n-N-1)/(N+1)]^2\}}{1-\exp(-\alpha^2)}
\end{equation}
with $\alpha\sim 5$ has proved to be appropriate in practical calculations. 

Let us introduce an approximation of the potential operator 
\begin{equation}
\label{sepfe2b}
  V_l = {\bf 1} V_l {\bf 1} \approx {\bf {1}}_{N} V_l 
  {\bf {1}}_{N} = V_l^N = 
  \sum_{n,n' =0}^N
 |\widetilde{n}\rangle  \;
 \underline{V}_{n n^\prime} \;\mbox{} \langle \widetilde{n^{\prime }}| \ ,
\end{equation}
where the matrix elements
\begin{equation}
\label{v2b}
  \underline{V}_{n n^\prime} =
  \sigma^N_{n} \langle n|
  V_l| n^{\prime } \rangle \sigma^N_{n^{\prime}}\ ,  
\end{equation}
in general, are to be calculated numerically.
This approximation is called separable expansion, because the operator
$V_l^N$, { e.g.}\ in coordinate representation, takes the form 
\begin{equation}
  \langle r | V^N| r^{\prime } \rangle = \sum_{n, n' =0}^N
  \langle r |\widetilde{n }\rangle  \;
  \underline{V}_{n n'} \;\mbox{} \langle \widetilde{n^{\prime }
  }| r^{\prime} \rangle\ , 
\end{equation}
i.e.\  the dependence on $r$ and $r^{\prime}$ appears in a separated 
functional form.

With this  separable potential Eqs.\ (\ref{LS1}) and  (\ref{LS})
are  reduced to
\begin{equation}
\label{LSapp2}
  |\Psi_{l } \rangle = \sum_{n,n^\prime =0}^N
  G_l^{\rm C}(E ) |\widetilde{n}\rangle  \;
  \underline{V}_{n n'}\;\mbox{} \langle \widetilde{n^{\prime }}
  |\Psi_{l} \rangle\   
\end{equation}
and 
\begin{equation}
\label{LSapp1}
  |\Psi^{(\pm)}_l \rangle =|\Phi^{(\pm)}_l \rangle + \sum_{n,n^\prime =0}^N
  G^{\rm C}_l (E\pm {\rm i}0)  |\widetilde{n}\rangle  \;
  \underline{V}_{n n'} \;\mbox{} \langle \widetilde{n^{\prime }}
  |\Psi^{(\pm)}_l \rangle\ ,  
\end{equation} 
respectively. To derive equations for the 
coefficients $\underline{\Psi}_l^{(\pm)} =
\langle \widetilde{n^{\prime }}
|\Psi_l^{(\pm)} \rangle$ and $\underline{\Psi }_l 
=\langle \widetilde{n^{\prime }}
|\Psi_{l} \rangle$, we have to
act with states $\langle \widetilde{n''}|$
from the left. Then the following homogeneous and inhomogeneous   
algebraic equations are obtained, for bound  and scattering state  
problems, respectively: 
\begin{equation}
\label{eq18b}
  \lbrack (\underline{G}_l^{\rm C} (E))^{-1}-
  \underline{V}_l]\underline{\Psi }_l= 0 
\end{equation}
and
\begin{equation}
\label{eq18a}
  \lbrack (\underline{G}^{\rm C}_l (E \pm {\rm i}0))^{-1}-
  \underline{V}_l]\underline{\Psi}_l^{(\pm)}=
  \underline{\Phi}_l^{(\pm)},  
\end{equation}
where the overlap
${\Phi}_{n l}^{(\pm)}=\langle  \widetilde{n}|\Phi_l^{(\pm)} \rangle $
can also be calculated analytically \cite{papp2}.
The homogeneous equation (\ref{eq18b}) is solvable if and only if 
\begin{equation}
\label{determinans}
  \det [(\underline{G}_l^{\rm C}(E))^{-1}-\underline{V}_l]=0 
\end{equation}
holds, which is an implicit nonlinear equation for the bound and
resonant state energies.  
As far as the scattering states are concerned the solution of (\ref{eq18a}) 
provides the overlap $\langle \widetilde{n}|\Psi_l \rangle$. 
From this quantity
any scattering information can be inferred, for example the
scattering amplitude corresponding to potential $V_l$ is given by 
\cite{newton}
\begin{equation}
\label{scamp}
  A^{V} = \langle 
  \Phi^{(-)}_l | V_l | \Psi^{(+)}_l  \rangle = \underline{\Phi}_l^{(-)}
  \underline{V}_l \;\underline{\Psi}_l^{(+)}.
\end{equation}
Note that also the Green's matrix of the total Hamiltonian,
which is equivalent with the complete solution to the physical system,
can be constructed as 
\begin{equation}
\label{totgrm}
  \underline{G}_l(z)=\lbrack (\underline{G}_l^{\rm C} (z))^{-1}-
  \underline{V}_l ]^{-1} \ .
\end{equation}

Finally, it should also be emphasized, that in this approach only the
potential operator is approximated, but the asymptotically important 
$H^{\rm C}$ term remains intact.
The properties of the short-range potential is buried into the numerical values of
the matrix elements. Thus the method is  applicable to all types of
potentials, as long as we can calculate their matrix elements somehow. 
Beside usual potentials this equally applies to complex, 
momentum-dependent, non-local, etc.\  potentials relevant to 
practical problems of atomic, nuclear and particle physics. 
Furthermore, the present formalism 
is equally suited to problems including attractive or repulsive 
long-range Coulomb-like  and short-range potentials. 
The solutions are defined on the whole Hilbert 
space, not only on a finite subspace. The wave functions are not
linear combinations of the basis functions,
but rather, as Eqs.\ (\ref{LSapp1})
and (\ref{LSapp2}) indicate, linear combinations of the states
$G_l^{\rm C}(E)  |\widetilde{n}\rangle$, which have been shown to possess 
correct Coulomb asymptotics \cite{papp3}.

\subsection{Bound states}

First we  consider only the nuclear part of potential (\ref{be8pot}) 
and switch off the Coulomb interaction by setting $Z=0$.
According to Ref.\ \cite{schmid}, this potential supports altogether 
four bound states: three with $l=0$ and one with $l=2$. 
However, it is 
known that the first two $l=0$ and the single $l=2$ state are unphysical, 
since they are forbidden because of the Pauli principle. 
This fact is not 
taken into account in this simple potential model. 
Although from the physical point of view 
these Pauli-forbidden states have to be dismissed as unphysical, 
they are legitimate solutions to our simple model
potential. The proper inclusion of the Pauli principle into the model
would turn the potential into a non-local one. This problem has been 
considered within the present method in Ref.\ \cite{papp2}.

In Table \ref{boundstates}  we present the results of our calculations 
for the three $l=0$ states showing the convergence of the method
with respect to $N$, the number of basis 
states used in the expansion. We determined the energies of these states 
from Eq.\ (\ref{determinans}), using the CS parameter $b=4$ fm$^{-1}$. 
It can be seen that the method is very accurate,
convergence up to 12 digits can easily been reached. 
We note that according to Ref.\ \cite{schmid}, the energy of the two 
lowest (i.e.\ the unphysical) $l=0$ states is $E=-76.903\ 6145$ and 
$E=-29.000\ 48$ MeV in the uncharged case, which is in reasonable agreement 
with our results.

\begin{table}
\begin{center}
\begin{tabular}{rccc}
$N$ & $E_{00}$ (MeV) & $E_{10}$ (MeV) & $E_{20}$ (MeV) \\
\hline \\
 8 & $-$76.903\ 557\ 1529 & $-$29.005\ 234\ 9134 & $-$1.739\ 478\ 2626 \\  
10 & $-$76.903\ 609\ 9717 & $-$29.000\ 352\ 3141 & $-$1.637\ 269\ 0831 \\  
15 & $-$76.903\ 614\ 3090 & $-$29.000\ 469\ 8249 & $-$1.608\ 824\ 6403 \\  
18 & $-$76.903\ 614\ 3254 & $-$29.000\ 470\ 2338 & $-$1.608\ 742\ 5166 \\ 
20 & $-$76.903\ 614\ 3263 & $-$29.000\ 470\ 2566 & $-$1.608\ 741\ 0685 \\  
25 & $-$76.903\ 614\ 3265 & $-$29.000\ 470\ 2623 & $-$1.608\ 740\ 8256 \\  
28 & $-$76.903\ 614\ 3265 & $-$29.000\ 470\ 2625 & $-$1.608\ 740\ 8216 \\ 
30 & $-$76.903\ 614\ 3265 & $-$29.000\ 470\ 2625 & $-$1.608\ 740\ 8213 \\  
35 & $-$76.903\ 614\ 3265 & $-$29.000\ 470\ 2626 & $-$1.608\ 740\ 8214 \\
40 & $-$76.903\ 614\ 3265 & $-$29.000\ 470\ 2626 & $-$1.608\ 740\ 8214  \\
\\
\end{tabular}
\end{center}
\caption{Convergence of the $l=0$ bound state energy eigenvalues $E_{nl}$ 
in $V_{\alpha-\alpha}(r)$ in the uncharged ($Z=0$) case. $N$ denotes  
the number of basis states used in the expansion. 
\label{boundstates}}
\end{table}

\subsection{Resonance states}

Switching on the repulsive Coulomb interaction ($Z=2$) the bound states 
are shifted to higher energies. The most spectacular effect is that 
the third $l=0$ state, which is located at $E=-1.608\ 740\ 8214$ MeV in 
the uncharged case, moves to positive energies and becomes a resonant state. 
This is in agreement with 
the observations: the $\alpha-\alpha$ system (i.e.\ the $^8$Be nucleus) 
does not have a stable ground state, rather it decays  with a half life 
of $7\times 10^{-17}$ sec.

In our calculations we determined the energies corresponding to this 
resonance and to other ones as well by the same techniques we used before 
to find  bound states. In fact, we used the same 
computer code and the same CS parameter ($b=4$ fm$^{-1}$) as we used in the 
analysis of bound states. The method, again, requires locating the poles 
of the Green's matrix, but not on the real energy 
axis, rather on the complex energy plane.
 
\begin{figure}
\begin{center}
\resizebox{10cm}{!}{
\rotatebox{-90}{
\includegraphics{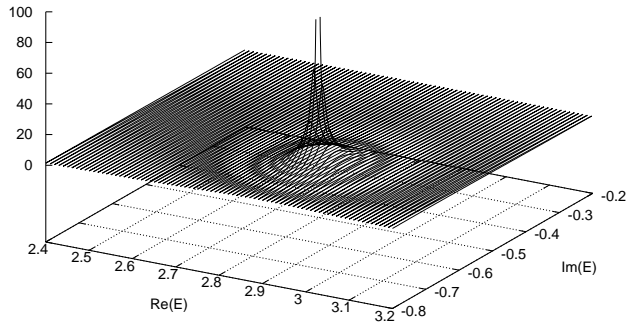}
}} 
\end{center}
\caption{ The modulus of the determinant of the 
$ \underline{G}(E)$ Green's matrix for 
the $\alpha-\alpha$ potential on the complex energy plane for $l=2$.
The pole at $E=2.889\ 25-{\rm i\ } 0.620\ 61$ MeV  corresponds 
to a resonance.  \label{fig:be8res}}
\end{figure}

In Table \ref{konv_resonances} we demonstrate the convergence of our method 
with respect to $N$ for the lowest $l=0$ and $l=2$ resonance states. 
In Fig.\ \ref{fig:be8res} we plotted the modulus of the 
determinant of the (\ref{totgrm}) Green's matrix  
over the complex energy plane for $l=2$. 
The resonance is located at the pole of this function. 
Finally, we mention that there is a resonance state for $l=4$ at 
$E_{{\rm res},4} =11.791\ 038-{\rm i\ } 1.788\ 957$ MeV.  

Although it is not our aim here to reproduce experimental data 
with this simple potential model, we note that the corresponding 
experimental values \cite{ajzenberg}
are $E_{{\rm res}, l=0}=$0.09189  MeV, 
$E_{{\rm res}, l=2}=$3.132 $\pm$ $0.030$ MeV, $E_{{\rm res}, l=4}=$11.5 $\pm$ 
0.3 MeV, 
and $\Gamma_{l=0}/2 = (3.4 \pm 0.9)\times 10^{-6}$ MeV, 
$\Gamma_{l=2}/2$=0.750 $\pm$ 0.010 MeV, 
$\Gamma_{l=4}/2\simeq$1.75  MeV.

\begin{table}
\begin{center}
\begin{tabular}{rccc}
$N$ & $E_{{\rm res}, 0}$  (MeV) & $E_{{\rm res}, 2}$ (MeV) \\
\hline \\
 8 & $-$0.000\ 854\ 9596  $+$i\ 0.000\ 000\ 0000  & 2.807\ 21 $-$i\ 0.607\ 11  \\  
10 &  \phantom{$-$}0.063\ 364\ 2503  $-$i\ 0.000\ 000\ 0681  & 2.866\ 30  $-$i\ 0.628\ 56
\\  
15 &  \phantom{$-$}0.091\ 785\ 0787  $-$i\ 0.000\ 002\ 8092  & 2.889\ 68  $-$i\ 0.620\ 99 
\\  
18 &  \phantom{$-$}0.091\ 963\ 0277  $-$i\ 0.000\ 002\ 8572  & 2.889\ 34  $-$i\ 0.620\ 53 
\\ 
20 &  \phantom{$-$}0.091\ 969\ 7296  $-$i\ 0.000\ 002\ 8588  & 2.889\ 24  $-$i\ 0.620\ 56  
\\  
25 &  \phantom{$-$}0.091\ 971\ 8479  $-$i\ 0.000\ 002\ 8592  & 2.889\ 23  $-$i\ 0.620\ 62  
\\  
28 &  \phantom{$-$}0.091\ 971\ 9788  $-$i\ 0.000\ 002\ 8592  & 2.889\ 25  $-$i\ 0.620\ 62  
\\ 
30 &  \phantom{$-$}0.091\ 972\ 0064  $-$i\ 0.000\ 002\ 8592  & 2.889\ 25  $-$i\ 0.620\ 61 
\\  
35 &  \phantom{$-$}0.091\ 972\ 0258  $-$i\ 0.000\ 002\ 8592  & 2.889\ 24  $-$i\ 0.620\ 61  
\\
40 &  \phantom{$-$}0.091\ 972\ 0290  $-$i\ 0.000\ 002\ 8592  & 2.889\ 25  $-$i\ 0.620\ 61 
\\
\\
\end{tabular}
\end{center}
\caption{
Convergence of the energy eigenvalues $E_{{\rm res},l}$ for the $l=0$ 
and $l=2$ resonances in the $V_{\alpha-\alpha}(r)$ potential. 
$N$ denotes the number of basis states used in the expansion. 
\label{konv_resonances}}
\end{table}

\subsection{Scattering states}

In order to demonstrate the performance of our approach for scattering 
states we calculated scattering phase shifts $\delta_l(E)$ for 
${V}_{\alpha-\alpha}(r)$ in (\ref{be8pot}). As described previously in this 
subsection, phase shifts can be extracted from the scattering amplitude 
given in Eq.\ (\ref{scamp}). Specifying this formula for the Coulomb--like 
case and for a given partial wave $l$ we have 
\begin{equation}
a_l=\frac{1}{k}\exp({\rm i}(2\eta_l+\delta_l))\sin \delta_l\ ,
\label{coulphsh}
\end{equation}
where $a_l$ is the Coulomb--modified scattering amplitude corresponding
to the short-range potential, 
$\eta_l={\rm arg} \Gamma(l+i\gamma+1))$ is the phase shift of the Coulomb 
scattering with $\gamma=Z^2{\rm e}^2m/\hbar^2k$ 
and $\delta_l$ is the phase shift due to the short-range potential. 

The convergence of the phase shifts
 with respect to $N$ is demonstrated in Table 
\ref{konv_phaseshift}, where $\delta_0(E)$ is displayed at 
three different energy 
values. As in our calculations for the bound and the resonance states, 
we used $b=4$ fm$^{-1}$here too.  

\begin{figure}
\begin{center}
\resizebox{6.5cm}{!}{
\rotatebox{-90}{
\includegraphics{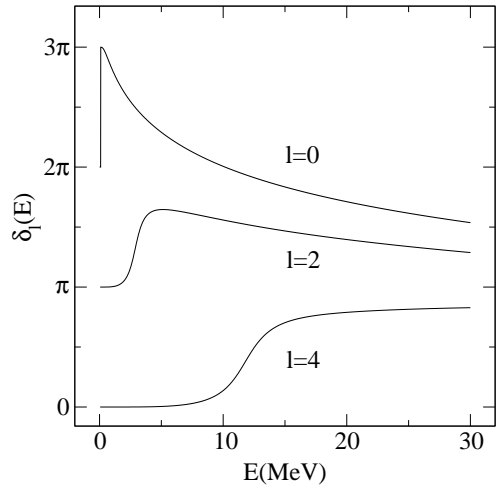}
}} 
\end{center}
\caption{Scattering phase shifts $\delta_l(E)$ (in radians) 
in the $\alpha-\alpha$ 
potential for $l=0$, 2 and 4. The resonances in these partial waves appear 
as sharp rises in the corresponding phase shifts. In these calculations 
a basis with $N=35$ was used.\label{fig:phsh}}
\end{figure}

In Fig.\ \ref{fig:phsh} we plotted the scattering phase 
shifts $\delta_l(E)$ for 
$l=0$, 2 and 4 up to $E=30$ MeV. 
In all three plots in Fig.\ \ref{fig:phsh} the location of the corresponding 
resonance is clearly visible as a sharp rise of the phase around the 
resonance energy $E_{\rm res}$. This rise is expected to be more sudden 
for sharp resonances, and this is, in fact, the case 
here too. 
The phase changes with an abrupt jump of $\pi$ for the sharp $l=0$ resonance, 
while it is slower for the broader $l=2$ and $l=4$ resonances. 
We also note that the phase shifts plotted in Fig.\ \ref{fig:phsh} are also 
in accordance with the Levinson theorem, which states 
that $\delta_l(0)=m\pi$, where $m$ is the number of bound states in the 
particular angular momentum channel. Indeed, as we have discussed earlier, 
there are two bound states for $l=0$, one for $l=2$ and none for $l=4$. 

\begin{figure}
\begin{center}
\resizebox{7.cm}{!}{
\rotatebox{-90}{
\includegraphics{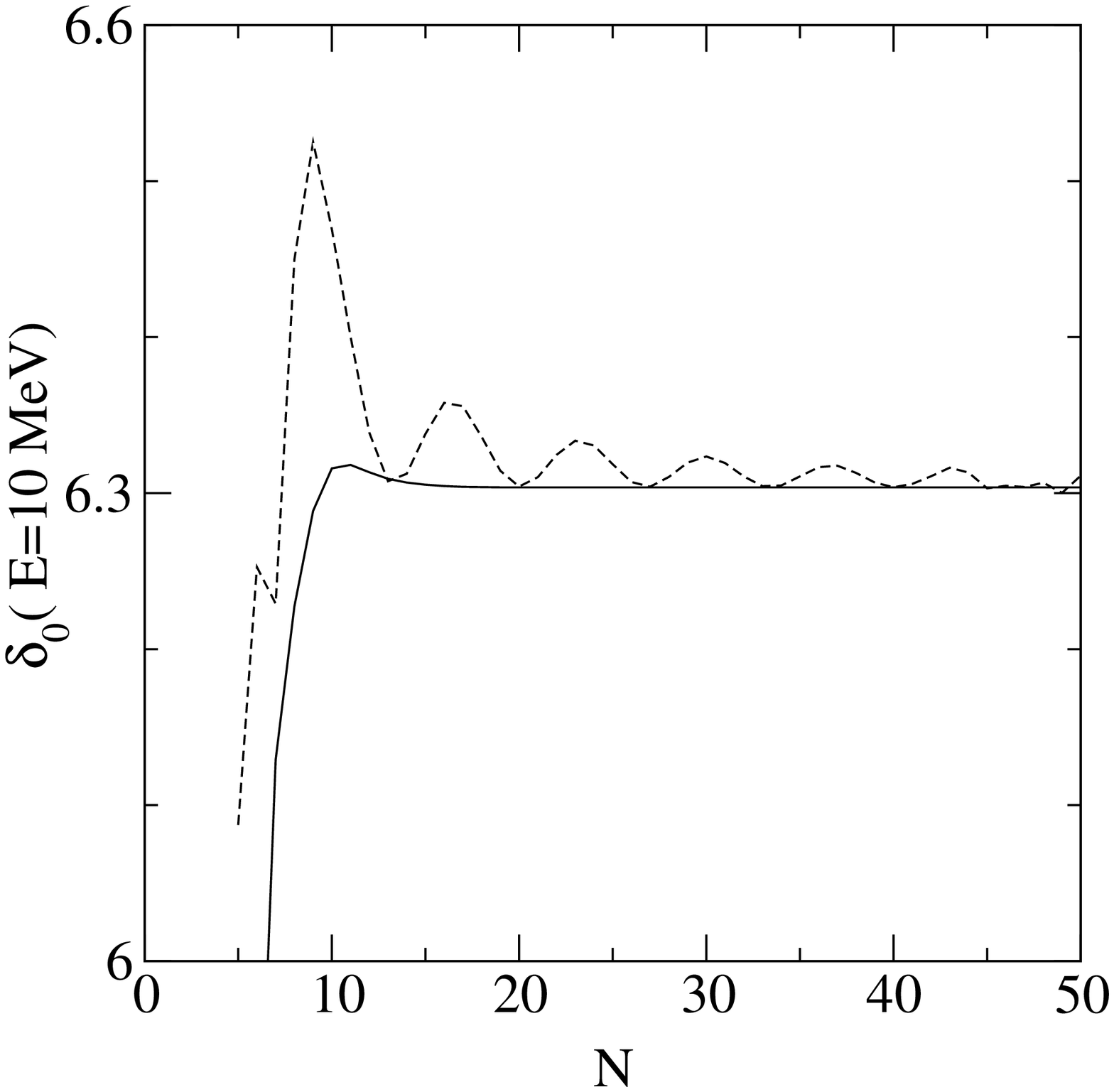}
}} 
\end{center}
\caption{Convergence of the scattering phase shift $\delta_0(E)$ 
(in radians) at $E=10$ 
MeV calculated with (solid line) and without
(dashed line) smoothing factors. \label{fig:simicska}}
\end{figure}

As an illustration of the importance of the smoothing factors 
we show in Fig.\ \ref{fig:simicska} the convergence of the 
phase shift $\delta_0(E)$ at a specific energy $E= 10$ MeV with and without 
the smoothing factors $\sigma_n^N$ in (\ref{1N}), and consequently 
in (\ref{v2b}). (Here and everywhere 
else the $\alpha$ parameter of the \eqref{sigma} $\sigma$ factors
was chosen to be 5.2.) 
Clearly, the convergence is much poorer without 
the smoothing factors. We note that this also applies to  other 
quantities calculated for bound and resonance states.

\begin{table}
\begin{center}
\begin{tabular}{rccc}
$N$ & $E=0.1$ MeV & $E=1$ MeV & $E=30$ MeV \\
\hline \\
 8 & 6.283\ 230  & 8.817\ 731 & 7.783\ 217 \\  
10 & 9.424\ 059  & 8.862\ 581 & 4.817\ 163 \\  
15 & 9.424\ 018  & 8.859\ 651 & 4.835\ 479 \\  
18 & 9.424\ 022  & 8.859\ 467 & 4.829\ 861 \\ 
20 & 9.424\ 023  & 8.859\ 441 & 4.828\ 882 \\  
25 & 9.424\ 024  & 8.859\ 419 & 4.828\ 563 \\  
28 & 9.424\ 024  & 8.859\ 414 & 4.828\ 555 \\ 
30 & 9.424\ 024  & 8.859\ 412 & 4.828\ 554 \\  
35 & 9.424\ 024  & 8.859\ 411 & 4.828\ 552 \\
40 & 9.424\ 024  & 8.859\ 411 & 4.828\ 552 \\
\\
\end{tabular}
\end{center}
\caption{
Convergence of the $\delta_0(E)$ phase shift (in radians) 
in the $V_{\alpha-\alpha}(r)$ potential at three different energies. 
$N$ denotes the number of basis states used in the expansion.
\label{konv_phaseshift}}
\end{table}

\subsection{Conclusions}

In this section we used the 
Coulomb--Sturmian representation of the Coulomb Green's operator 
to solve   the Lippmann--Schwinger integral
equation with a model nuclear potential describing the interaction of two 
$\alpha$ particles.  In practice this means 
the approximation of the potential term on a finite subset of this basis. 
This is the only stage where approximations are made (remember the Green's
operator is calculated analytically), otherwise 
this method is exact, analytic and provides asymptotically correct 
solutions. Consequently, bound, resonance and 
scattering problems can be treated on an equal footing, while 
these phenomena are usually discussed in rather 
different ways in conventional quantum mechanical approaches. 
The unified treatment is also reflected by the fact that all the 
calculations are made using the same discrete basis, containing 
also the same basis and other parameters.

Finally, we should call the attention upon the fact that this method is very
accurate. Reasonable accuracy is reached already at relatively small
basis, around $N=20$. The accuracy gained in larger bases 
is beyond most of the practical requirements. 
Test calculations have been performed on a linux PC (Intel PII, 266 MHz)
using double precision arithmetic. 
The calculation of a typical bound or resonant state 
energy requires the evaluation of the 
potential matrix by Gauss--Laguerre quadrature and finding  the zeros
of the determinant (\ref{determinans}), which incorporates 
the evaluation of the
Coulomb Green's matrix and the  the calculation of a determinant by 
performing an LU decomposition in each steps. 
The determination of the energy value in the first
column and last row of Table \ref{boundstates}, which meant $6$ steps in the 
zero search and handling of $40\times 40$ matrices,
took $0.06$ sec. The corresponding resonance energy value in 
Table \ref{konv_resonances} required $12$ steps in the
zero search on the complex 
energy plane and  $0.8$ sec. 
The evaluation of the three phase shift values in the last row
of Table \ref{konv_phaseshift} took $0.19$ sec., $0.11$ sec.\ and $0.08$ sec.,
respectively. 
So, this method is not only  accurate but also very fast.

\section{An atomic three-body problem}

The bound state problem of the Helium atom is 
investigated as an example of
the application of the Faddeev--Merkuriev (FM)
 integral equations to the atomic
three-body problems. The FM equations describing  three  charged
particles interacting through the long-range Coulomb 
potential are solved using the
Coulomb--Sturmian discrete Hilbert space basis 
representation, which transforms
the integral equations into a matrix equation.  The solution of the matrix
equation is  possible
due to the analytic representation of a  three-body Green's 
operator constructed from two independent two-body Green's operators
by performing a convolution integral (Section \ref{sec:convolution}).

We recall here that for three-body systems the Faddeev equations 
\cite{faddeev} are the fundamental equations.
Their homogeneous form is fully equivalent to 
the Schr\"odinger equation and
after one iteration they possess connected kernels,  consequently they are,
in fact, Fredholm integral equations of second kind.
Therefore the Fredholm
alternative applies: at certain energy either the homogeneous or 
the inhomogeneous
equations have solutions. Three-body bound states correspond to
the solutions of the homogeneous Faddeev equations at real energies.

The Hamiltonian of an interacting three-body system can be written as
\begin{equation}
  H=H^0 + v_\alpha+ v_\beta + v_\gamma ,
\label{H}
\end{equation}
where $H^0$ is the three-body kinetic energy
operator and $v_\alpha$ denotes the
interaction in subsystem $\alpha$ (i.e.\ the pair interaction of particles
$\beta$ and $\gamma$).
Here the usual configuration space 
Jacobi coordinates  ${x}_\alpha$, ${y}_\alpha$ 
are used, ${x}_\alpha$ is the coordinate of
the $(\beta,\gamma)$ pair and ${y}_\alpha$ is the
coordinate of the particle $\alpha$ relative to the center of mass
of the pair $(\beta,\gamma)$.
Therefore  $v_\alpha=v_\alpha (x_\alpha)$ represents the
interaction of the pair $(\beta,\gamma)$ and  depends only on the $x_\alpha$
relative coordinate of this pair.
In Jacobi coordinates the $H^0$ three-body kinetic energy operator
is given as a sum of two-body
kinetic energy operators
\begin{equation}
  H^0 = h^0_{x_\alpha} + h^0_{y_\alpha}=
  h^0_{x_\beta} + h^0_{y_\beta}= h^0_{x_\gamma} + h^0_{y_\gamma}.
\label{H0}
\end{equation}

Suppose the $v_{\alpha},v_{\beta},v_{\gamma}$ pair interactions are 
short-range-type potentials.  In this case the Faddeev procedure 
leads to  mathematically sound integral equations for the
three-body system. The three-body wave  function $\ket{\Psi}$ is 
decomposed as
\begin{equation}
  |\Psi \rangle = |\psi_{\alpha} \rangle  +|\psi_{\beta} \rangle +
  |\psi_{\gamma} \rangle,
\end{equation}
where the components are defined by
\begin{equation}
  |\psi_{i} \rangle =   G^{0} v_i  |\Psi \rangle
  \qquad \mbox{for} \quad i=\alpha,\beta,\gamma,
\label{f_component}
\end{equation}
with the $G^{0}(z)=(z-H^0)^{-1}$  being the free Green's operator.
 For bound states the set of homogeneous Faddeev equations
appear as
\begin{eqnarray}
\label{feqs}
  |\psi_{\alpha} \rangle&=& G_\alpha (E ) [ v_\alpha
  |\psi_{\beta}\rangle + v_\alpha |
  \psi_{\gamma}\rangle ]  \\
  |\psi_{\beta} \rangle&=& G_\beta (E ) [ v_\beta
  |\psi_{\gamma}\rangle + v_\beta |
  \psi_{\alpha}\rangle ]  \nonumber \\
  |\psi_{\gamma} \rangle&=& G_\gamma (E ) [ v_\gamma
  |\psi_{\alpha}\rangle + v_\gamma |
  \psi_{\beta}\rangle ]   \nonumber
\end{eqnarray}
where $G_\alpha (z) =
(z-H_\alpha)^{-1}$ with $H_\alpha  = H^{0} + v_\alpha$. It is proven that
the above set of
coupled integral equations for short-range potentials 
have an unique solution.

Unfortunately in the case of the scattering of three charged particles
the situation becomes more complicated since the Coulomb potential
enters into the game. The Faddeev equations originally were derived 
for short range
interactions and if we simply plug in a Coulomb-like potential they become
singular.
The solution has been formulated by  Faddeev and Merkuriev 
\cite{fm-book} in a mathematically sound and elegant way
via integral equations with connected (compact) kernels and
configuration space differential equations with asymptotic boundary
conditions.

In practice, however only this latter version of the theory has been applied,
since the  FM integral equation formulation
was too complicated for practical use.
Moreover, in bound-state problems only the original version of the Faddeev
equations were applied \cite{fonseca,Schout,kvitsinsky}
which, in sound mathematical sense, are not well-behaved for Coulomb case
and only the bound state nature of
the problems helped to overcome the difficulties, however  slow convergence 
in partial wave channels was reported \cite{fonseca,Schout}. 
In order to find a remedy,
the equations were solved in total
angular momentum representation, which led to
three-dimensional equations \cite{kvitsinsky}.
So, due to these problems and difficulties,
the belief spread 
that the Faddeev equations are not well-suited for
treating atomic three-body problems and other techniques
can perform much better, at least for bound states.

Recently, a novel method was proposed by Papp
for treating the three-body Coulomb problem
via solving the set of Faddeev--Noble and Lippmann--Schwinger
integral equations in Coulomb--Sturmian discrete Hilbert space basis 
representation.
The method was elaborated first
for bound-state problems \cite{pzwp} with repulsive Coulomb plus nuclear
potential, then it was extended for analyzing $p-d$
scattering at energies below the breakup threshold \cite{pzsc}.
In these calculations  excellent agreements with the results of other well
established methods were found and
the efficiency and the accuracy of the method were demonstrated.
This approach has also been applied to
atomic  bound-state problems \cite{pzatom}.
The Coulomb interactions
were split, \`{a} la Noble \cite{noble},
into long-range and short-range terms and the Faddeev
procedure was applied only to the short-range potentials.
By studying benchmark bound-state problems, contradictory to the conventional
approaches, a fast convergence with respect to
angular momentum channels was observed. This approach, however, has a
limitation which is related to Noble's splitting.
The Noble splitting  does not separate the
asymptotic channels, so the equations are applicable only in a restricted
energy range of low-lying states.

In this section, instead of the Noble splitting, we make use of  
the mathematically sound Merkuriev splitting of the 
long-range Coulomb potential \cite{fm-book}, and by  doing so we solve 
the Faddeev--Merkuriev integral equations.

\subsection{Faddeev--Merkuriev integral equations}

The fundamental  equations for the Coulomb three-body problem 
is discussed here following Merkuriev ideas.
We recall that the Faddeev procedure 
is based on the observation that the
Hamiltonian of the three-body system can be written as the sum of the
asymptotically relevant $H^{(l)}$ and irrelevant short-range terms
\begin{equation}
\label{H_split}
  H=H^{(l)} + v_\alpha^{(s)}+ v_\beta^{(s)} + v_\gamma^{(s)},
\end{equation}
where
\begin{equation}
  H^{(l)}=H^0 + v_\alpha^{(l)}+ v_\beta^{(l)} + v_\gamma^{(l)}.
\end{equation}
In equation \eqref{H_split} we assumed that the long-range 
Coulomb potential can be split as
\begin{equation}
  v^C =v^{(l)} +v^ {(s)}
\end{equation}
by using an appropriate cut-off procedure.

By applying the Faddeev procedure for the
short-range part of the potentials we can derive connected kernel integral
equations. For this, however it is essential that the asymptotically relevant
Hamiltonian $H^{(l)}$ possesses only continuous spectrum.
This property guarantees the asymptotic filtering behavior of the
Faddeev decomposition \cite{vanzani}, 
and thus the asymptotic orthogonality of the
Faddeev components. A potential with an attractive Coulomb tail
has infinitely many bound states accumulated at the lower edge of
the continuous spectrum, thus  the long-range part of an attractive Coulomb
potential always has infinitely many bound states. Fortunately, this
is true only in the two-body Hilbert space. In a broader space, like
the three-body Hilbert space, this statement is not necessarily valid.
In the three-body Hilbert space, where we have to consider our potentials,
we simply have an extra kinetic energy term,
which can modify  the character of the  spectrum.

\begin{figure}[tbp]
{\centering \begin{tabular}{|c|c|}
\hline 
\resizebox*{6cm}{6cm}{\rotatebox{-90}
{\includegraphics{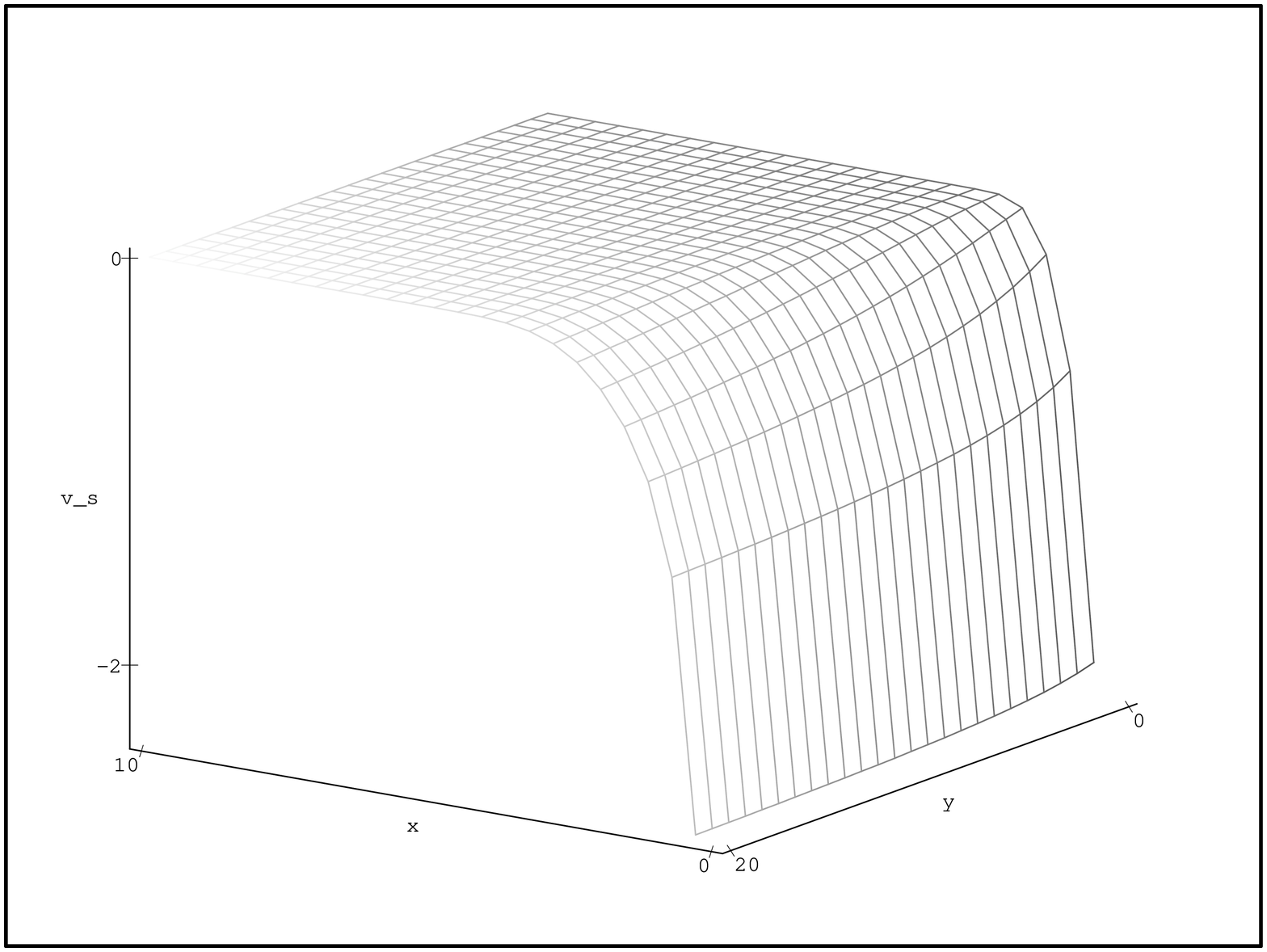}}} &
\resizebox*{6cm}{6cm}{\rotatebox{-90}
{\includegraphics{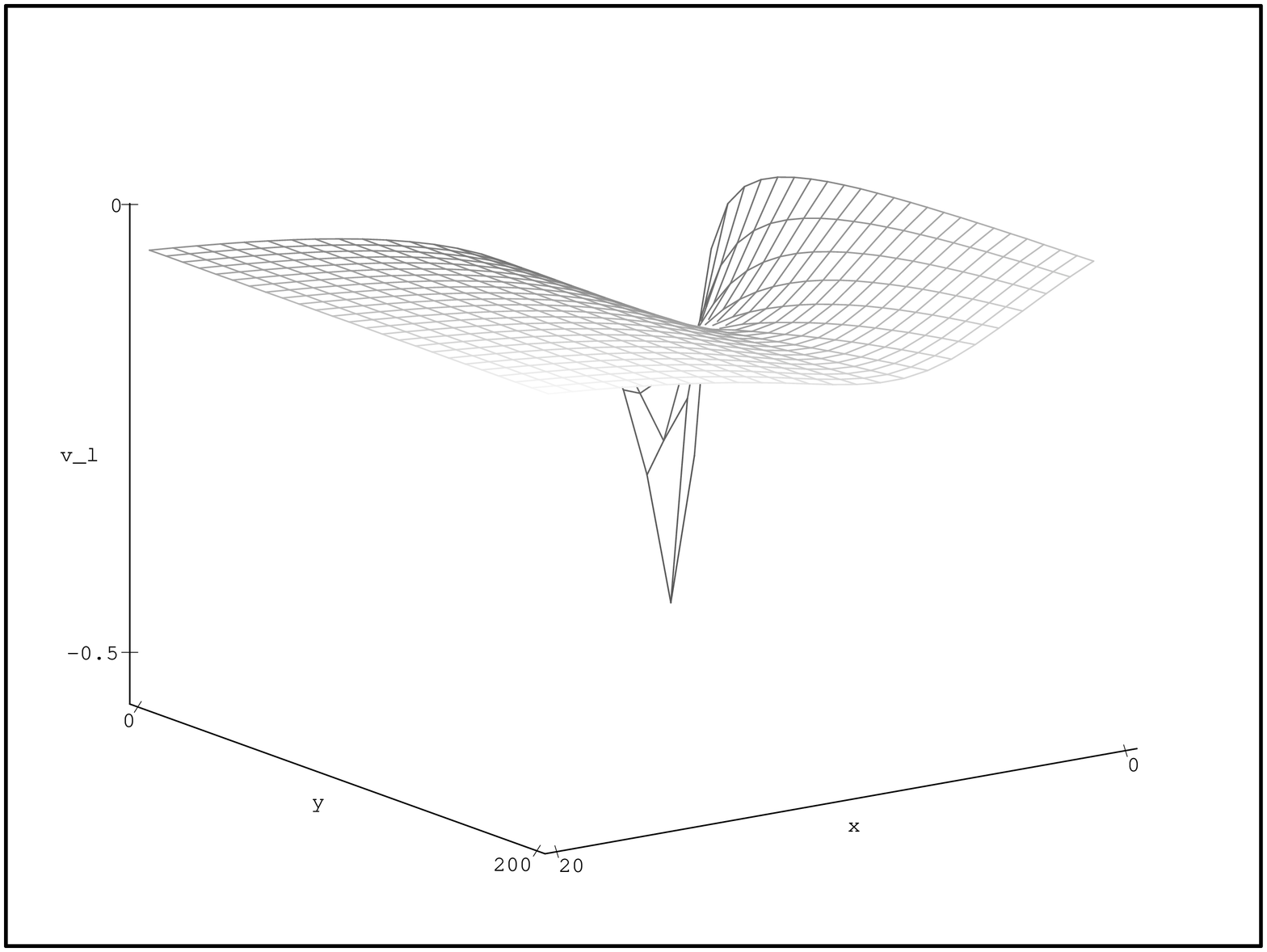}}}   \\
$v^{(s)}$ &
$v^{(l)}$ \\
\hline 
\end{tabular}\par}
\caption{The short- and the long-range part of an attractive 
Coulomb potential
in the $(x,y)$ three-body configuration space 
with parameters $x_0=1$, $y_0=5$ and $\nu=2.2$ of $\zeta$ defined in
Eq.\ \eqref{zeta}.
\label{fig:merkurcut}}
\end{figure}

Merkuriev proposed to split the Coulomb potential $v_\alpha^C(x_\alpha)$
into a short-range and a long-range
part by introducing a cut-off function 
$\zeta_{\alpha}=\zeta_{\alpha}(x_\alpha,y_\alpha)$  defined 
on the three-body configuration space leading to 
\begin{equation}
\label{mvs}
  v_\alpha^{(s)}(x_\alpha,y_\alpha)=
  v_\alpha^C(x_\alpha) \zeta_\alpha(x_\alpha,y_\alpha),
\end{equation}
and
\begin{equation}
\label{mvl}
  v_\alpha^{(l)}(x_\alpha,y_\alpha)=
  v_\alpha^C(x_\alpha) [1- \zeta_\alpha(x_\alpha,y_\alpha) ].
\end{equation}
The function $\zeta_\alpha$ is constructed  
in such a way, that it separates the
asymptotic two-body sector $\Omega_\alpha$ from the rest
of the three-body configuration space.
On the region of $\Omega_\alpha$
the splitting function $\zeta_\alpha$
asymptotically tends to $1$ and  on the complementary asymptotic region
of the configuration space it tends
to $0$. Rigorously, $\Omega_\alpha$ is defined as a part of the
three-body configuration
space where the condition
\begin{equation}
\label{oma}
  |x_\alpha| < {y_\alpha}_0 ({x_\alpha}_0+|y_\alpha|/{y_\alpha}_0)^{1/\nu},
  \mbox{with} \quad {x_\alpha}_0,{y_\alpha}_0>0,\ \nu > 2,
\end{equation}
is satisfied.
So, in $\Omega_\alpha$ the short-range part
$v_\alpha^{(s)}$ coincides with the
original  Coulomb--like potential $v_\alpha^C$
and in the complementary region vanishes, whereas
the opposite holds true for $v_\alpha^{(l)}$.
A possible functional form for  $\zeta$ is given by
\begin{equation}
\label{zeta}
  \zeta(x,y)= 2 \left\{
  1+\exp \left [\frac {(x/x_0)^\nu} {1+ y/y_0 }\right ] \right\}^{-1},
\end{equation}
where the parameters $x_0$ and $y_0$ are rather arbitrary. 
Fig.\ \ref{fig:merkurcut} shows a typical example for the
short- and long-range part of an attractive
Coulomb potential, respectively.
Merkuriev proved that $H^{(l)}$ of Eq.\ \eqref{H_split}
with long-range potentials defined in this way
possesses only continuous spectrum if the ${x_\alpha}_0$ 
and ${y_\alpha}_0$ parameters
for all $\alpha$ fragmentations
are chosen big enough (see p. 248 of Ref.\ \cite{fm-book}).

Now, if we follow the Faddeev procedure with the Merkuriev split of 
\eqref{mvs} and \eqref{mvl} we obtain the Faddeev--Merkuriev integral equations
\begin{equation}
\label{fmeqs}
  |\psi_{\alpha} \rangle= G_\alpha^{(l)} (E ) [ v_\alpha^{(s)}
  |\psi_{\beta}\rangle + v_\alpha^{(s)} |
  \psi_{\gamma}\rangle ],
\end{equation}
where the $G_\alpha^{(l)} $ channel Green's operator
is defined as
\begin{equation}
\label{channelgreen}
  G_\alpha^{(l)}(z)=(z-H_\alpha^{(l)})^{-1},
\end{equation}
with
\begin{equation} 
  H_\alpha^{(l)}  = H^{(l)} + v_\alpha^{(s)}
  =H^0 + v^{C}_\alpha
  + v_\beta^{(l)} + v_\gamma^{(l)}.
\end{equation}

It is proven \cite{fm-book} that Eqs.\ (\ref{fmeqs}) below the threshold
of the continuous spectrum of the Hamiltonian $H$ allow the nontrivial
solutions only for discrete set of energy corresponding to the binding 
energies of an atomic three-body system.

The solution of the Faddeev equations \eqref{feqs} and\eqref{fmeqs}
necessitates
the determination of the asymptotically relevant
Green's operator in some basis
representation. This can be done using  convolution integral technique
of Section \ref{sec:convolution}.

\subsection{Solution in Coulomb--Sturmian space representation}

The three-body Hilbert space is a direct product of two-body Hilbert
spaces, so an appropriate discrete basis in the three-body Hilbert space
can be constructed 
as the angular momentum coupled direct product of the \eqref{csf}
Coulomb--Sturmian basis  functions as
\begin{equation}
\label{cs3}
  | n \nu l \lambda \rangle_\alpha =
  [ | n l \rangle_\alpha \otimes | \nu
  \lambda \rangle_\alpha ] , \ \ \ \ (n,\nu=0,1,2,\ldots),
\end{equation}
where $l$ and $\lambda$ denote the angular momenta associated
with  the coordinates $x$ and $y$, respectively, and
the bracket stands for angular momentum coupling.
With this basis the completeness relation
takes the form (with angular momentum summation implicitly included)
\begin{equation}
  {\bf 1} =\lim\limits_{N\to\infty} \sum_{n,\nu=0}^N |
  \widetilde{n \nu l
  \lambda} \rangle_\alpha \;\mbox{}_\alpha\langle
  {n \nu l \lambda} | =
  \lim\limits_{N\to\infty} {\bf 1}^{N}_\alpha.
\end{equation}
It should be noted that in the three-body Hilbert space
three equivalent bases belonging to fragmentation
$\alpha$, $\beta$ and $\gamma$ are possible.

In Eqs.\ (\ref{fmeqs})  we introduce the approximation
\begin{equation}
\label{feqsapp}
  |\psi_{\alpha} \rangle= G_\alpha^{(l)} (z)
  {\bf 1}^{N}_\alpha v^{(s)}_\alpha \sum_{\gamma\neq\alpha}
  {\bf 1}^{N}_\gamma |\psi_{\gamma} \rangle,
\end{equation}
i.e.\ the short-range potential
$v_\alpha^{(s)}$ in the three-body
Hilbert space is taken in the separable form
\begin{equation}
\label{sepfe}
  v_\alpha^{(s)}\approx \sum_{n,\nu ,n^{\prime },
  \nu ^{\prime }=0}^N|\widetilde{n\nu l\lambda }\rangle _\alpha \;
  \underline{v}_{\alpha \beta }^{(s)}\;
  \mbox{}_\beta \langle \widetilde{n^{\prime }
  \nu ^{\prime }l^{\prime }\lambda^{\prime }}|, 
\end{equation}
where $\underline{v}_{\alpha \beta}^{(s)}=
\mbox{}_\alpha \langle n\nu l\lambda |
v_\alpha^{(s)}|n^{\prime }\nu ^{\prime
}l^{\prime }{\lambda }^{\prime }\rangle_\beta$.
These matrix elements can be calculated numerically by making use
of the transformation of Jacobi coordinates.
The ket and bra states are defined
for different fragmentation, depending on the
environment of the potential operators in the equations.

Now, with this approximation, the solution of the homogeneous
Faddeev--Mer\-ku\-ri\-ev integral equations
turns into the solution of a matrix equation for the component vector
$\underline{\psi}_{\alpha}=
 \mbox{}_\alpha \langle \widetilde{ n\nu l\lambda} | \psi_\alpha  \rangle$
\begin{equation}
\label{feqm}
  \underline{\psi}_{\alpha} = \underline{G}_\alpha^{(l)} (z)
  \underline{v}^{(s)}_\alpha \sum_{\gamma\neq\alpha}
  \underline{\psi}_{\gamma},
\end{equation}
where $\underline{G}_\alpha^{(l)}=\mbox{}_\alpha \langle \widetilde{
n\nu l\lambda} |G_\alpha^{(l)}|\widetilde{n^{\prime}\nu^{\prime}
l^{\prime}{\lambda}^{\prime} }\rangle_\alpha.$ A unique solution exists if
and only if
\begin{equation}
\label{faddeevdet}
  \det \{ [ \underline{G}^{(l)}(z)]^{-1} - \underline{v}^{(s)} \} =0.
\end{equation}

The solution of the above matrix equation requires the determination of
the Coulomb--Sturmian representation of the three-body
Green's operator ${G}_\alpha^{(l)}$ of Eq.\ \eqref{channelgreen}.
Normally this task necessitates the solution of Faddeev-type integral
equations or  a triad of Lippmann--Schwinger integral equations \cite{glockle}.
However, the Hamiltonian $H_\alpha^{(l)}$ has the peculiar
property, that it supports bound state only in the subsystem $\alpha$, and
thus there is  only one kind of asymptotic channel, the $\alpha$ channel.
For such a system, if the asymptotics is treated properly,
one single Lippmann--Schwinger equation is sufficient for
an unique solution \cite{sandhas}.
Merkuriev proposed  an appropriate  Lippmann-Schwinger equation
which provides a unique solution for ${G}_\alpha^{(l)}$
\begin{equation}
\label{mls}
  G_\alpha^{(l)}(z)=G_\alpha^{as}(z) + G_\alpha^{as}(z) V^{as}_\alpha
  G_\alpha^{(l)}(z),
\end{equation}
where the three-body potential $V_{\alpha}^{as}$ is defined such that it
decays faster than the Coulomb potential
in all direction of the three-body configuration space.
The operators $G_\alpha^{as}$ and $V^{as}_\alpha$ are complicated three-body
operators, and to construct them,
Merkuriev used different approximation schemes in different 
regions of the configuration space.

It is important to realize that in our approach to get the solution only
matrix elements of ${G}_\alpha^{(l)}$ between finite number
of square integrable CS functions are needed.
That is the reason why the matrix elements of the channel
Green's operator  can be obtained as
\begin{equation}
\label{LSm}
  \underline{G}_\alpha^{(l)}(z)=\underline{\widetilde{G}}_\alpha (z) +
  \underline{\widetilde{G}}_\alpha(z) \underline{U}^\alpha
  \underline{G}_\alpha^{(l)}(z),
\end{equation}
where
\begin{equation}
\label{matrixelements}
  \underline{\widetilde{G}}_\alpha=\mbox{}_\alpha \langle \widetilde{
  n\nu l\lambda} |\widetilde{G}_\alpha |\widetilde{n^{\prime}\nu^{\prime}
  l^{\prime}{\lambda}^{\prime} }\rangle_\alpha, \quad
  \underline{U}^\alpha=\mbox{}_\alpha \langle {
  n\nu l\lambda} |U^\alpha |{n^{\prime}\nu^{\prime}
  l^{\prime}{\lambda}^{\prime} }\rangle_\alpha,
\end{equation}
with $\widetilde{G}_\alpha (z)=(z-\widetilde{H}_\alpha)^{-1}$
and $U^\alpha=H_\alpha^{(l)}-\widetilde{H}_\alpha$.
The channel-distorted long-range Hamiltonian
$\widetilde{H}_\alpha$ is defined as
\begin{equation}
\label{htilde}
 \widetilde{H}_\alpha=h_{x_\alpha}^C + h_{y_\alpha}^{(l)}, 
\end{equation}
where $h_{x_\alpha}^C=h_{x_\alpha}^0 +v_\alpha^C$ is the two-body Coulomb
Hamiltonian.

Since $H_\alpha^{(l)}$ does not generate rearrangement channels,
we should define $h_{y_\alpha}^{(l)}$ such that $\widetilde{H}_\alpha$
also preserve this property. If we are dealing with
repulsive Coulomb interactions, i.e.
$e_\alpha (e_\beta+e_\gamma) \ge 0$,
this requirement can be easily  fulfilled and $h_{y_\alpha}^{(l)}$ is 
defined by
\begin{equation}
\label{hyl}
  h_{y_\alpha}^{(l)} =  h_{y_\alpha}^C  = h_{y_\alpha}^{0} + 
  {e_\alpha (e_\beta+e_\gamma) }/{y_\alpha}
\end{equation}
and $U^\alpha$ takes the form
\begin{equation}
\label{ualpha}
  U^\alpha= v_\beta^{(l)}+v_\gamma^{(l)}-
  {e_\alpha (e_\beta+e_\gamma) }/{y_\alpha}.
\end{equation}
On the other hand, in the case of an attractive interaction,
like the Helium atom where $e_\alpha (e_\beta+e_\gamma) < 0$,
the attractive Coulomb tail of $h_{y_\alpha}^C$
generate infinitely many bound states which has to be
pushed  away from the spectrum. In practice this is achieved by
introducing a repulsive Gaussian term into \eqref{ualpha}
\begin{equation}
\label{ualpha2}
  U^\alpha= v_\beta^{(l)}+v_\gamma^{(l)}
  -\Lambda \exp (-\kappa y_{\alpha}^2)
  -{e_\alpha (e_\beta+e_\gamma) }/{y_\alpha},
\end{equation} 
where $\Lambda$ and  $\kappa$ are free parameters.
Now the \eqref{matrixelements} matrix elements of $U^\alpha$ can easily 
be calculated numerically applying  basis functions from the same
fragmentations on both sides.

We note here that an alternative and  mathematically more sound procedure
for removing the bound states generated
by the attractive Coulomb potential is being under development.
The basic idea of the new procedure is the application of projection
operators constructed as   convolution integrals of Green's operators.

\begin{figure}[tbp]
\begin{center}
\resizebox{10cm}{!}{
\includegraphics{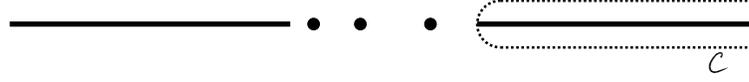}} 
\end{center}
\caption{Contour $C$ for the convolution integral of 
Eq.\ \eqref{contourint}. The contour $C$
encircles the (continuous) spectrum of
$g^{(l)}_{y_\alpha}$ and avoids the singularities of
$g^C_{x_\alpha}$. 
\label{fig:boundcont}}
\end{figure}

The most crucial point in the solution of Eq.\ \eqref{LSm}  is the
calculation of the matrix elements
$\underline{\widetilde{G}}_{\alpha}$.
The operator $\widetilde{G}_\alpha$
is a resolvent of the sum of two commuting Hamiltonians,
$ h^C_{x_\alpha}$ and $h_{y_\alpha}^{(l)}$ according to \eqref{htilde},
which act in different two-body Hilbert spaces.
As it was discussed in Section \ref{sec:convolution},
the three-body Green's operator
$\widetilde{G}_\alpha$ equates to
a convolution integral of two-body Green's operators, i.e.
\begin{equation}
\label{contourint}
  \widetilde{G}_\alpha (z)=
  \frac 1{2\pi \mathrm{i}}\oint_C
  dz^\prime \,g^C_{x_\alpha }(z-z^\prime)\;
  g_{y_\alpha}^{(l)}(z^\prime),
\end{equation}
where
$g^C_{x_\alpha}(z)=(z-h^C_{x_\alpha})^{-1}$  and
$g_{y_\alpha}^{(l)}(z)=(z-h_{y_\alpha}^{(l)})^{-1}$.
The contour $C$ should be taken such that it
encircles the continuous spectrum of $h_{y_\alpha}^{(l)}$
so that $g^C_{x_\alpha }$ is analytic on the domain encircled
by $C$. For bound state problems this requirement can easily
be fulfilled (see Fig.\ (\ref{fig:boundcont})).
The matrix elements $\underline{\widetilde{G}}_\alpha$ of 
Eq.\ \eqref{matrixelements} can be cast into
the convolution integral of the outer product of the corresponding matrices 
\begin{equation}
  \underline{\widetilde{G}}_\alpha (z)=
  \frac 1{2\pi \mathrm{i}}\oint_C
  dz^\prime \,  
  _\alpha \bra{\widetilde{nl}} g^C_{x_\alpha }(z-z^\prime) 
  \ket{\widetilde{n'l'}}_\alpha
   {} _{\alpha}\bra{\widetilde{\nu \lambda}} g_{y_\alpha}^{(l)}(z^\prime)
  \ket{\widetilde{\nu' \lambda'}}_\alpha  .
\end{equation}
The great advantage of using CS basis is that on this basis the matrix elements
of the two-body Green's operators 
are  given analytically  on the whole complex plane 
(see Section \ref{sec:coulomb}), thus the convolution
integral can be performed in practice. 

As an application,  the binding energy of the Helium atom  (with an infinitely
massive core) is determined as the three-body energy to be obtained 
from equation \eqref{faddeevdet}. The convergence of the binding energy
with respect to $N$, the number of basis states in the approximation of
Eq.\ \eqref{feqsapp}, is shown in Table \ref{conv_helium}. Excellent  
convergence is 
achieved  up to $6-7$ significant digits with  $N \sim 19$
applied for $n$ and $\nu$. In Table \ref{conv_helium}, for comparison,
we also quote the result
of a very accurate variational calculation.

\begin{table}
\begin{center}
\begin{tabular}{rcc}
$N$ & $l=0$ & $l=4$    \\
\hline \\
10 & 2.897586	& 2.903741  \\  
11 & 2.897591	& 2.903746  \\  
12 & 2.897593   & 2.903746  \\  
13 & 2.897593   & 2.903746  \\ 
14 & 2.897593   & 2.903746  \\  
15 & 2.897593   & 2.903746  \\  
16 & 2.897593   & 2.903746  \\ 
17 & 2.897593   & 2.903745  \\  
18 & 2.897593   & 2.903745  \\
19 & 2.897593   & 2.903745  \\
\\
\hline \\
\multicolumn{2}{l}{
Variational calculation in ref. \cite{helium_acc}}   & 2.903724376984\\ 
\end{tabular}
\end{center}
\caption{
Convergence of the binding energy of the Helium atom.
$N$ denotes the maximum value of  $n$ and $\nu$, i.e. the number  of employed
basis states. The quoted energy values are given 
in atomic units $(\hbar=m_e=e^2=1)$.
Angular momentum states have been taken into account up to $l=0$ and 
$l=4$. The values
$b=5$, $\kappa=2$, $\Lambda=5$ are chosen for the CS basis and 
the \eqref{ualpha2} repulsive Gaussian  parameters respectively.
The Merkuriev parameters of the  splitting function of Eq.\ \eqref{zeta}
are $x_0=0.7$, $y_0=15$ and $\nu=2.2$ respectively.  
\label{conv_helium}
}
\end{table}

\chapter*{Summary}
\lhead[\thepage]{}
\rhead[SUMMARY ]{\thepage}
\addcontentsline{toc}{chapter}{\numberline{}Summary}

In this work we have presented a rather general and easy-to-apply
method for discrete Hilbert space representation of quantum mechanical
Green's operators. We have shown that if in some discrete 
Hilbert space basis representation
the Hamiltonian takes an infinite symmetric tridiagonal, i.e. Jacobi-matrix
form  the corresponding Green's matrix can be calculated on the whole
complex energy plane by a continued fraction. The procedure necessitates only
the analytic calculation of the Hamiltonian matrix elements, which are used to
construct the coefficients of the continued fraction. This continued
fraction representation of the Green's operator was shown to be convergent for
the bound state energy region. The theory of analytic continuation of continued
fractions  was utilized to extend the representation to the whole complex
energy plane.  The presented method provides a simple, easily applicable and
analytically correct recipe for calculating discrete basis representation of
Green's operators. 

The general procedure was applied  to determine matrix representation of
specific Green's operators. The D-dimensional Coulomb Hamiltonian was shown to
possess a Jacobi-matrix structure on the Coulomb--Sturmian basis. The
tridiagonal matrix elements of the Hamiltonian were used to construct the
continued fraction representation of the Coulomb Green's operator. Numerical
tests on the convergence of the continued fraction and on the effect of
the analytic continuation   were presented on this example. Continued fraction
representation of relativistic Green's operators corresponding to the
Klein--Gordon and the second order Dirac equation could be determined because of
the Jacobi-matrix structure of the relativistic Hamiltonians on the relativistic
Coulomb--Sturmian basis. The relativistic energy spectra of the hydrogen-like
atoms are calculated as the poles of the relativistic Coulomb Green's operator
and demonstrate the high precession numerical accuracy of our continued fraction
representation. The method has been applied for the D-dimensional harmonic
oscillator, as well. As a non-trivial example the generalized Coulomb potential,
which is a member of the exactly solvable Natanzon confluent  potential class,
is considered. The radial Hamiltonian containing this potential exhibits a
tridiagonal form with analytically known matrix elements on the generalized
Coulomb--Sturmian basis, so our procedure is also applicable here.

Once the representation of the Green's operator in a discrete basis is available
we can proceed to solve few-body integral equations which provide the real test
field for our Green's operator in respect of its practical applicability and
importance. 

The continued fraction representation of the Coulomb--Sturmian space Coulomb
Green's operator is used for giving a unified solution of the two-body
Lippmann--Schwinger equation for the bound, resonant and scattering states.
The performance of this approach is illustrated by the detailed investigation of
a model nuclear potential describing the interaction of two $\alpha$ particles.
 
As a second application the Faddeev--Merkuriev equations are solved for an
atomic three-body bound state problem. The solution method requires the
evaluation of a three-body Green's operator, wich is done by performing a
convolution integral of two-body Coulomb Green's operators.
This convolution integral represents the real test for our Green's matrices
calculated by continued fractions over the whole complex plane.

We can conclude that our general, readily computable and numerically exact
continued fraction method for determining discrete Hilbert space representation
of Green's operators is turned out to be  valuable  in solving few-body
problems.

\chapter*{\"Osszefoglal\'as\\ 
{\normalfont\normalsize (The Hungarian summary of the thesis.)}}

\selectlanguage{magyar}
\rhead[\"OSSZEFOGLAL\'AS]{\thepage}
\addcontentsline{toc}{chapter}{\numberline{}\"Osszefoglal\'as}

\section*{Előzmények}

A bennünket körülvevő fizikai világ mikroszkopikus leírását két különböző
irányból kísérelhetjük meg.  Soktest vagy térelméleti módszereket
követve a sok, illetve  végtelen szabadsági fokkal rendelkező 
fizikai rendszereket mint statisztikus sokaságokat kezelhetjük. Ezzel szemben
a néhánytest fizika célja az olyan kevés szabadsági fokkal
rendelkező rendszerek vizsgálata, ahol az egymással kölcsönható objektumok
viselkedésének minél teljesebb fizikai megismerését még reális célként
fogalmazhatjuk meg.  
A kvantummechanikai néhánytest rendszerek elméleti tanulmányozása központi 
szerepet tölt be mind az atom-,
a mag- és a részecskefizika fejlődésében, hisz 
az alapvető természeti törvények
megismeréséhez nélkülözhetetlenek a néhánytest modellek. 

A kvantummechanikai néhánytest probléma alapvető egyenletei, mint 
például a
Lippmann--Schwinger- és a Fagyejev-egyenletek, általában 
integrálegyenletek. Néhánytest
fizikában az integrálegyenletek használata a differenciálegyenletek
helyett többek közt
azzal az előnnyel jár, hogy a megoldandó egyenletek automatikusan tartalmazzák
a  dinamika szempontjából alapvető határfeltételeket.
Éppen ez az
oka annak, hogy a bonyolult aszimptotikus határfeltételekkel 
jellemezhető szórási
problémák tanulmányozása 
során az integrálegyenletek jelentős előnyt élveznek.   
Azonban  a legtöbb gyakorlati alkalmazás esetében az integrálegyenleteket,
 azok
előnyös tulajdonságai  ellenére, mellőzik és helyettük
inkább a megfelelő Schrödinger-egyenlet
valamely alkalmas közelítését használják. 
Ennek oka abban keresendő, hogy az integrálegyenletek a Hamilton-operátor
helyett annak rezolvensét a Green-operátort tartalmazzák. A Green-operátor
meghatározása pedig jóval bonyolultabb feladat 
mint a megfelelő Hamilton-operátor közelítése.

Egy adott kvantummechanikai néhánytest rendszer 
teljes Green-ope\-rá\-to\-rá\-nak meghatározása
ekvivalens a probléma teljeskörű megoldásával, ugyanis 
a Green-operátor hordozza a fizikai rendszerről
nyerhető összes információt. Többek között a  rendszer energiasajátértékei,
kötött, rezonancia- és szórási állapotai, 
hullámfüggvénye, állapotsűrűsége valamint időbeli fejlődése határozható
meg a Green-operátor valamely reprezentációja ismeretében.

A legalapvetőbb néhánytest rendszerek, mint pl.\ 
a szabad részecske, a harmonikus
oszcillátor vagy a Coulomb-térben mozgó töltött részecske, Green-operátorának
valamely reprezentációja az irodalomból már ismert
 \cite{newton,taylor,ho_green,papp1}.
 Ezen operátorok számolását az
alkalmasan választott reprezentáció (momentum saját vektorok, harmonikus
oszcillátor függvények, Coulomb--Strum-függvények) valamint a matematika
speciális függvényeinek bizonyos feltételek melletti intenzív használata tette
lehetővé. 

A Green-operátorok analitikus ismerete képezi az alapját egy, az utóbbi időben
kifejlesztett kvantummechanikai közelítő módszernek is \cite{pse}.
A módszer során a 
Hamilton-operátor  aszimptotikusan meghatározó tagjainak
(kinetikus energia operátor,
hosszú hatótávolságú kölcsönhatások)  megfelelő Green-operátort egzaktul
analitikusan számolják, és csupán az aszimptotikusan irreleváns rövid
hatótávolságú potenciált közelítik egy szeparábilis kifejtéssel 
a Hilbert-tér egy csonkolt bázisán 
(a módszer ezért
kapta a PSE, Potential Separable Expansion elnevezést). A hosszú hatótávolságú
tagok egzakt kezelése biztosítja a megoldások aszimptotikusan korrekt 
voltát. A módszert sikeresen alkalmazták a kéttest rendszereket leíró
Lippmann--Schwinger- \cite{hopse,papp1,papp2,papp3},
valamint a háromtest rendszerek alapvető egyenleteit
jelentő Fagyejev-egyenletek megoldására \cite{pzwp}. 
A módszer alkalmazása során a Fagyejev-egyenletek esetében a legnagyobb
kihívást   egy háromtest Green-operátor
számolása jelenti,
melyet kéttest Green-operátorok konvolúciós integráljaként állítanak
elő.      

Mindezek miatt  állíthatjuk, hogy valamely
rendszer Green-operátorának analitikus előállítása  a
kvantummechanikai néhánytest problémák megoldása szempontjából
nagy jelentőséggel bír.

\section*{Eredm\'enyeim}

Doktori munkám során
kidolgoztam egy általános, egyszerűen megvalósítható, numerikusan stabil
analitikus módszert a kvantummechanikai Green-operátorok diszkrét 
Hilbert-térbeli reprezentációjának lánctörtes előállítására. 
A módszer alapját az a felismerés képezi, hogy 
amennyiben a rendszer Hamilton-operátora valamely
diszkrét Hilbert-tér bázison végtelen szimmetrikus tridiagonális mátrix, azaz
Jacobi-mátrix, akkor a megfelelő Green-mátrix egy tetszőleges véges
almátrixa  a tridiagonális Hamilton mátrixelemekből és egy lánctörtből
számolható. A lánctört együtthatóit szintén a Jacobi-mátrixelemekből
nyerhetjük \cite{jacobi_jmp}.

Továbbá a kötöttenergia-tartományban a  Green-operátor mátrixelemei
egy háromtagú rekurziós
reláció minimális megoldásaiként adódnak, ami Pincherle tétele
alapján biztosítja 
a Green-operátor lánctörtes előállításának konvergens voltát
ezen a tartományon.

A komplex energia-sík szórási és rezonancia tartományaiban a Green-operátor
lánctörtes reprezentációját a kötött tartományon konvergens  lánctört
analitikus folytatásával adtam meg. Erre azért volt szükség, mert
ezen energiaértékeknél
a Green-mátrixelemek már nem minimális megoldásai a rekurziónak, s így az
eredeti lánctört nem, hanem csak annak analitikus folytatása konvergens.
Az analitikusan folytatott lánctörttel a Green-operátor teljes
komplex sikon vett analitikus reprezentációját állítottam elő.
  
Numerikus szempontból fontos 
felismerést jelentett az, hogy az 
analitikusan folytatott lánctört konvergenciájának sebességét és
numerikus stabilitását a lánctörtek
Bauer-Muir transzformációjának \cite{lorentzen}
többszörös alkalmazása jelentősen javítja.

Felhasználva, hogy a
megfelelő bázison az alábbi rendszerek Hamilton
operátora tridiagonális szerkezetű, a kifejlesztett általános
módszert alkalmaztam a következő
Green-operátorok  diszkrét bázison történő 
lánctörtes reprezentációjának előállítására:

\begin{itemize}
\item
D-dimenziós Coulomb-probléma Green-operátora a Coulomb--Sturm-bázison,

\item
 D-dimenziós harmonikus oszcillátor Green-operátora 
a harmonikus oszcillátor függvények bázisán,

\item
Általánosított Coulomb-potenciál Green-operátora az általánosított
Coulom--Sturm-bázison,

\item
Relativisztikus Coulomb--Green-operátorok, azaz a Klein--Gordon- 
és a másodrendű
Dirac-egyenlet, a relativisztikus Coulomb--Sturm-bázison.

\end{itemize}

Az analitikusan számolt lánctörtes Green-operátorokat 
a két- illetve háromtest
problémák integrálegyenleteinek megoldására szolgáló
kvantummechanikai közelítő
módszerekben  teszteltem, ezáltal  bizonyítva
a lánctörtes Green-operátorok gyakorlati alkalmazhatóságát:

\begin{itemize}

\item
A Coulomb--Green-operátor Coulomb--Sturm reprezentációjának lánctörtes
előállítását használtam fel a Coulomb--Sturm-bázisú PSE módszer alkalmazása 
során egy, az alfa
részecskék kölcsönhatását leíró magfizikai modellpotenciált tartalmazó  
kéttest Lippmann--Schwinger-integrálegyenlet megoldására \cite{continued_prc}. 

\item
Egy, a  Fagyejev--Merkuriev háromtest integrálegyenletek megoldására szolgáló
közelítő módszerben alkalmaztam a lánctörtes Coulomb--Green-operátort
egy atomfizikai háromtest probléma kötött állapoti megoldása során.

\end{itemize}

Munkám során a matematikai fizika és a néhánytest kvantummechanika
módszereit alkalmaztam, és eredményeim a néhánytest módszerekkel leírható
atom-,  mag- és részecskefizika területén alkalmazhatóak. 
Lehetőséget látok arra
is, hogy a tridiagonális mátrixok és a lánctörtek 
használatát jelentő alapötlet a
fizika más területein is sikerrel alkalmazható legyen.

\selectlanguage{english}

\end{document}